\documentclass[12pt]{article}
\pdfoutput=1
\usepackage{amssymb,amsmath,dsfont}
\usepackage{slashed}
\usepackage{epsfig}
\usepackage{epstopdf}
\usepackage{latexsym}
\usepackage{graphicx}
\usepackage{booktabs}
\usepackage{bbm}
\usepackage{cancel}
\usepackage[utf8]{inputenc}
\usepackage[babel]{csquotes}
\usepackage{enumitem}
\usepackage[numbers,compress]{natbib}
\usepackage[T1]{fontenc}
\usepackage[margin=20pt,small]{caption}
\usepackage{subcaption}
\usepackage[toc]{appendix}
\usepackage{color}
\usepackage{datetime}
\usepackage[
      colorlinks=false,
      linkcolor=darkblue,  
      urlcolor=blue,    
      filecolor=blue,     
      citecolor=red,
linktocpage=true,
      pdfstartview=FitV,
      bookmarksopen=true,
	  hidelinks
      ]{hyperref}
\numberwithin{equation}{section}

\ifpdf
\fi
\newcommand*{\boxedcolor}{red}
\makeatletter
\renewcommand{\boxed}[1]{\textcolor{\boxedcolor}{%
  \fbox{\normalcolor\m@th$\displaystyle#1$}}}
\makeatother

\definecolor{cardinal}{rgb}{0.6,0,0}
\definecolor{darkgreen}{rgb}{0,0.5,0}
\definecolor{golden}{rgb}{0.92, 0.7, 0}
\definecolor{midnight}{rgb}{0, 0, 0.5}
\definecolor{darkblue}{rgb}{0.2, 0, 0.8}

\def\Re{{\rm Re}} \def\Im{{\rm Im}}

\DeclareMathOperator{\bZ}{\mathbb{Z}}

\def\Tr{{\rm Tr}\,}

\newcommand{\bigzero}{\mbox{\normalfont\Large\bfseries 0}}
\newcommand{\rvline}{\hspace*{-\arraycolsep}\vline\hspace*{-\arraycolsep}}

\def\sl{\frak{sl}}

\newcommand{\be}{\begin{equation}}
\newcommand{\ee}{\end{equation}}
\newcommand{\bea}{\begin{eqnarray}}
\newcommand{\eea}{\end{eqnarray}}

\begin{document}  

\begin{titlepage}
 
\medskip
\begin{center} 
{\Large \bf 3d Abelian Gauge Theories at the Boundary}

\bigskip
\bigskip
\bigskip

{\bf Lorenzo Di Pietro$^1$, Davide Gaiotto$^1$, Edoardo Lauria$^{2}$ and Jingxiang Wu$^1$\\ }
\bigskip
\bigskip
${}^{1}$
Perimeter Institute for Theoretical Physics, \\
31 Caroline St N, Waterloo, ON N2L 2Y5, Canada
\vskip 5mm
${}^{2}$
Centre for Particle Theory, Department of Mathematical Sciences\\Durham University, DH1 3LE, UK
\vskip 5mm

\texttt{ldipietro@perimeterinstitute.ca,~dgaiotto@perimeterinstitute.ca,\\
~edoardo.lauria@durham.ac.uk,~jingxiang.wu@perimeterinstitute.ca} \\
\end{center}

\bigskip
\bigskip

\begin{abstract}

\noindent  
\end{abstract}
A four-dimensional Abelian gauge field can be coupled to a 3d CFT with a $U(1)$ symmetry living on a boundary. This coupling gives rise to a continuous family of boundary conformal field theories (BCFT) parametrized by the gauge coupling $\tau$ in the upper-half plane and by the choice of the CFT in the decoupling limit $\tau \to \infty$. Upon performing an $SL(2,\mathbb{Z})$ transformation in the bulk and going to the decoupling limit in the new frame, one finds a different 3d CFT on the boundary, related to the original one by Witten's $SL(2, \mathbb{Z})$ action \cite{Witten:2003ya}. In particular the cusps on the real $\tau$ axis correspond to the 3d gauging of the original CFT. We study general properties of this BCFT. We show how to express bulk one and two-point functions, and the hemisphere free-energy, in terms of the two-point functions of the boundary electric and magnetic currents. We then consider the case in which the 3d CFT is one Dirac fermion. Thanks to 3d dualities this BCFT is mapped to itself by a bulk $S$ transformation, and it also admits a decoupling limit which gives the $O(2)$ model on the boundary. We compute scaling dimensions of boundary operators and the hemisphere free-energy up to two loops. Using an $S$-duality improved ansatz, we extrapolate the perturbative results and find good approximations to the observables of the $O(2)$ model. We also consider examples with other theories on the boundary, such as large-$N_f$ Dirac fermions --for which the extrapolation to strong coupling can be done exactly order-by-order in $1/N_f$-- and a free complex scalar.

\noindent

\end{titlepage}


\setcounter{tocdepth}{2}

\tableofcontents

\section{Introduction}
\label{sec:intro}

The objective of this paper is to study conformal invariant boundary conditions for free Abelian  
gauge theory in four-dimensions. A striking property of these BCFTs is that they are typically well-defined 
on some open patch in the space of the four-dimensional gauge coupling. 

The simplest way to produce such boundary conditions is to couple the four-dimensional gauge fields 
to a three-dimensional CFT with a $U(1)$ global symmetry. This is sometimes called a ``modified Neumann'' 
boundary condition \cite{Gaiotto:2008sa}. Assuming that certain mild conditions are satisfied, one obtains a BCFT which is well-defined as long as the four-dimensional gauge coupling is sufficiently small \cite{Teber:2012de, Kotikov:2013kcl, Teber:2014ita, Teber:2016unz, Kotikov:2016yrn, Herzog:2017xha,Dudal:2018pta}. 
The conformal data of the BCFT can be computed from the data of the original CFT by perturbation theory  
in the four-dimensional gauge coupling. 

Conversely, there is a general expectation that any BCFT $B$ defined at arbitrarily small 4d gauge coupling
will be either a Dirichlet boundary condition or a modified Neumann boundary condition associated to some 3d CFT
$T_\infty[B]$ with a $U(1)$ symmetry. Because of electric-magnetic duality, the same statement applies to any other ``cusp'' $C$ in the space of the complexified gauge coupling, where some dual description of the four-dimensional gauge field becomes arbitrarily weakly coupled. If the BCFT $B$ is defined around the cusp $C$, we can associate to it another 3d CFT $T_C[B]$, which is obtained from $T_\infty[B]$ by applying the $SL(2,\mathbb{Z})$ transformation \cite{Witten:2003ya} that maps the cusp at infinity to $C$. Therefore, the theories living at the other cusps can be thought of as 3d Abelian gauge theories obtained by gauging the $U(1)$ global symmetry of $T_\infty[B]$.

In the absence of phase transitions, a given BCFT $B$ can be defined on the whole space of 4d gauge couplings 
and is thus associated to an infinite family $T_*[B]$ of 3d CFTs. The conformal data of the BCFT will admit a similar collection of perturbative expansions in the neighbourhood of each cusp. 

In the first part of this paper we study general properties of this family of BCFT's. A universal feature is the presence in the spectrum of boundary operators of two conserved $U(1)$ currents, the electric and the magnetic currents, that arise as a consequence of the electric and magnetic one-form symmetries in the bulk \cite{Gaiotto:2014kfa}. The endpoints of bulk line operators carry charge under this $U(1)\times U(1)$ symmetry, while all the local boundary operators are neutral. By matching the bulk and boundary OPE expansions of correlators of the bulk field strength, we show that several BCFT observables --including non-local ones such as the free-energy on a hemisphere background-- can be obtained in terms of the coefficients $c_{ij}$ in the two-point correlators of these currents, and of the coefficient $C_{\hat{D}}$ of the two-point function of the displacement operator. The latter relations hold for any $\tau$, provided $B$ exists. We also show that the leading perturbative corrections to $c_{ij}$ and $C_{\hat{D}}$ around a cusp are captured universally in terms of the two-point function of the $U(1)$ current of the 3d CFT living at the cusp, in the decoupling limit.

In the second part of this paper we turn these abstract considerations into a very concrete computational strategy: if some $T_C$ is simple enough for perturbative computations to be feasible, we may study the properties of other $T_*$ theories by re-summing the perturbation theory. If we happen to know, or conjecture, that 
there are two cusps $C$ and $C'$ such that $T_C$ and $T_{C'}$ are both simple, 
we may be able to implement an enhanced re-summation which uses both piece of data to 
predict the properties of the other $T_*$ theories. 

This approach gives a new approximation scheme, orthogonal to previously known perturbative approaches to 3d Abelian gauge theories such as the $\epsilon$-expansion \cite{DiPietro:2015taa, Giombi:2015haa, Chester:2015wao, Janssen:2017eeu, DiPietro:2017kcd, DiPietro:2017vsp, Ji:2018emi, Zerf:2018csr} or the large-$N$ expansion (see e.g. \cite{Giombi:2016fct, Chester:2017vdh, Gracey:2018fwq, Benvenuti:2018cwd,Benvenuti:2019ujm,2018arXiv181202720B,PhysRevB.98.035137} for recent results and the review \cite{Gracey:2018ame}). We will apply this strategy to a very nice boundary condition for a $U(1)$ gauge theory, which is conjecturally associated to a free Dirac fermion at two distinct cusps 
and to the $O(2)$ model at two other cusps \cite{Seiberg:2016gmd, Metlitski:2015eka, Wang:2015qmt, Hsiao:2017lch, Hsiao:2018fsc}. The fact that these theories appear at the cusps can be seen as a consequence of the recently discovered 3d dualities \cite{Seiberg:2016gmd, Aharony:2015mjs, Karch:2016sxi}, and it entails the existence of a $\mathbb{Z}_2$ action on $\tau$ that leaves $B(\tau,\bar{\tau})$ invariant. We will do a two-loop calculation at the free-fermion cusp and then extrapolate to the $O(2)$ cusp, finding good agreement with the known data of the $O(2)$ model. 

We also consider other applications: Taking the boundary degrees of freedom to be an even number $2N_f$ of free Dirac fermions, setting the gauge coupling to $g^2 = \lambda / N_f$ and taking $N_f$ to infinity with $\lambda$ fixed, we argue that the theory admits a $1/N_f$-expansion, which interpolates between the free theory at $\lambda = 0$ and large-$N_f$ QED$_3$ at $\lambda = \infty$. The exact $\lambda$ dependence can be easily obtained order-by-order in the $1/N_f$ expansion. Applying the general strategy to compute the hemisphere partition function to this case, and taking the limit $\lambda \to \infty$, we obtain the $1/N_f$ correction to the sphere partition function of large-$N_f$ QED$_3$.  Another example with a $\mathbb{Z}_2$ duality acting on $\tau$ is conjecturally obtained in the case where the theory on the boundary is a free complex scalar, or equivalently the $U(1)$ Gross-Neveu model \cite{Rosenstein:1990nm, ZinnJustin:1991yn}. We consider perturbation theory around the free-scalar cusp, and show the existence of a stable fixed point for the classically marginal sextic coupling on the boundary at large $\tau$. We also discuss an example with two bulk gauge fields coupled to two distinct Dirac fermions on the boundary. We show how to obtain QED$_3$ with 2 fermionic flavors starting with this setup, using the extended electric-magnetic duality group $Sp(4,\mathbb{Z})$ that acts on the two bulk gauge fields. 

\subsection{Structure of the paper} 
We start in section \ref{sec:mixedDef} by reviewing the non-interacting boundary conditions for a Maxwell field in four dimensions. We then define the family of interacting boundary conditions $B(\tau,\bar{\tau})$. We derive the general relations that we described above for the bulk two- and three-point functions of the field strength, and obtain the leading corrections in perturbation theory around the cusps in the $\tau$ plane. In section \ref{sec:FreeEnergy} we obtain similar results for a different observable, the hemisphere partition function of $B(\tau,\bar{\tau})$. In particular we show how to recover the $S^3$ partition function for the 3d CFTs in the decoupling limit. In section \ref{sec:MinPhTr} we put this machinery at work in the example of the boundary condition defined by the $O(2)$ model / a free Dirac fermion. Section \ref{sec:exa} contains the other applications that we consider: large-$N_f$ fermions, a complex scalar, and two bulk gauge fields coupled to two Dirac fermions. We conclude in section \ref{sec:conc} by discussing some future directions. Several appendices include the details of calculations, and some supplementary material, e.g. a calculation of the anomalous dimension of the boundary stress-tensor using multiplet recombination in appendix \ref{app:fakestress}, and an explanation of the technique that we used to evaluate the two-loop integrals in appendix \ref{app:FeynInt}.

\section{Boundary Conditions for 4d Abelian Gauge Field}
\label{sec:mixedDef}

\subsection{Generalities}

Boundary Conformal Field Theories for a free $d$-dimensional bulk quantum field theory are interesting theoretical objects.
On one hand, the correlation functions of bulk local operators are controlled by the free equations of motion.
In particular, they are fully determined by their behaviour near the boundary, which is encoded in some very simple 
bulk-to-boundary OPE for the bulk free fields. 

The free bulk-to-boundary OPE essentially identifies some special boundary local operators 
as the boundary values of the bulk free fields and their normal derivatives. The correlation functions of these 
boundary operators determine all correlation functions of bulk operators. These boundary correlation functions, though,
can in principle be as complicated as those of any CFT in $(d-1)$ dimensions. 

The case of four-dimensional free Abelian gauge theory (with compact gauge group) 
is particularly interesting because the bulk theory has an exactly marginal gauge coupling.\footnote{If the gauge group is compact, say $U(1)$, the gauge field has an intrinsic 
normalization and thus the coefficient in front of the bulk Lagrangian is canonically defined 
even if the bulk theory is free. Local interactions between the gauge fields 
and any other degrees of freedom localized in non-zero co-dimension obviously cannot renormalize
the bulk gauge coupling. Furthermore, the strength of the interactions between the gauge fields 
and such other degrees of freedom is controlled by the bulk gauge coupling 
and by quantized gauge charges and thus cannot get renormalized. The only possible beta functions 
involve gauge-invariant boundary local operators. This fact is often obfuscated in 
perturbative treatments  and then proven with the help of Ward identities, in a manner analogous to the non-renormalization of gauge charges in QED \cite{Teber:2012de, Kotikov:2013kcl, Teber:2014ita, Teber:2016unz, Kotikov:2016yrn, Herzog:2017xha,Dudal:2018pta}.}
Furthermore, a BCFT defined for some value of the bulk gauge coupling can typically be 
deformed to a BCFT defined at a neighbouring value of the bulk gauge coupling by conformal perturbation theory 
in the gauge coupling. The leading order obstruction is the presence of marginal boundary operators 
in the bulk-to-boundary OPE of the bulk Lagrangian operators $F^2$ and $F \wedge F$, 
which can lead to a logarithmic divergence as the bulk perturbation approaches the boundary. 
Generically, no such operators will be present and the BCFT can be deformed.

In this section, we will discuss the properties of some standard BCFT's 
which can be defined in an arbitrarily weakly-coupled gauge theory, starting with 
free boundary conditions and then including interacting degrees of freedom
at the boundary. On general grounds, we expect that any BCFT
which can be defined at arbitrarily weak coupling will take this form.

\subsection{Free Boundary Conditions and $SL(2,\mathbb{Z})$ Action}
\label{sec:Free}

Consider a $U(1)$ gauge field $A_\mu$ on $\mathbb{R}^3 \times \mathbb{R}_+$. We adopt Euclidean signature, and use coordinates $x = (\vec{x},y)$ where $x^4\equiv y \geq 0$ is the coordinate on $\mathbb{R}_+$, and $\vec{x}$ are the coordinates on $\mathbb{R}^3$. We denote the components of $x$ as $x^\mu$, $\mu = 1,2,3,4$, and those of $\vec{x}$ as $x^a$, $a=1,2,3$. The field strength is $F_{\mu\nu} = \partial_\mu A_\nu - \partial_\nu A_\mu$, its Hodge dual is $\tilde{F}_{\mu\nu} = \tfrac{1}{2}\epsilon_{\mu\nu}^{~~\rho\sigma}F_{\rho\sigma}$ and the self-dual/anti-self-dual components are $F_{\mu\nu}^{\pm} = \tfrac{1}{2}(F_{\mu\nu} \pm \tilde{F}_{\mu\nu})$. They satisfy $\tfrac{1}{2}\epsilon_{\mu\nu}^{~~\rho\sigma} F^{\pm}_{\rho\sigma} = \pm F^{\pm}_{\mu\nu}$. 

In the absence of interactions with boundary modes, by varying the action
\begin{align}
S[A, \tau] & = \int_{y\geq 0} dy\,d^3\vec{x} \left(\frac{1}{4 g^2} F_{\mu\nu}F^{\mu\nu} + \frac{i \theta}{32 \pi^2}\epsilon_{\mu\nu\rho\sigma}F^{\mu\nu}F^{\rho\sigma}\right) \\
& = -\frac{i}{8\pi}\int_{y\geq 0} dy\,d^3\vec{x}\left(\tau F^-_{\mu\nu}F^{-\,\mu\nu} - \bar{\tau} F^+_{\mu\nu}F^{+\,\mu\nu}\right)
~,\label{eq:actau}
\end{align}
we find the bulk equation of motion $\frac{1}{g^2}\partial_\mu F^{\mu\nu} = 0$ and the boundary term
\begin{align}
\delta S_{\partial}  & = - \int_{y = 0} d^3\vec{x} \, \delta A^a \left(\frac{1}{g^2} F_{ya} + i \frac{\theta}{4 \pi^2}\tilde{F}_{y a} \right) \\ & = \frac{i}{2\pi} \int_{y = 0} d^3\vec{x} \, \delta A^a(\tau F^-_{ya} - \bar{\tau} F^+_{ya}) ~.\label{eq:bdtau}
\end{align}
Our convention for the orientation is $\epsilon_{abc y} = \epsilon_{abc}$.  In equations \eqref{eq:actau}-\eqref{eq:bdtau} we combined $g$ and $\theta$ in the complex coupling $\tau = \frac{\theta}{2\pi} + \frac{2\pi i }{g^2}$. From eq. \eqref{eq:bdtau} we see that the possible boundary conditions for the gauge field when no boundary modes are present are
\begin{itemize} 
\item{{Dirichlet:} $\delta A_a\vert_{y=0}=0$, which is equivalent to 
\begin{equation}
(F^-_{ya} -  F^+_{ya})\vert_{y=0} = - \tilde{F}_{ya}\vert_{y=0}=0~;\label{eq:Dbc}
\end{equation}}
\item{{Neumann:} 
\begin{equation}
(\tau F^-_{ya} - \bar{\tau} F^+_{ya})\vert_{y=0} = 0~.\label{eq:Nbc}
\end{equation}
Equivalently, introducing 
\begin{equation}
\gamma = \frac{\mathrm{Re}\tau}{\mathrm{Im}\tau}= \frac{\theta\,g^2}{4\pi^2}\in \mathbb{R}~,\label{eq:defgamma}
\end{equation} 
we can write this condition as
$
(F_{ya} +i \gamma \tilde{F}_{ya})\vert_{y=0}=0
$, in particular for $\gamma = 0$ it simplifies to the standard Neumann condition
$
F_{ya}\vert_{y=0}=0
$.
}
\end{itemize}
It is convenient to introduce the boundary currents  
\begin{equation}
\begin{array}{ll}
2 \pi i\hat{J}_a & = \tau F^-_{ya}(\vec{x},y=0) -\bar{\tau} F^+_{ya}(\vec{x},y=0) ~, \cr 2\pi i \hat{I}_a & = F^-_{ya}(\vec{x},y=0)-F^+_{ya}(\vec{x},y=0)~.
\end{array}\label{eq:IJ}
\end{equation}
in terms of which the Dirichlet condition is $\hat{I} = 0$, and the Neumann condition is $\hat{J} = 0$.

On $\mathbb{R}^4$ this theory enjoys an $SL(2,\mathbb{Z})$ duality group
\begin{equation}
\tau \to \tau' = \frac{a \tau + b}{c \tau + d}\,,\quad a,b,c,d \in \mathbb{Z}\,,~ad-bc = 1~.
\end{equation}
The duality group acts on the fields as 
\begin{equation}
\begin{array}{ll}
F^-_{\mu\nu} & \to F^{'-}_{\mu\nu}=( c \tau + d) F^-_{\mu\nu}~,\cr
 F^+_{\mu\nu} & \to F^{'+}_{\mu\nu} =(c \bar{\tau} + d) F^+_{\mu\nu}~.
\end{array}\label{eq:ftr}
\end{equation}
When the boundary is introduced, the group $SL(2,\mathbb{Z})$ also acts on the boundary conditions. From \eqref{eq:ftr} we see that the action on the boundary currents is
\begin{equation}
\begin{array}{ll}
\hat{J}_a & \to a \hat{J}_a + b \hat{I}_a~,\cr
\hat{I}_a & \to c \hat{J}_a + d \hat{I}_a~.
\end{array}\label{eq:IJtr}
\end{equation}

The Dirichlet and Neumann boundary conditions above are exchanged under the $S$ transformation $\tau\to-\frac{1}{\tau}$, i.e. electric-magnetic duality. Indeed, the $S$ transformation exchanges $\hat{J}$ and $\hat{I}$. 

However, comparing eq.s \eqref{eq:Dbc}-\eqref{eq:Nbc} and eq.s \eqref{eq:ftr}-\eqref{eq:IJtr} we see that the general $SL(2,\mathbb{Z})$ transformation does not act within the set of boundary conditions that we described above. This is because we assumed that no degrees of freedom are present on the boundary, while the generic $SL(2,\mathbb{Z})$ transformation requires the introduction of topological degrees of freedom on the boundary, namely 3d gauge-fields with Chern-Simons (CS) actions, coupled to the bulk gauge field through a topological $U(1)$ current \cite{Witten:2003ya, Gaiotto:2008ak, Kapustin:2009av}. Note that even in the presence of these topological degrees of freedom the theory is still free, because the action is quadratic. Taking this into account, one finds that the most general free boundary condition for the $U(1)$ gauge field is
\begin{equation}
p \hat{J}_a + q \hat{I}_a = 0~, \label{eq:Rbc}
\end{equation}
where $p,q \in \mathbb{Z}$. This set of boundary conditions is closed under the action \eqref{eq:IJtr} of $SL(2,\mathbb{Z})$. We will refer to this more general free boundary condition as ``$(p,q)$ boundary condition''. The $(0,1)$ and $(1,0)$ boundary conditions correspond to the Dirichlet and Neumann boundary conditions above, respectively.

When we impose the $(p,q)$ condition, the unconstrained components of the gauge fields give a current operator on the boundary
\begin{equation}
p^\prime \hat{J}_a + q^\prime \hat{I}_a  
\end{equation}
with $pq^\prime - p^\prime q =1$, whose correlators are just computed by Wick contraction, i.e. the boundary theory is a mean-field theory for this current. We can always shift $(p',q')$ by a multiple of $(p,q)$, and this gives rise to the same current thanks to the boundary condition.

Since the above boundary conditions preserve conformal symmetry, we can regard this system as a free boundary conformal field theory, and rephrase the boundary conditions in terms of a certain bulk-to-boundary OPE of the field strength $F_{\mu\nu}$. Using the equation of motion and the Bianchi identity one finds that the only primary boundary operators that can appear in the bulk-to-boundary OPE of $F_{\mu\nu}$ are conserved currents, see appendix \ref{app:detailsBulkDefOPE} for a derivation. The free boundary conditions described above correspond to having only one conserved current in this OPE, that can be identified with $p^\prime \hat{J}_a + q^\prime \hat{I}_a$. For instance, for the Dirichlet $(0,1)$ boundary condition 
\begin{equation}
F_{\mu\nu}(\vec{x}, y) \underset{y\to 0}{\sim} - g^2 \hat{J}^a(\vec{x})2\delta_{a[\mu }\delta_{\nu] y}+\dots~,\label{eq:freeOPE}
\end{equation}
where the dots denote subleading descendant terms, and the square brackets denote antisymmetrization. The general $(p,q)$ case can be obtained from the Dirichlet case by acting with an $SL(2,\mathbb{Z})$ transformation \eqref{eq:ftr}-\eqref{eq:IJtr}. 

\subsection{Two-point Function in the Free Theory}
\label{sec:FreeTwoP}

In this section we compute the two-point function $\langle F_{\mu\nu}(x_1) F_{\rho\sigma}(x_2) \rangle$ on $\mathbb{R}^3\times \mathbb{R}_+$ in the free theory. We use that the two-point function is a Green function, i.e. it satisfies the equations of motion
\begin{equation}
\frac{1}{g^2}\partial_\mu \langle F_{\mu\nu}(x_1) F_{\rho\sigma}(x_2) \rangle = (\delta_{\nu\sigma} \partial_\rho - \delta_{\nu\rho} \partial_\sigma) \delta^4(x_{12})~,
\end{equation}
and the Bianchi identity 
\begin{equation}
\epsilon_{\tau\lambda\mu\nu}\partial_\lambda \langle F_{\mu\nu}(x_1) F_{\rho\sigma}(x_2) \rangle = 0~.
\end{equation}
on $y \geq 0$, and it also satisfies the boundary conditions at $y=0$. We are denoting  $x_{12} \equiv x_1 - x_2$.

To start with, the Green function on $\mathbb{R}^4$ (i.e. without a boundary) is
\begin{align}\label{eq:noboundaryF}
\langle F_{\mu\nu}(x_1) F_{\rho\sigma}(x_2) \rangle_{\mathbb{R}^4} & = \frac{g^2}{\pi^2} G_{\mu\nu,\rho\sigma}(x_{12})~,\\
G_{\mu\nu,\rho\sigma}(x) & \equiv  \frac{I_{\mu\rho}(x)I_{\nu\sigma}(x)- I_{\nu\rho}(x)I_{\mu\sigma}(x) }{(x^2)^2}~,
\end{align}
where $I_{\mu\nu}(x)=\delta_{\mu\nu}- \frac{2x_\mu x_\nu}{x^2}$. Starting from \eqref{eq:noboundaryF} and using the method of images we can easily write down the two-point function in the presence of the boundary. The calculation is showed in the appendix \ref{app:Imag}. 

In the case $\gamma = 0$ we find
\begin{align}
\langle F_{\mu\nu}(x_1) F_{\rho\sigma}(x_2) \rangle_{\mathbb{R}^3 \times \mathbb{R}_+}  & = \frac{g^2}{\pi^2}\left[(1 - s \,v^4 )\, G_{\mu\nu,\rho\sigma}(x_{12}) + s \,v^4 \,H_{\mu\nu,\rho\sigma}(\vec{x}_{12},y_1,y_2)\right]~,\label{eq:NDIm}\\
H_{\mu\nu,\rho\sigma}(\vec{x}_{12},y_1,y_2) & \equiv  2\frac{1}{(x^2)^2}\left[X_{1\,\mu}X_{2\,\rho}I_{\nu\sigma}(x_{12}) + X_{1\,\nu}X_{2\,\sigma}I_{\mu\rho}(x_{12}) \right. \nonumber \\ & \left. - X_{1\,\mu}X_{2\,\sigma}I_{\nu\rho}(x_{12}) - X_{1\,\nu}X_{2\,\rho}I_{\mu\sigma}(x_{12})\right]~,
\end{align}
for Dirichlet ($s=1$) and Neumann ($s=-1$) conditions. Here $X_{i\,\mu}$ are the conformally covariant vectors \cite{McAvity:1995zd}
\begin{equation}
X_{i\,\mu} \equiv y_i \frac{v}{\xi} \partial_{i\,\mu} \xi = v\left(2 \,\frac{y_i \,s_i \,x_{12\,\mu}}{x_{12}^2}- n_\mu\right)~,\quad i =1,2~,\quad s_1 = - s_2 = 1~,
\end{equation}
and $\xi$ is the conformally invariant cross-ratio
\begin{equation}
\xi \equiv \frac{x_{12}^2}{4y_1y_2}\equiv \frac{v^2}{1-v^2}~.
\end{equation}

For the more general Neumann boundary condition with $\gamma \neq 0$ we find
\begin{align}
\label{eq:imagesFF}
\langle F_{\mu\nu}(x_1) F_{\rho\sigma}(x_2) \rangle_{\mathbb{R}^3 \times \mathbb{R}_+}  & = \frac{g^2}{\pi^2} \left[\left(\delta_{[\rho}^{\rho'}\delta_{\sigma]}^{\sigma'} + \,v^4\left( \frac{1-\gamma^2}{1+\gamma^2} \delta_{[\rho}^{\rho'}\delta_{\sigma]}^{\sigma'} - i \frac{\gamma}{1+\gamma^2} \epsilon_{\rho\sigma}^{~~~\rho'\sigma'}\right) \right)\, G_{\mu\nu,\rho'\sigma'}(x_{12})\right.\nonumber\\ & -\,v^4 \,\left.\left( \frac{1-\gamma^2}{1+\gamma^2} \delta_{[\rho}^{\rho'}\delta_{\sigma]}^{\sigma'} - i \frac{\gamma}{1+\gamma^2} \epsilon_{\rho\sigma}^{~~~\rho'\sigma'}\right) H_{\mu\nu,\rho'\sigma'}(\vec{x}_{12},y_1,y_2)\right]~.
\end{align}
Even though not manifest, it can be verified that Bose symmetry is satisfied in this expression. From now on we will drop the subscript $\mathbb{R}^3 \times \mathbb{R}_+$.

It is also useful to rewrite this two point function in terms of the selfdual/antiselfdual components. The selfdual/antiselfdual projectors are
\begin{equation}
P_{\mu\nu}^{\pm~\rho\sigma} = \tfrac 12 (\delta_{[\mu}^{\rho}\delta_{\nu]}^{\sigma} \pm \tfrac 12 \epsilon_{\mu\nu}^{~~~\rho\sigma})~.
\end{equation}
We introduce the following notation
\begin{align}
G^{\pm,\pm}_{\mu\nu,\rho\sigma} & \equiv P_{\mu\nu}^{\pm~\mu'\nu'}P_{\rho\sigma}^{\pm~\rho'\sigma'}G_{\mu'\nu',\rho'\sigma'}~,\\
G^{\pm,\mp}_{\mu\nu,\rho\sigma} & \equiv P_{\mu\nu}^{\pm~\mu'\nu'}P_{\rho\sigma}^{\mp~\rho'\sigma'}G_{\mu'\nu',\rho'\sigma'}~,
\end{align}
and similarly for the structure $H$. The following  identities hold
\begin{align}
G^{\pm,\pm} & = 0~, \\
G^{\pm,\mp} - H^{\pm,\mp} & = 0~.
\end{align}
Recalling the definition \eqref{eq:defgamma} of $\gamma$, we obtain
\begin{align}
\langle F_{\mu\nu}^+(x_1) F_{\rho\sigma}^+(x_2)\rangle & =  \frac{2}{\pi\,\mathrm{Im}\tau}\frac{\tau}{\bar{\tau}} \, v^4 H^{++}_{\mu\nu,\rho\sigma}(\vec{x}_{12},y_1,y_2)~,\\
\langle F_{\mu\nu}^-(x_1) F_{\rho\sigma}^-(x_2)\rangle & =\frac{2}{\pi\,\mathrm{Im}\tau}\frac{\bar{\tau}}{\tau} \, v^4 H^{--}_{\mu\nu,\rho\sigma}(\vec{x}_{12},y_1,y_2)~,\\
\langle F_{\mu\nu}^+(x_1) F_{\rho\sigma}^-(x_2)\rangle & = \frac{2}{\pi\,\mathrm{Im}\tau} G^{+-}_{\mu\nu,\rho\sigma}(x_{12})~,\\
\langle F_{\mu\nu}^-(x_1) F_{\rho\sigma}^+(x_2)\rangle & = \frac{2}{\pi\,\mathrm{Im}\tau}G^{-+}_{\mu\nu,\rho\sigma}(x_{12})~.
\end{align}
The result above is the field-strength two-point function in the free theory with Neumann boundary conditions. As we argued in section \ref{sec:Free}, the result for the $(p,q)$ boundary conditions \eqref{eq:Rbc} simply follows from an $SL(2,\mathbb{Z})$ transformation \eqref{eq:ftr}-\eqref{eq:IJtr}. As an example, for Dirichlet boundary conditions one finds
\begin{align}\label{FFDirich}
\langle F_{\mu\nu}^+(x_1) F_{\rho\sigma}^+(x_2)\rangle & =  \frac{2|\tau|^2}{\pi\,\mathrm{Im}\tau}{}{} \, v^4 H^{++}_{\mu\nu,\rho\sigma}(\vec{x}_{12},y_1,y_2)~,\\
\langle F_{\mu\nu}^-(x_1) F_{\rho\sigma}^-(x_2)\rangle & =\frac{2|\tau|^2}{\pi\,\mathrm{Im}\tau}{}{} \, v^4 H^{--}_{\mu\nu,\rho\sigma}(\vec{x}_{12},y_1,y_2)~,\\
\langle F_{\mu\nu}^+(x_1) F_{\rho\sigma}^-(x_2)\rangle & = \frac{2 |\tau|^2}{\pi\,\mathrm{Im}\tau} G^{+-}_{\mu\nu,\rho\sigma}(x_{12})~,\\
\langle F_{\mu\nu}^-(x_1) F_{\rho\sigma}^+(x_2)\rangle & = \frac{2|\tau|^2}{\pi\,\mathrm{Im}\tau}G^{-+}_{\mu\nu,\rho\sigma}(x_{12})~.
\end{align}

\subsection{Coupling to a CFT on the Boundary}
\label{sec:Interaction}

Consider now a 3d CFT living on the boundary at $y=0$. We assume that the CFT has a $U(1)$ global symmetry, with associated current $\hat{J}_{\text{CFT}\,a}$. We take the Neumann boundary condition for the gauge field, which corresponds to a mean-field current operator $\hat{I}_a$ on the boundary. The two sectors can be coupled in a natural way, simply by gauging the $U(1)$ symmetry via the $y\to 0$ limit of the bulk gauge field. This amounts to adding the boundary coupling
\begin{equation}
\int_{y=0} d^3 \vec{x} \, \hat{J}_{\text{CFT}}^a A_a + \text{seagulls}~,
\end{equation}
and restricting the spectrum of local boundary operators to the $U(1)$ invariant ones. Charged boundary operators can be made gauge-invariant by attaching to them bulk Wilson lines. Therefore, it still makes sense to consider them after the gauging, but as endpoints of line operators rather than as local boundary operators. 

The boundary coupling modifies the boundary condition of the gauge field to the ``modified Neumann'' condition
\begin{equation}\label{eq:modNeu}
\hat{J}_a \equiv \hat{J}_{\text{CFT}\,a} ~.
\end{equation}
Hence as a consequence of the interactions both $\hat{I}_a$ and $\hat{J}_a$ are nontrivial operators.

As we explained above $\tau$ is an exactly marginal coupling, but we should worry about quantum effects breaking the boundary conformal symmetry by generating beta functions for boundary interactions. If the original 3d CFT has no marginal operators, these boundary beta functions start at linear order in the coupling and can be cancelled order-by-order in perturbation theory by turning on extra boundary interactions of order $\tau^{-1}$.\footnote{E.g. if the theory on the boundary is a free scalar field, loop corrections can generate the operator $\phi^2$ on the boundary with coefficient $\sim\tau^{-1}\Lambda^2_{UV}$, where $\Lambda_{UV}$ is the cutoff, but the only implication of this term is that the tuning of $m^2$ needs to be adjusted at order $\tau^{-1}$.} Barring other non-perturbative 
phenomena such as the emergence of a condensate, we expect a BCFT to exist for sufficiently large $\tau$, with conformal data perturbatively close to that of the original CFT. We denote this BCFT with $B(\tau, \bar{\tau})$.

If the original 3d CFT has marginal operators the situation is more subtle: turning on boundary couplings $\hat{\lambda}$ will produce a 
beta function of order $\hat{\lambda}^2$ for the marginal operators. This may or not have the correct sign to cancel the 
$\tau^{-1}$ contributions. If it does not, we do not expect any unitary BCFT to exist, though one may be able to 
produce some non-unitary ``complex'' BCFT with complex couplings. 

Conversely, suppose that we are given a BCFT $B(\tau,\bar \tau)$ defined continuously for arbitrarily weak gauge coupling. 
If $B(\tau,\bar \tau)$ is an interacting boundary condition, we expect that if we take the gauge coupling to $0$
the properties of $B(\tau,\bar \tau)$ will approach those of a 3d CFT with a $U(1)$ global symmetry.

As we will discuss later in this section, the bulk correlation functions are determined by the boundary correlation functions of the two conserved boundary current $\hat{I}_a$ and $\hat{J}_a$ defined in eq. \eqref{eq:IJ}. Due to the boundary condition \eqref{eq:modNeu}, at weak coupling, $\hat{J}_a$ is inherited from the boundary degrees of freedom and the corresponding charge is carried by the endpoints of bulk Wilson lines ending at the boundary. On the other hand, $\hat{I}_a$ is analogue to the ``topological'' charge in three-dimensional $U(1)$ gauge theories and the corresponding charge is carried by the endpoints of bulk 't Hooft lines ending at the boundary. 

When the coupling is turned off, the conformal dimension of endpoints of 't Hooft lines blows up and the $\langle \hat{I}_a \hat{I}_a \rangle$ correlation functions go to zero. The $\hat{I}_a$ current decouples from the BCFT correlation functions as they collapse to the correlation functions of the underlying 3d CFT $T_{0,1}[B]$ (this is the CFT that we denoted with $T_\infty[B]$ in the introduction).

\subsection{Boundary Propagator of the Photon}

In order to compute corrections to boundary correlators and beta functions of boundary couplings in perturbation theory at large $\tau$, we need the propagator of the gauge field between two points on the boundary. Since we are perturbing around the decoupling limit, this can be readily obtained from the knowledge of the two-point function in the free theory \eqref{eq:imagesFF}. Recall from the discussion around eq. \eqref{eq:freeOPE} that in the free theory $F_{\mu\nu}$ has a non-singular bulk-to-boundary OPE. So the boundary two-point function of the operator $F_{ab}$ is obtained by specifying the indices to be parallel in eq. \eqref{eq:imagesFF}, and then taking the limit in which both insertion points approach the boundary. When taking this limit, we need to pay attention to possible contact terms that can arise due the following nascent delta-functions
\begin{equation}
\frac{y}{(y^2 + \vec{x}^2)^2}\underset{y\to 0}{\longrightarrow} \pi^2 \delta^3(\vec{x})~,
\end{equation}
and its derivatives. Even though usually we only compute correlators up to contact terms, these kind of contact terms in the two-point functions of 3d currents do actually contain physical information \cite{Closset:2012vp}. In this context, they encode the $\theta$-dependence of the boundary two-point function of $F_{ab}$. Relatedly, they are also needed to obtain the correct boundary propagator of the photon.

To obtain the $(ab,cd)$ components of the two-point function \eqref{eq:imagesFF} we need the components $(ab,cd)$ and $(y a^\prime, cd)$ of the structures $G$ and $H$. The structure $G$ gives
\begin{align}
\left(1 + \,v^4\left(\frac{1-\gamma^2}{1+\gamma^2} \right)\right)\, G_{ab,cd}(x_{12})& \underset{y_{1,2}\to 0}{\longrightarrow} \frac{2}{1+\gamma^2}\,G^{\rm 3d}_{ab,cd}(\vec{x}_{12})~,\\
- 2 v^4 \frac{\gamma}{1+\gamma^2}i\epsilon_{ab}^{~~y a'}\, G_{ya',cd}(x_{12}) &\underset{y_{1,2}\to 0}{\longrightarrow}  -\frac{2\gamma}{1+\gamma^2} i  \,\pi^2\epsilon_{ab [c}(\partial_{\vec{x}_{12}})_{d]}\delta^3(\vec{x}_{12})~.
\end{align}
Here $G^{\rm 3d}_{ab,cd}$ denotes the same structure as in eq. \eqref{eq:noboundaryF} with the replacement of $I_{\mu\nu}$ by the 3d analogue 
\begin{equation}\label{eq:I3d}
I^{\rm 3d}_{ab}(\vec{x}) \equiv \delta_{ab} -\frac{2 x^a x^b}{\vec{x}^2}~.
\end{equation} 
 On the other hand the only non-zero component of the structure $H$ in the limit $y_{1,2}\to 0$ is $H_{ya,yb}$, hence the $H$ structure completely drops in the calculation of the propagator. The result is
\begin{equation}
\langle F_{ab}(\vec{x}_1,0)  F_{cd}(\vec{x}_2,0)\rangle = \frac{g^2}{\pi^2}\left[\frac{2}{1+\gamma^2}\,G^{\rm 3d}_{ab,cd}(\vec{x}_{12})-\frac{2\gamma}{1+\gamma^2} i  \,\pi^2\epsilon_{ab [c}(\partial_{\vec{x}_{12}})_{d]}\delta^3(\vec{x}_{12})\right]~.\label{eq:twopbpos}
\end{equation}
 
It is convenient to go to momentum space, by applying a Fourier transform with respect to the boundary coordinates
\begin{equation}
\langle F_{ab}(\vec{x}_1,0)  F_{cd}(\vec{x}_2,0)\rangle \equiv \int \frac{d^3 \vec{p}}{(2\pi)^3} \langle F_{ab}(\vec{p},0)  F_{cd}(-\vec{p},0)\rangle e^{i \vec{p}\cdot \vec{x}_{12}}~.
\end{equation}
We obtain
\begin{equation}
\langle F_{ab}(\vec{p},0)  F_{cd}(-\vec{p},0)\rangle = \frac{2g^2}{1+\gamma^2}\left[|\vec{p}\,|\left( \frac{\delta_{a[c}p_{d]}p_b}{\vec{p\,}^2} -\frac{\delta_{b[c}p_{d]}p_a}{\vec{p\,}^2}\right) + \gamma \epsilon_{ab[c} p_{d]}\right]~.\label{eq:twopbmom}
\end{equation}
We can finally determine the propagator of the gauge field between two-points in the boundary by imposing that the exterior derivative reproduces the two-point function \eqref{eq:twopbmom}. The result is
\begin{equation}
\langle A_{a}(\vec{p},0)  A_{b}(-\vec{p},0)\rangle \equiv \Pi_{ab}(\vec{p}\,) = \frac{g^2}{1+\gamma^2}\left[\frac{\delta_{ab} -(1-\xi)\frac{p_ap_b}{\vec{p\,}^2}}{|\vec{p\,}|} + \gamma \epsilon_{abc} \frac{p^{c}}{\vec{p\,}^2}\right]~.\label{eq:prop}
\end{equation}
The parameter $\xi$ is not fixed by requiring consistency with eq. \eqref{eq:twopbmom}, and parametrizes a choice of gauge. From the structure of the propagator we see that the natural perturbative limit is $g^2 \to 0$ with $\gamma$ fixed, which means $\tau \to \infty$ with a fixed ratio $\gamma$ between the real and the imaginary part. Observables are expressed as a power series in $\frac{g^2}{1+\gamma^2}$ with coefficients that are themselves polynomials in $\gamma$, more precisely the coefficient of the  order $\mathcal{O}\left(\left(\frac{g^2}{1+\gamma^2}\right)^n\right)$ is a polynomial in $\gamma$ of degree $n$.

\subsubsection{Relations to Large-$k$ and Large-$N_f$ Perturbation Theories}\label{sec:largeNlargek}
Recall that a 3d Abelian gauge field $a$ with CS action $i \frac{k}{4\pi}\int a \wedge da$ has propagator (up to gauge redundancy)
\begin{equation}
\langle a_{a}( \vec{p})  a_{b}( -\vec{p})\rangle = \frac{2\pi}{k}\epsilon_{abc} \frac{p^{c}}{\vec{p\,}^2}~.
\end{equation}
We see that the contact term in eq. \eqref{eq:twopbmom} produced a term in the boundary propagator \eqref{eq:prop} that is identical to the CS one. In particular, from the perturbation theory that we will consider one can immediately recover results for large-$k$ perturbation theory in Abelian 3d gauge theories, simply by setting (recall that $\gamma = \frac{g^2 \theta}{4\pi^2}$)
\begin{align}
\left(\frac{g^2}{1+\gamma^2}\right)^n \gamma^m & \longrightarrow 0~,~~\text{if $m<n$} \\
\left(\frac{g^2}{1+\gamma^2}\right)^n \gamma^n & \longrightarrow \left(\frac{2\pi}{k}\right)^n~.
\end{align} 
Indeed, in the limit $g^2 \to \infty$ only the $\theta$-term is left in the bulk action, and the model that we are considering is equivalent to a CS theory on the boundary, with $k = \frac{\theta}{2\pi}$. The only role played by the bulk in this case is to allow generic real values of the CS coupling.

We can also compare to the limit of large number of matter flavors $N_f$, in which observables at the IR fixed point of 3d Abelian gauge theories can be computed perturbatively in $1/N_f$. In this regime, after resumming bubble diagrams, one finds the following ``effective'' propagator (again, up to gauge redundancy)
\begin{equation}
\langle a_{a}(\vec{p})  a_{b}(-\vec{p})\rangle \sim \frac{1}{N_f}\frac{\delta_{ab}}{|\vec{p}\,|}~.
\end{equation}
The proportionality constant depends on the details of the theory. The resulting ``non-local'' propagator has precisely the same form of the boundary propagator \eqref{eq:prop} in the case $\gamma = 0$.\footnote{The two types of non-locality have different physical origins, in our setup the non-locality on the boundary is due to the existence of the bulk, while in the large-$N_f$ limit it emerges due to the resummation of infinitely-many Feynman diagrams. The fact that the resulting two-point functions of the field strength have the same power of momentum is of course no surprise, because that is just fixed by the scaling dimension of conserved currents in 3d.} Hence, once again, the two types of perturbation theories inform each other, and results for one case can be applied in the other case as well. Compared to the large-$k$ perturbation theory, here additional care is needed, because the order at which we are computing a certain observable in the $1/N_f$-expansion does not coincide with the number of internal photon lines in the corresponding diagram, owing to the fact that diagrams with a larger number of internal photon lines can get an enhancement by a positive power of $N_f$ from loops of matter fields. Nevertheless, single diagrams computed in one context can be used in the other context, and we will see an application of this observation later. A generalization of the large-$N_f$ limit is obtained by taking both $N_f$ and $k$ large, with a fixed ratio, and was studied recently in \cite{Chester:2017vdh}. In this case one finds a propagator that contains both terms in eq. \eqref{eq:prop}, and the same comments about the relation of the two types of perturbation theory apply.

\subsection{Exploring Strong Coupling}\label{sec:Strong}

As the coupling is increased, the two currents $\hat{I}_a$ and $\hat{J}_a$ should be treated on an even footing. Indeed, they are rotated into each other by the  $SL(2,\mathbb{Z})$ group of electric-magnetic dualities of the bulk theory. Assuming no phase transitions, as we approach cusps $\tau \to -\frac{q}{p}$ where the dual gauge coupling becomes weak in some alternative duality frame, we expect dual statements to be true: the $p \hat{J} + q \hat{I}$ current should decouple from the BCFT correlation functions as they collapse to the correlation functions of a new 3d CFT $T_{p,q}[B]$, which gives the dual weakly coupled description of the original BFCT. 

Using the notion of duality walls \cite{Gaiotto:2008ak, Kapustin:2009av}, one can argue that $T_{p,q}$ should be obtained from $T_{0,1}$ by Witten's $SL(2,\mathbb{Z})$ action on 3d CFTs equipped with a $U(1)$ global symmetry \cite{Witten:2003ya}. This involves coupling $T_{0,1}$ to a certain collection of 3d Abelian gauge fields with appropriate Chern-Simons couplings. This statement requires some care and several caveats about the absence of phase transitions as we vary $\tau$. 

\begin{figure}
\centering
\includegraphics[clip, height=7cm]{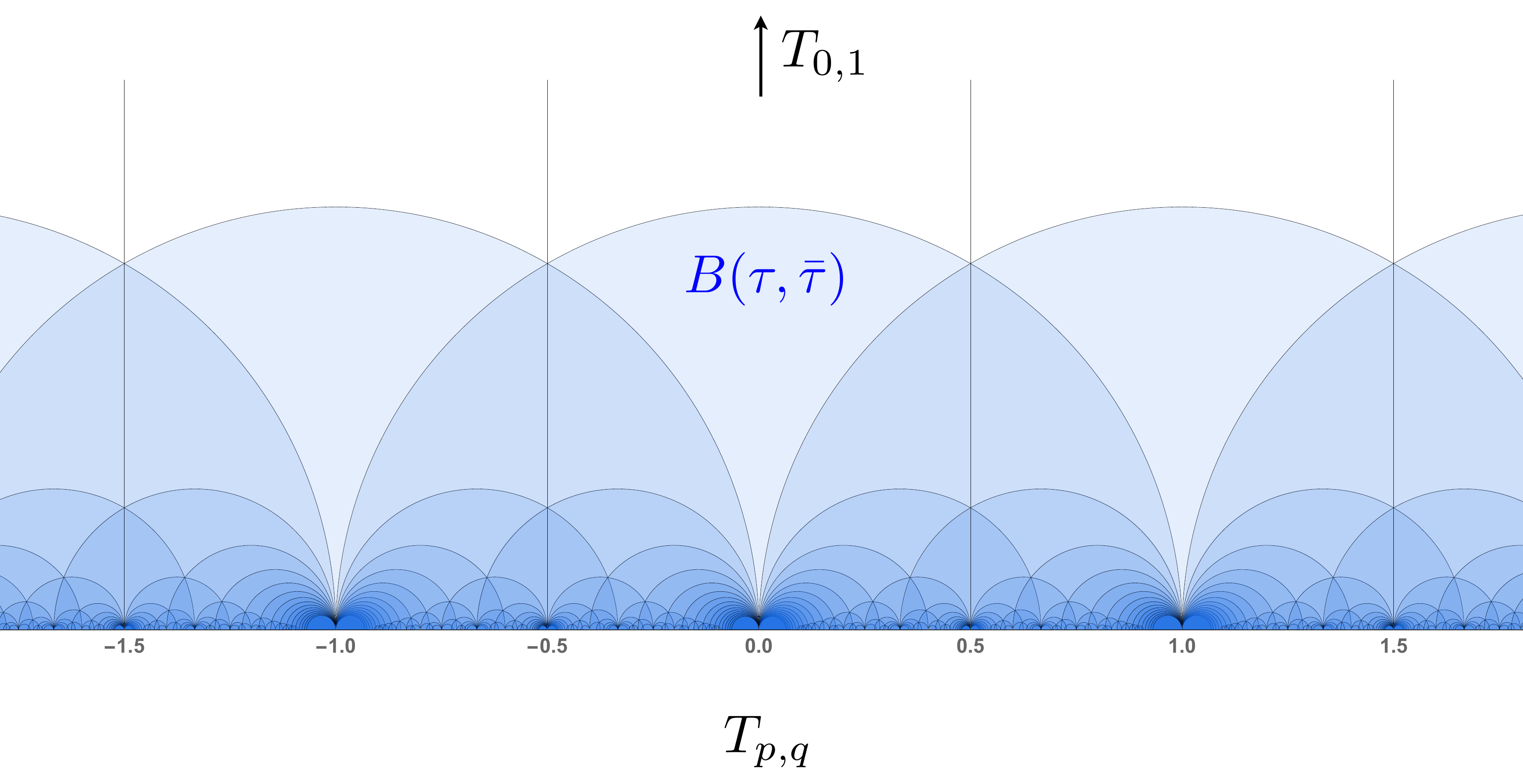}
\caption{The family of conformal boundary conditions $B(\tau,\bar{\tau})$ labeled by the variable $\tau$ in the upper-half plane and by a 3d CFT $T_{0,1}$ with $U(1)$ global symmetry. At the cusp at infinity the current $\hat{I}^a$ decouples and we are left with the local 3d theory $T_{0,1}$ on the boundary, with $U(1)$ current $\hat{J}^a$. Approaching this cusp from $T$-translations of the fundamental domain amounts to adding a CS contact term to the 3d theory, or equivalently to redefine the current $\hat{J}^a$ by multiples of the current $\hat{I}^a$ that is decoupling. This is the $T$ operation on $T_{0,1}$ in the sense of \cite{Witten:2003ya}. In the favorable situation in which no phase transitions occur, the BCFT continuously interpolate to the cusps at the rational points of the real axis $\tau = -q/p$, where again the bulk and the boundary decouple and we find new 3d CFTs $T_{p,q}$. These theories are obtained from $T_{0,1}$ with a more general $SL(2,\mathbb{Z})$ transformation, that involves coupling the original $U(1)$ global symmetry to a 3d dynamical gauge field. \label{fig:Tpq}}
\end{figure}

In an optimal situation where these phase transitions are absent, this picture implies that the data of $B(\tau,\bar \tau)$ will approach the data of an infinite collection of 3d CFTs $T_{p,q}$ as $\tau \to - \frac{q}{p}$, sitting in the same universality classes as certain 3d Abelian gauge theories coupled to $T_{0,1}$. This is depicted in fig. \ref{fig:Tpq}. If we ``integrate out'' the bulk and restrict our attention to the 3d boundary, what we just described can be stated as the existence of a family of non-local 3d conformal theories (i.e. with no stress-tensor in the spectrum) that continuously interpolate between different local 3d CFTs. More precisely, in the decoupling limit the 3d theory is a direct product of a 3d CFT and a non-local sector associated to the boundary condition of the free bulk field. This is reminiscent of the construction of \cite{Paulos:2015jfa, Behan:2017emf, Behan:2017dwr} in the context of the long-range Ising model.

\begin{figure}
\centering
\includegraphics[clip, height=7cm]{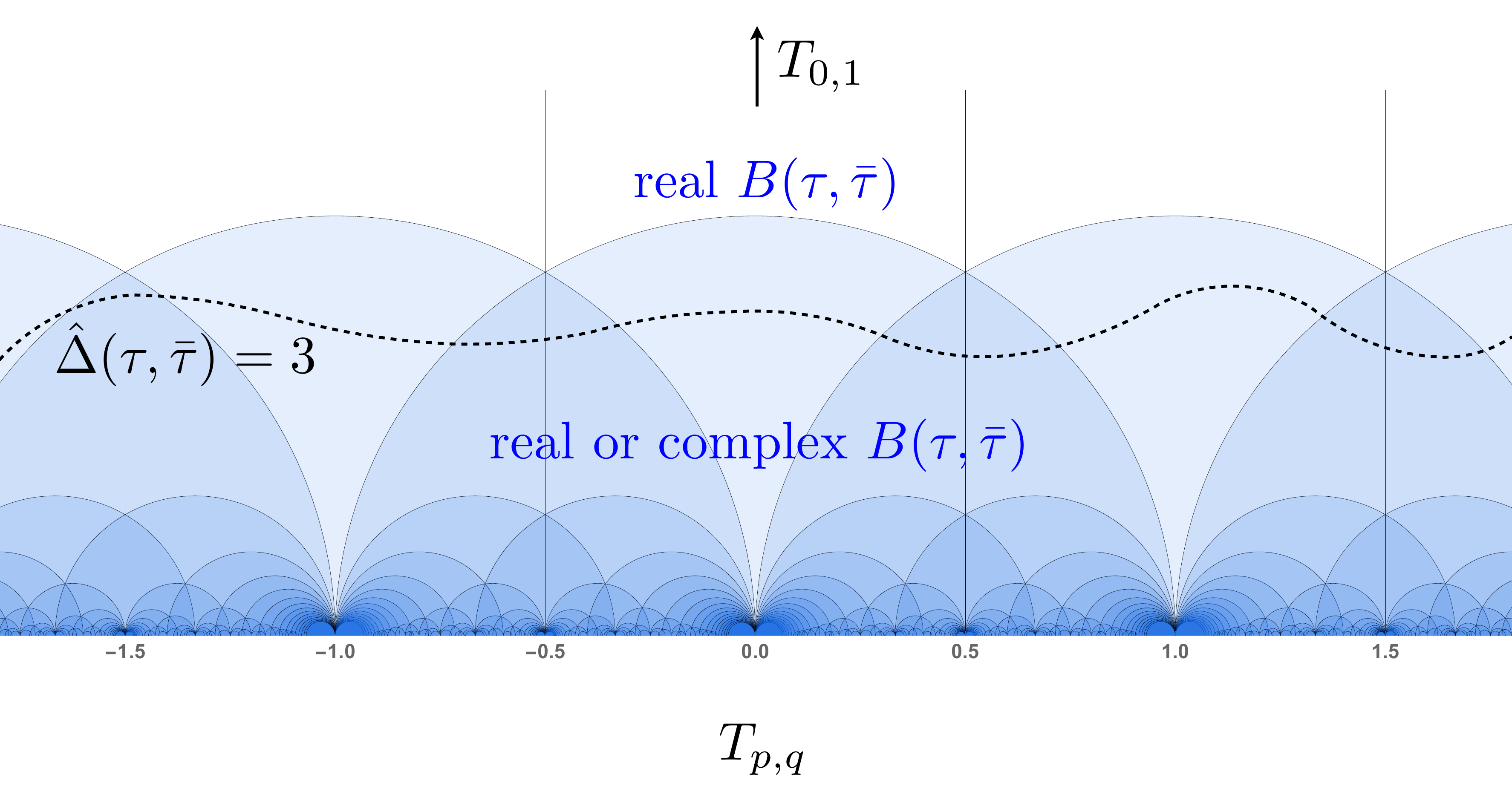}
\caption{A cartoon of a possible phase transition at strong coupling. A scalar boundary operator becomes marginal at a certain curve in the $\tau$ plane, i.e. setting $\hat{\Delta}(\tau,\bar{\tau}) = 3$ we find solutions in the upper-half plane. In conformal perturbation theory from a point on the curve, the beta function takes the form \eqref{eq:betamarg}. We might be unable to find real fixed points for the marginal coupling. In such a situation, $B(\tau,\bar{\tau})$ can only be defined as a complex BCFT. Assuming that we were able to define $B(\tau,\bar{\tau})$ as a real BCFT in perturbation theory around $\tau \to \infty$ by continuity such a real BCFT is ensured to exist in the full region above the wall, but we might be unable to continue it beyond the wall without introducing complex couplings (or breaking conformality). \label{fig:Tpqphasetransition}}
\end{figure}

Let us mention a possible mechanism for a phase transition. As we change continuously $\tau$ from the neighbourhood of the ``ungauged cusp'' $T_{0,1}$ towards the ``gauged cusps'' $T_{p,q}$, the dimension of boundary operators are nontrivial functions of $\tau$. A scalar boundary operator $\hat{O}$ might become marginal at a certain codimension 1 wall in the $\tau$-plane. This possibility is depicted in fig. \ref{fig:Tpqphasetransition}. 
In perturbation theory in the vicinity of the wall, we can repeat the logic that we used in the subsection \ref{sec:Interaction} when discussing perturbation theory around $T_{0,1}$ in presence of boundary marginal operators. Namely, the boundary marginal coupling $\hat{\lambda}$ will generically have a non-trivial beta function, which depends both on $\hat{\lambda}$ and $\tau$, and whose leading contributions are
\footnote{Note that this expression for the beta function is valid also in the decoupling limit $\tau\to\infty$. Indeed in that limit $b_{(F^-)^2, \hat{O}}\propto \tau^{-2}$ and $b_{(F^+)^2, \hat{O}}\propto\bar{\tau}^{-2}$, from which we recover that the leading contributions from the bulk gauge fields are of order $\tau^{-1}$ and $\bar{\tau}^{-1}$.}
\begin{equation}\label{eq:betamarg}
\beta_{\hat{\lambda}}(\tau,\bar{\tau},\hat{\lambda}) = b_{(F^-)^2, \hat{O}}\, \delta\tau + b_{(F^+)^2, \hat{O}}\, \delta{\bar{\tau}} + C_{\hat{O}\hat{O}\hat{O}} \hat{\lambda}^2 + \dots ~.
\end{equation}
Here we are perturbing around a point $\tau_0$ on the wall, the coefficient $b$'s and $C$ are (up to numerical factors) the bulk-to-boundary OPE coefficients \cite{Karch:2018uft}, and the OPE coefficient of the boundary conformal theory, respectively. These OPE coefficients are functions of $\tau_0$. Depending on $\tau_0$ and on the various OPE coefficients, setting $\beta_{\hat{\lambda}} =0$ one might or might not be able to find a real solution for $\hat{\lambda}$. If a real solution can be found perturbing away from the wall in a certain direction, by continuity $B(\tau,\bar{\tau})$ defines a real BCFT in a region of the $\tau$ plane on that side of the wall. Otherwise, on a side of the wall $B(\tau,\bar{\tau})$ exists only as a non-unitary ``complex'' BCFT.

\subsection{Two-point Function from the Boundary OPE}
\label{sec:twopointint}
In section \ref{sec:FreeTwoP} we computed the two-point function of the field strength in free theory using the method of images. We will now compute it in the more general case with interactions on the boundary. We will see that it can be fixed completely in terms of the coefficient of the two-point function of the boundary currents. The method that we will use is an explicit resummation of the bulk-to-boundary OPE.

As a consequence of the interaction, the bulk-to-boundary OPE of the field strength contains two independent primary boundary operators, both of them conserved currents, rather than just one like in the free case. The leading terms in this OPE are
\begin{equation}\label{dOPEF}
F_{\mu\nu}(\vec{x}, y) \underset{y\to 0}{\sim}\hat{V}_1^a(\vec{x})2\delta_{a[\mu }\delta_{\nu] y}- i\epsilon^{abc}\hat{V}_{2\,c}(\vec{x})\delta_{a[\mu}\delta_{\nu] b}+\dots~.
\end{equation}
The complete form of the above (including all descendants) can be found in \eqref{defectOPE}. The boundary currents $\hat{V_1}$ and $\hat{V}_2$ can be expressed in terms of $\hat{J}^a$ and $\hat{I}^a$ as follows
\begin{align}
\hat{V}_1^a & = -g^2\left(\hat{J}^a-\frac{\theta}{2\pi}\hat{I}^a\right)~,\\
\hat{V}_2^a & = -2\pi \hat{I}^a ~.
\end{align}
If the 3d CFT that the gauge field couples to has parity symmetry (i.e. symmetry under reflection of one of the coordinates) then the full boundary CFT $B(\tau,\bar{\tau})$ admits such a symmetry when restricted to $\theta = 0$. Under this symmetry $V_1$ transforms like an ordinary vector, while $V_2$ transforms like an axial vector. We can extend this symmetry to the more general case $\theta\neq 0$ by viewing it as a spurionic symmetry that flips the sign of $\theta$. 

Plugging the bulk-to-boundary OPE in the two-point function, one obtains the \emph{boundary channel} decomposition. In this case, since only two boundary primaries appear in the OPE, we can explicitly resum the contributions from all the descendants. The result can be written in terms of the structures defined in \eqref{eq:imagesFF}
\begin{align}\label{eq:FFfull}
\langle F_{\mu\nu}(x_1) F_{\rho\sigma}(x_2) \rangle  & = \left(\alpha_1\delta_{[\mu}^{\mu'}\delta_{\nu]}^{\nu'} - \,v^4\left(\alpha_2\,\delta_{[\mu}^{\mu'}\delta_{\nu]}^{\nu'} + i \frac{\alpha_3}{2} \,\epsilon_{\mu\nu}^{~~\mu'\nu'}\right) \right)\, G_{\mu'\nu',\rho\sigma}(x_{12})\nonumber\\& +\,v^4 \,\left(\alpha_2\,\delta_{[\mu}^{\mu'}\delta_{\nu]}^{\nu'} + i \frac{\alpha_3}{2} \epsilon_{\mu\nu}^{~~\mu'\nu'}\right) H_{\mu'\nu',\rho\sigma}(\vec{x}_{12},y_1,y_2)~.
\end{align}
with coefficients
\begin{align}\label{eq:alphadef}
\alpha_1= \tfrac{1}{2}{ \left( c_{11}(\tau,\bar{\tau})+ c_{22}(\tau,\bar{\tau})\right)},\quad\alpha_2= \tfrac{1}{2}\left( c_{11}(\tau,\bar{\tau})- c_{22}(\tau,\bar{\tau})\right),\quad 
\alpha_3= - c_{12}(\tau,\bar{\tau})~.
\end{align}
where 
\begin{equation}\label{eq:VVco}
 \langle \hat{V}_i^a(\vec{x}) \hat{V}_j^b(0) \rangle = c_{ij}(\tau,\bar{\tau})\frac{ I^{\rm 3d\,ab}(\vec{x})}{|\vec{x}|^{4}} + \text{contact term}~.
 \end{equation}
We see that eq. \eqref{eq:FFfull} is written explicitly in terms of data of the boundary conformal theory. For the time being we can ignore the contact term in the current two-point function because it cannot contribute to the two-point function of $F_{\mu\nu}$ at separated points.

To make the action of $SL(2,\mathbb{Z})$ more transparent we will also rewrite the above results in the selfdual/antiselfdual components. The bulk-to-boundary OPE takes the following form
\begin{equation}
F^{\pm}_{\mu\nu} (\vec{x}, y) \underset{y\to 0}{\sim} \hat{V}_{\pm\,a}(\vec{x}) 4 P_{\mu\nu}^{\pm~ay}+\dots~,
\end{equation}
where 
\begin{align}
\hat{V}_+ & = \frac{1}{2}(\hat{V}_1 -i \hat{V}_2) = -\frac{2\pi}{{\rm Im}\tau}(\hat{J} -\tau \hat{I})~,\\
\hat{V}_- & = \frac{1}{2}(\hat{V}_1 +i \hat{V}_2) = -\frac{2\pi}{{\rm Im}\tau}(\hat{J} -\bar{\tau} \hat{I})~.
\end{align}
An $SL(2,\mathbb{Z})$ transformation acts on $\hat{V}_\pm$ in the same way as it acts on $F^\pm$. In particular under an $S$ transformation $\hat{V}_+ \to \bar{\tau} \,\hat{V}_+$ and $\hat{V}_- \to \tau \,\hat{V}_-$. Using the structures introduced in section \ref{sec:FreeTwoP}, the result \eqref{eq:FFfull} can be rewritten in more compact form
\begin{align}\label{FFfulldual}
\langle F_{\mu\nu}^+(x_1) F_{\rho\sigma}^+(x_2)\rangle & =   (\alpha_2+i \alpha_3)\, v^4 H^{++}_{\mu\nu,\rho\sigma}(\vec{x}_{12},y_1,y_2)~,\\
\langle F_{\mu\nu}^-(x_1) F_{\rho\sigma}^-(x_2)\rangle & =(\alpha_2-i \alpha_3)\, v^4 H^{--}_{\mu\nu,\rho\sigma}(\vec{x}_{12},y_1,y_2)~,\\
\langle F_{\mu\nu}^+(x_1) F_{\rho\sigma}^-(x_2)\rangle & = \alpha_1 G^{+-}_{\mu\nu,\rho\sigma}(x_{12})~,\\
\langle F_{\mu\nu}^-(x_1) F_{\rho\sigma}^+(x_2)\rangle & = \alpha_1 G^{-+}_{\mu\nu,\rho\sigma}(x_{12})~.
\end{align}
Note that $\alpha_2\pm i \alpha_3 = 2 c_{\pm\pm}$ while $\alpha_1 = 2 c_{+-} = 2 c_{-+}$. In this basis the $SL(2,\mathbb{Z})$ action on the above two-point functions can be immediately read from \eqref{eq:ftr}. 

While in this subsection we discussed the two-point function of $F_{\mu\nu}$, clearly a similar computational strategy could be used for an arbitrary $n$-point function, therefore reducing any such bulk correlation functions to correlators of the boundary currents $\hat{J}$, $\hat{I}$. Of course generically for $n>2$ these correlation function are not just captured by the coefficients $c_{ij}$, because they are sensitive to the full spectrum of boundary operators entering in the OPE of the currents. 
 
\subsection{One-point Functions from the Bulk OPE}
\label{sec:resultsBoot}
When $x_{12}^2 \ll y^2$ we can expand the two-point function \eqref{eq:FFfull} in the bulk OPE limit, which is controlled by the OPE of free Maxwell theory
\begin{align}\label{FFope}
F_{\mu\nu}(x) F_{\rho\sigma}(0)\underset{x\to 0}{\sim} \frac{g^2}{\pi^2}G_{\mu\nu,\rho\sigma}(x)+\frac{1}{12}(\delta_{\mu\rho}\delta_{\nu\sigma}-\delta_{\nu\rho}\delta_{\mu\sigma})F^2(0)+\frac{1}{12}\epsilon_{\mu\nu\rho\sigma}F\tilde{F}\,(0)+\dots,
\end{align}
where we neglected spinning bulk primaries (since they do not acquire vev) and descendants, and we used the shorthand notation $F^2 \equiv F_{\mu\nu}F^{\mu\nu}$ and $F\tilde{F} \equiv F_{\mu\nu}\tilde{F}^{\mu\nu}$.

Plugging the bulk OPE into the l.h.s. of \eqref{eq:FFfull} one obtains the following {\it bulk channel decomposition} of the two-point function
\begin{align}\label{FFbulk}
\langle F^{\mu\nu}(x_1) F^{\rho\sigma}(x_2) \rangle&\underset{x_1\to x_2}{\sim}{\frac{g^2}{\pi^2} G_{\mu\nu,\rho\sigma}(x_{12})}{}\nonumber\\
&+\frac{1}{12}(\delta^{\mu\rho}\delta^{\nu\sigma}-\delta^{\nu\rho}\delta^{\mu\sigma})\frac{a_{F^2}(\tau,\bar{\tau})}{y_2^4}+\frac{1}{12}\epsilon^{\mu\nu\rho\sigma}\frac{a_{F\tilde{F}}(\tau,\bar{\tau})}{y_2^4}\nonumber\\
&+\dots~,
\end{align}
where $\dots$ denote subleading descendant terms, and we parametrized bulk one-point functions as
\begin{equation}
\langle {\mathcal{O}}(\vec{x},y)\rangle=a_{\mathcal{O}}{y^{-\Delta_{\mathcal{O}}}}~.
\end{equation}
Comparing \eqref{FFbulk} and \eqref{eq:FFfull} (see appendix \ref{bootapp} for details) we obtain a constraint from the contribution of the identity
\begin{equation}\label{eq:Idconstr}
c_{11}(\tau,\bar{\tau})+ c_{22}(\tau,\bar{\tau}) =  \frac{4}{\pi \,{\rm Im} \tau} ~,
\end{equation}
and the following expressions for the one-point functions\footnote{Note that $a_{F^2}\in\mathbb{R}$ while $a_{F\tilde{F}} \in i \mathbb{R}$. To see this, it is useful to think about these coefficients in radial quantization, as the overlap between the state defined by the local operator $F^2$/$F\tilde{F}$ and the state defined by the conformal boundary condition. Applying an inversion, the overlap gets conjugated. Hence the reality conditions stated above simply follow from the fact that the operator $F^2$/$F\tilde{F}$ is even/odd under inversion.}
\begin{align}\label{eq:Bootsres}
 \quad a_{F^2}(\tau,\bar{\tau})& = \frac38\,\left(c_{22}(\tau,\bar{\tau})- c_{11}(\tau,\bar{\tau})\right) = \frac34\left(c_{22}(\tau,\bar{\tau}) -  \frac{2}{\pi \,{\rm Im} \tau}\right) ~,\\
a_{F\tilde{F}}(\tau,\bar{\tau}) & = i\frac34\,c_{12}(\tau,\bar{\tau})~.
\end{align}
This shows that the bulk one-point functions of $F^2$ and $F\tilde{F}$ are determined by the constants $c_{ij}$. 
Note that these relations are compatible with the (spurionic) parity symmetry, because $a_{F\tilde{F}}$ and $c_{12}$ are odd, while all the other coefficients are even. What we discussed here is a very simple example of the use of the crossing symmetry constraint on bulk two-point functions to determine data of BCFTs \cite{Liendo:2012hy}. The constraint can be solved exactly in this case because the bulk theory is gaussian.

Equivalently, in selfdual/antiselfdual components
\begin{equation}
 \quad a_{F_\pm^2}(\tau,\bar{\tau})= \frac{3}{16}\,\left(c_{22}(\tau,\bar{\tau}) \pm 2 i\,c_{12}(\tau,\bar{\tau})- c_{11}(\tau,\bar{\tau})\right) = -\frac{3}{4}c_{\pm\pm}(\tau,\bar{\tau})~.
 \end{equation}
 
Note that due to the constraint in eq. \eqref{eq:Idconstr}, the three entries of the matrix $c_{ij}$ actually only contain two independent functions of the coupling. In the appendix \ref{app:current2ptfunctions} we show how to express $c_{ij}$ (and also the possible contact terms in \eqref{eq:VVco}) in terms of two real functions $c_J$ and $\kappa_J$ of $(\tau, \bar{\tau})$, which are the coefficients in the two-point function of $\hat{J}$.

\subsection{$c_{ij}(\tau,\bar{\tau})$ in Perturbation Theory}

Having derived the bulk one-point and two-point functions in terms of the coefficients $c_{ij}(\tau,\bar{\tau})$ in the two-point function of the boundary currents, we will now give the leading order results for these coefficients in perturbation theory in $\tau^{-1}$. 

Note that thanks to the modified Neumann condition, at leading order $\hat{J}$ is identified with the U(1) current $\hat{J}_{\text{CFT}}$, whose two-point function can be parametrized as
\begin{equation}\label{eq:JCFT}
\langle\hat{J}^a_{\text{CFT}}(\vec{x}_1)\hat{J}^b_{\text{CFT}}(\vec{x}_2)\rangle = c^{(0)}_J \frac{I^{\rm 3d}_{ab}(\vec{x}_{12})}{|\vec{x}_{12}|^4} - i \frac{\kappa^{(0)}_J}{2\pi} \epsilon_{abc}\partial_{1}^c \delta^3(\vec{x}_{12})~.
\end{equation}

Using the expression for $c_{ij}(\tau,\bar{\tau})$ in appendix \ref{app:current2ptfunctions}, and plugging $c_J = c^{(0)}_J +\mathcal{O}(\tau^{-1})$ and $\kappa_J = \kappa^{(0)}_J +\mathcal{O}(\tau^{-1})$, we obtain
\begin{align}
c_{22}(\tau,\bar{\tau}) & =\frac{4}{\pi}\frac{{\rm Im}\tau}{|\tau|^2}-4\frac{({\rm Im}\tau^2 - {\rm Re}\tau^2)\, \pi^2 c^{(0)}_J +4 \,{\rm Im}\tau\,{\rm Re}\tau\,\frac{ \kappa^{(0)}_J}{2\pi}}{|\tau|^4} +\mathcal{O}(|\tau|^{-3})~,\label{eq:c22pert}\\
c_{12}(\tau,\bar{\tau}) & =-\frac{4}{\pi}\frac{{\rm Re}\tau}{|\tau|^2}+\frac{\,{\rm Im}\tau\,{\rm Re}\tau \,\pi^2 c^{(0)}_J -({\rm Im}\tau^2 - {\rm Re}\tau^2)\,\frac{ \kappa^{(0)}_J}{2\pi}}{|\tau|^4} +\mathcal{O}(|\tau|^{-3})~.\label{eq:c12pert}
\end{align}
$c_{11}(\tau,\bar{\tau})$ is obtained by $c_{22}(\tau,\bar{\tau})$ using \eqref{eq:Idconstr}. Note the compatibility with the (spurionic) parity symmetry, under which both ${\rm Re} \tau$ and $\kappa_J^{(0)}$ flip sign, and $c_{22}$ ($c_{12}$) is even (odd, respectively).

We observe that, to this order, 
\begin{equation}
\frac{\partial c_{22}}{\partial {\mathrm{Re}}\tau } + \frac{\partial c_{12}}{\partial {\mathrm{Im}}\tau } = 0~.\label{eq:idder}
\end{equation}
An explanation of this relation, and also a reason why it must hold to all orders in perturbation theory, will be provided in section \ref{sec:FreeEnergy}.

Going to higher orders in $\tau^{-1}$, the correlators of $\hat{J}$, and in particular the coefficients $c_J$ and $\kappa_J$, will start deviating from those of the CFT. When the CFT is free, these corrections can be computed by ordinary Feynman diagrams on the boundary. We will see examples of this in the following. In the more general case of an interacting CFT, these correction can be computed in conformal perturbation theory, by lowering an insertion of the bulk Lagrangian \eqref{eq:actau} integrated over the region $y\geq 0$. It would be interesting to characterize the CFT observables that enter the subleading orders of this perturbation theory. We leave this problem for the future.

\subsection{Displacement Operator}

In every BCFT with $d$-dimensional bulk there exists a boundary scalar operator of protected scaling dimension $d$, the so-called displacement operator. It can be defined as the only scalar primary boundary operator that appears in the bulk-to-boundary OPE of the bulk stress tensor
\begin{equation}\label{eq:TOPED}
T_{\mu\nu}(\vec{x},y) \underset{y\to 0}{\sim}\frac{d}{d-1} \left( \delta_{\mu y}\delta_{\nu y}-\frac1d\delta_{\mu\nu}\right)\hat{D}(\vec{x}) + \dots~.
\end{equation}
There is a Ward Identity associated to this operator, namely
\begin{equation}\label{eq:DWI}
\int d^d \vec{x} \langle \hat{D}(\vec{x}) O_1(\vec{x}_1,y_1) \dots O_n(\vec{x}_n,y_n) \rangle = (\partial_{y_1} +\dots +\partial_{y_n}) \langle O_1(\vec{x}_1,y_1) \dots O_n(\vec{x}_n,y_n) \rangle~,
\end{equation}
that fixes the normalization of the operator. In this normalization its two-point function is
\begin{equation}
\langle \hat{D}(\vec{x}_1) \hat{D}(\vec{x}_2)\rangle = \frac{C_{\hat{D}}}{|\vec{x}_{12}|^{2d}}~,
\end{equation}
and the quantity $C_{\hat{D}}$ is an observable of the BCFT. 

It follows from \eqref{eq:TOPED} that the displacement operator is the restriction of the component $T_{yy}$ of the stress-tensor to the boundary. In the theory that we are considering the bulk stress-tensor is the usual Maxwell stress-tensor
\begin{equation}
T_{\mu\nu} = \frac{{\rm Im}\tau}{2\pi} \left(F_{\mu\rho} F_\nu^{~\rho} -\frac 14 \delta_{\mu\nu} F_{\rho\sigma} F^{\rho\sigma}\right)~.\label{eq:strtensmax}
\end{equation}
Writing $T_{yy}(y=0)$ in terms of the currents $\hat{I}$ and $\hat{J}$ leads to the following expression for the displacement operator
\begin{equation}\label{eq:dispform}
\hat{D} = \frac{\pi}{\Im \tau}(|\tau|^2 \hat{I}^2+\hat{J}^2 - 2 {\rm Re}\tau\, \hat{I}\hat{J}) = \frac{{\rm Im}\tau}{4 \pi}(\hat{V}_1^2 + \hat{V}_2^2)~.
\end{equation} 
The right-hand side of \eqref{eq:dispform} contains products of two boundary operators at the same point, that are defined through a point-splitting procedure, similarly to the products on the right-hand side of \eqref{eq:strtensmax}. Such a point splitting makes sense for arbitrary $\tau$ even though generically the boundary currents are not generalized free fields. This is because their dimension and the dimension of $\hat{D}$ are protected, and the contribution of $\hat{D}$ in their OPE is non-singular, so after subtracting the contribution of the identity and possibly of additional operators of scaling dimension $<4$ we can always take the coincident-point limit.

\begin{figure}
	\centering
	\includegraphics[width=0.7\linewidth]{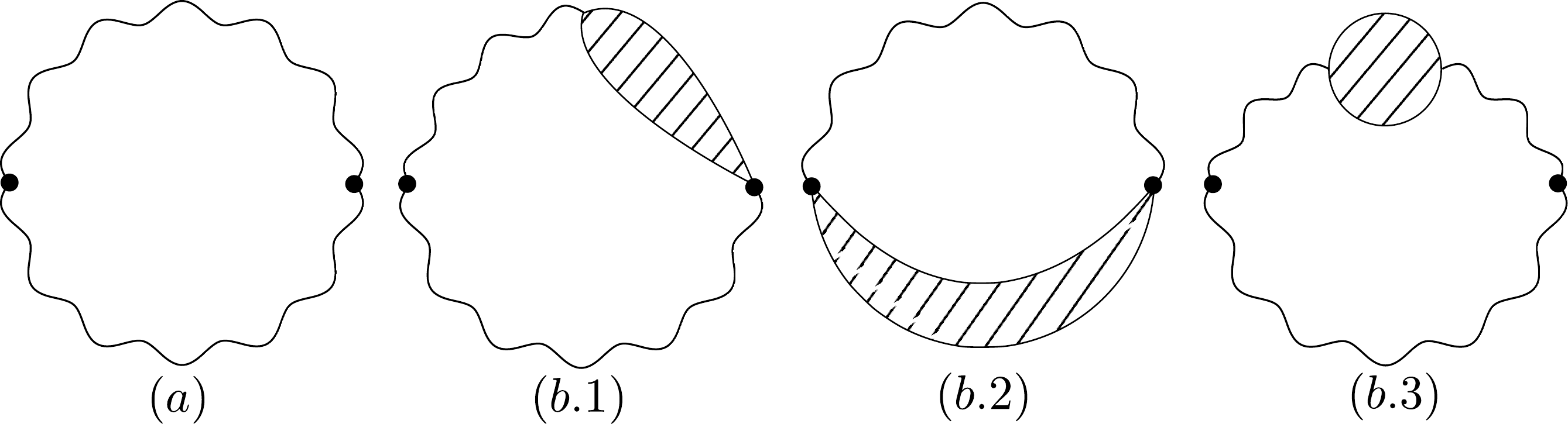}
	\caption{Diagrams for the two-point function of the displacement operator. The leading order contribution $(a)$ is the square of the two-point function of the topological current $\hat{I}$. At next-to-leading order we have the diagrams $(b.1)$-$(b.2)$-$(b.3)$ that are also sensitive to the electric current $\hat{J}$. The shaded blobs denote insertions/correlators of $\hat{J}$ in the undeformed CFT.\label{fig:cd}}
\end{figure}

We can use the expression \eqref{eq:dispform} to obtain the first two orders in the perturbative expansion of $C_{\hat{D}}$ universally in terms of the two-point function of the CFT current \eqref{eq:JCFT}. The leading order contribution to $C_{\hat{D}}$ at large $\tau$ comes from the contraction of the $\hat{I}$ currents in the $\hat{I}^2$ term, and is therefore proportional to the square of $c_{22}$ at leading order. At next-to-leading order there is a contribution from the correction to $c_{22}$, and a contribution from the $\hat{I}\hat{J}$ term. See fig. \ref{fig:cd}.
The result is
\begin{equation}
C_{\hat{D}} = \frac{6}{\pi^4} - \frac{12}{\pi} \frac{{\rm Im}\tau}{|\tau|^2}c_J^{(0)} + \mathcal{O}(|\tau|^{-2})~.\label{eq:CDpert}
\end{equation}
Even though the 3d CFT sector decouples from the bulk in the limit $\tau \to \infty$, and in particular it has a conserved 3d stress tensor, the displacement operator still exists within the sector of boundary operators coming from the free boundary condition of the bulk Maxwell field, and in particular $C_D$ is finite in this limit. Plugging ${\rm Re}\,\tau = 0$ and the value of $c_J^{(0)}$ for a theory of two Dirac fermions, namely $c_J^{(0)} = \frac{1}{4\pi^2}$, we find perfect agreement with \cite{Herzog:2017xha}. 

\subsection{Three-Point Function $\langle\hat{V}_i \hat{V}_j \hat{D}\rangle$}

Some of the distinctive features of the conformal theory living on the boundary of $B(\tau,\bar{\tau})$ are
\begin{itemize}
\item{the presence of a scalar operator of dimension 4, the displacement operator $\hat{D}$~; this feature is common to all conformal boundary conditions~;}
\item{the presence of the two $U(1)$ currents $\hat{V}_1$ and $\hat{V}_2$~.}
\end{itemize}
We will now show that the displacement operator $\hat{D}$ appears in the OPE of the currents, with a matrix of OPE coefficients that can be fixed in terms of the coefficients of the bulk one-point functions $a_{F^2}$ and $a_{F\tilde{F}}$, and the coefficient $C_{\hat{D}}$.

To show this, we consider the three-point correlator between the field strength and the displacement operator
\begin{align}
\langle F_{\mu\nu}(x_1)F_{\rho\sigma}(x_2)\hat{D}(\vec{x}_3)\rangle~.
\end{align}
We compute this three-point function in two OPE channels for $F_{\mu\nu}(x_1)F_{\rho\sigma}(x_2)$. In the boundary channel $y_{1,2} \to 0$, using the OPE \eqref{dOPEF} this three-point function can be fixed in terms of the OPE coefficients $\langle\hat{V}_i \hat{V}_j \hat{D}\rangle$ that we want to determine. On the other hand, in the bulk OPE channel $x_{12}\to 0$ this three point function can be computed in terms of the bulk-boundary two-point functions $\langle O(x_1) \hat{D}(\vec{x}_3) \rangle$ between the displacement and the operators $O$ in the bulk OPE of two $F$'s. The last step of the argument amounts to relating the latter two-point function to the one-point function of $O$ if $O$ is a scalar operator, or to $C_{\hat{D}}$ if $O$ is the stress-tensor.

The coefficients appearing in the three-point function are \cite{Costa:2011mg, Dymarsky:2017xzb}
\begin{align}\label{eq:VVD3main}
\langle \hat{V}_i^a(\vec{x}_1)\hat{V}_j^b(\vec{x}_2)\hat{D}(\infty)\rangle = \, \lambda_{ij\hat{D}+}^{(1)}\,\, \delta^{ab}+\lambda_{ij\hat{D}-}^{(1)}\,\, {\hat{x}_{12}^c}{}\epsilon^{abc}~.
\end{align}
For simplicity we placed the displacement at infinity. $\lambda_{ij\hat{D}+}^{(1)}$ and $\lambda_{ij\hat{D}-}^{(1)}$ are respectively the parity-even/odd OPE coefficients in the conventions of \cite{Dymarsky:2017xzb}, and $\hat{x}^a=x^a/|\vec{x}|$. Recall that under parity $\hat{V}_1$ is a vector while $\hat{V}_2$ is an axial vector, hence the coefficients $\lambda^{(1)}_{11\hat{D}-},\lambda^{(1)}_{22\hat{D}-},\lambda^{(1)}_{12\hat{D}+}$ are odd under a spurionic parity transformation, while the others are even. 

The details of the calculation are showed in the appendix \ref{app:VVD}, and here we will just give the final result
\begin{align}\label{eq:3ptfinal}
 \lambda_{11\hat{D}+}^{(1)}  & =-\frac{8}{3 \pi ^2}a_{F^2}+\frac{g^2}{3}C_{\hat{D}}~,\\
 \lambda_{22\hat{D}+}^{(1)}  &= \frac{8}{3 \pi ^2}a_{F^2}+\frac{g^2}{3}C_{\hat{D}}~,\\
\lambda_{12\hat{D}+}^{(1)}  & = -\frac{8}{3\pi^2}\, i a_{F\tilde{F}}~,\\
\lambda_{ij\hat{D}-}^{(1)}  & = 0~.
\end{align}
The parity-odd three-point structures are all set to zero. The spurionic parity symmetry is again satisfied because $\lambda_{12\hat{D}+}^{(1)}$ is proportional to the odd coefficient $a_{F\tilde{F}}$, while the formulas for $\lambda_{11\hat{D}+}^{(1)}$ and $\lambda_{22\hat{D}+}^{(1)}$ are even.

Going to the basis in which the matrix of current-current 2-pt functions is the identity
\begin{equation}
U_i^l U_j^k c_{lk} = \delta_{ij}~,
\end{equation}
the matrix of OPE coefficients  becomes
\begin{equation}
U\lambda_{D+}^{(1)}U^{T}=\frac{2}{\pi^2}\begin{pmatrix}
\frac{\mathcal{A} - \frac{\pi^2 C_{\hat{D}}}{8}}{\mathcal{A} - \frac{3}{4\pi^2}} & 0 \\
0 & \frac{\mathcal{A} + \frac{\pi^2 C_{\hat{D}}}{8}}{\mathcal{A} + \frac{3}{4\pi^2}}
\end{pmatrix}~,\label{eq:ldiag}
\end{equation}
where 
\begin{align}
\mathcal A \equiv \frac{1}{g^2}\sqrt{a_{F^2}^2 - a_{F\tilde{F}}^2}~.
\end{align}
Recall that $a_{F^2} \in \mathbb{R}$ and $a_{F\tilde{F}} \in i \mathbb{R}$, so $\mathcal{A}$ is real and $\geq 0$. Seemingly the upper entry has a pole at $\mathcal{A} = \frac{3}{4\pi^2}$, which corresponds to the value at the decoupling limit. However recall from \eqref{eq:CDpert} that $C_{\hat{D}} \to \frac{6}{\pi^4}$ in the decoupling limit, so that actually the entry is finite in the limit.

The upshot of this analysis is that the OPE coefficients between two currents and the displacement can be completely characterized in terms of the two positive quantities $\mathcal{A}$ and $C_{\hat{D}}$, that can be taken to effectively parametrize the position on the conformal manifold. It would be interesting to derive these relations from more standard analytic bootstrap techniques, along the lines of \cite{Bissi:2018mcq,Mazac:2018biw,Kaviraj:2018tfd}.

\section{Free Energy on a Hemisphere}
\label{sec:FreeEnergy}
In this section we study the hemisphere free energy for the conformal boundary conditions of the $U(1)$ gauge field.

Following \cite{Gaiotto:2014gha}, to any given conformal boundary condition for a CFT$_4$ we can assign a boundary free energy $F_\partial$, defined as 
\begin{align}\label{eq:defnRatio}
F_\partial=-\frac{1}{2}\log \frac{|Z_{HS^4}|^2}{Z_{S^4}}=-\Re \log Z_{HS^4}+\frac{1}{2}\log Z_{S^4}~.
\end{align}
$Z_{S^4}$ denotes the sphere partition function of the CFT, while $Z_{HS^4}$ denotes the partition function of the theory placed on an hemisphere, with the chosen boundary condition on the boundary $S^3$. In writing \eqref{eq:defnRatio} we discarded power-law UV divergences, and focused on the universal finite term. Conformal symmetry ensures that the coupling to the curved background can be defined via Weyl rescaling. 

In our setup the bulk theory is a $U(1)$ gauge-field with action \eqref{eq:actau}, so we have
\begin{align}\label{systemIR}
-8\pi\frac{\partial F_{\partial}}{\partial \Im\tau}=-\Re \int_{HS^4}d^4x \sqrt{g(x)}\langle F^2 (x)\rangle_{HS^4}+\frac{1}{2} \int_{S^4}d^4x\sqrt{g(x)}\langle F^2 (x)\rangle_{S^4},\nonumber\\
-8\pi\frac{\partial F_{\partial}}{\partial \Re\tau}=-\Re \int_{HS^4}d^4x\sqrt{g(x)} \langle i F\tilde{F}(x) \rangle_{HS^4}+\frac{1}{2} \int_{S^4}d^4x \sqrt{g(x)} \langle i  F\tilde{F}(x) \rangle_{S^4}.
\end{align}
Using a Weyl transformation the one-point functions can be obtained from those on $\mathbb{R}^3\times \mathbb{R}_+$ as  
\begin{align}
\langle F^2 (x) \rangle_{HS^4}=\Omega(x)^{-4}\frac{a_{F^2}}{u(x)^4}+\frac{1}{2}\mathcal{A}, \quad \langle F\tilde{F} (x)\rangle_{HS^4}=\Omega(x)^{-4}\frac{a_{F\tilde{F}}}{u(x)^4}+\frac{1}{2}\widetilde{\mathcal{A}}~.
\end{align}
Here $x$ is a point on the hemisphere, $\Omega(x)$ is the Weyl factor induced by the stereographic projection, and $u(x)$ denotes the chordal distance between the point $x$ and the boundary $S^3$. The shifts $\mathcal{A}$ and $\mathcal{\tilde{A}}$ stand for a scheme-dependent contribution to the one-point function, due to the ambiguity in the definition of the theory on the curved background: we can always add local counterterms given by a scalar density of dimension four built out of the background curvature, multiplied by the real or imaginary part of the marginal coupling $\tau$, and integrated in the interior of the hemisphere. On the other hand, if we compute the partition function on $S^4$ in the same scheme, the one-point functions on $S^4$ receive contribution only from those counterterms, because on $\mathbb{R}^4$ one-point functions must vanish, and there is a relative factor of two because in this case the counterterm is integrated over the whole sphere. Hence
\begin{align}
\langle F^2 \rangle_{S^4}=\mathcal{A}~,\quad \langle F\tilde{F} \rangle_{S^4}=\widetilde{\mathcal{A}}~,
\end{align}
such that the ambiguity precisely cancels in \eqref{systemIR}. Here we see the virtue of the choice of normalization in \eqref{eq:defnRatio}.

The remaining integral on $HS^4$ has a UV divergence when the point $x$ approaches the boundary $S^3$. We introduce a UV regulator $\epsilon \ll 1$ and restrict the integral to the region $u(x) > \epsilon$. The result is
\begin{equation}
\int_{u(x)>\epsilon}\sqrt{g(x)}\ \Omega(x)^{-4}\frac{1}{u(x)^4} = \frac{2\pi^2}{3\epsilon^3} -\frac{5\pi^2}{3\epsilon} +\frac{4\pi^2}{3} + \mathcal{O}(\epsilon)~.
\end{equation}
As implicit in the definition of $F_\partial$, we will neglect the power-law UV divergent term and focus on the universal finite piece. Hence we finally obtain
\begin{align}
\frac{\partial F_{\partial}}{\partial \Im\tau}& =\frac{\pi}{6} \,a_{F^2}=\frac{\pi}{8}c_{22}(\tau,\bar{\tau})-\frac{1}{4\, {\rm Im}\tau}~,\label{eq:onepoint1}\\  \frac{\partial F_{\partial}}{\partial \Re\tau}& =\frac{\pi}{6} \,i\,a_{F\tilde{F}}=-\frac{\pi}{8}c_{12}(\tau,\bar{\tau})~.\label{eq:onepoint2}
\end{align}
We used the relations \eqref{eq:Idconstr} to rewrite the result in terms of the two-point functions of the conserved currents.
A consequence of this equation is that the relation \eqref{eq:idder} must be valid to all orders in perturbation theory, or more generally whenever $F_{\partial}$ is well-defined. 

Plugging \eqref{eq:c22pert}-\eqref{eq:c12pert} in \eqref{eq:onepoint1}-\eqref{eq:onepoint2} and solving the equations we find the following leading behavior of $F_\partial$ at large $\tau$
\begin{equation}
F_\partial \underset{\tau\to\infty}{\sim} -\frac 14 \log\left[\frac{2 \,{\rm Im} \tau}{|\tau|^2}\right]+ C + \pi \frac{\frac{\pi^2}{2}  c^{(0)}_J {\rm Im} \tau + \frac{\kappa^{(0)}_J}{2\pi}{\rm Re} \tau}{|\tau|^2}+ \mathcal{O}(|\tau|^{-2})~.\label{eq:Fcuspinf}
\end{equation}
The first term, which diverges for $\tau \to \infty$, is the value of $F_\partial$ for a free Maxwell field with Neumann boundary conditions \cite{Gaiotto:2014gha}. Matching eq. \eqref{eq:Fcuspinf} with the value of $F_\partial$ for a decoupled system of a Maxwell field with Neumann conditions and a 3d CFT on the boundary, we find that the constant $C$, that remained undetermined by the differential constraint, is in fact the $S^3$ free energy $F_{0,1}$ of the theory $T_{0,1}$. 

Using an $SL(2,\mathbb{Z})$ transformation, the same asymptotic behavior holds in the vicinity of any cusp point, upon replacing $\tau$ with the transformed variable $\tau^\prime$ that goes to $\infty$ at the selected cusp, and identifying $C$ with the $S^3$ free energy of the decoupled 3d CFT living at the cusp. Near the cusp where the current $p\hat{J} + q \hat{I}$ decouples from the 3d theory $T_{p,q}$, we have
\begin{equation}
F_\partial \underset{\tau^\prime\to\infty}{\sim} -\frac 14 \log\left[\frac{2 \,{\rm Im} \tau^\prime}{|\tau^\prime|^2}\right]+ F_{p,q} + \mathcal{O}(|\tau^{\prime}|^{-1})~.\label{eq:Fcusppq}
\end{equation}
where $\tau^\prime = \frac{a \tau + b}{p \tau + q}$, with $a q - b p =1$, and $F_{p,q}$ is the $S^3$ free energy of $T_{p,q}$. Note that
\begin{align}
-\frac 14 \log\left[\frac{2 \,{\rm Im} \tau}{|\tau|^2}\right]  & \underset{\tau^\prime\to\infty}{\sim} -\frac 14 \log\left[\frac{2 \,{\rm Im} \tau^\prime}{|\tau^\prime|^2}\right] + \frac12 \log |q|~,\label{eq:transF}\\
-\frac 14 \log\left[2 \,{\rm Im} \tau\right]  & \underset{\tau^\prime\to\infty}{\sim} -\frac 14 \log\left[\frac{2 \,{\rm Im} \tau^\prime}{|\tau^\prime|^2}\right] + \frac12 \log |p|~.\label{eq:transF2}
\end{align}
Eq. \eqref{eq:transF} implies that the function
\begin{equation}
F_\partial +\frac 14 \log\left[\frac{2 \,{\rm Im} \tau}{|\tau|^2}\right]~,
\end{equation}
attains the finite value 
\begin{equation}
\frac12 \log |q| + F_{p,q}~,
\end{equation}
at all the cusps with $|q| \neq 0$. For the cusp with $q=0$ we can simply use \eqref{eq:transF2} to derive that 
\begin{equation}
F_\partial +\frac 14 \log\left[ 2 \,{\rm Im} \tau \right]~,
\end{equation}
approaches 
\begin{equation}
\frac12 \log |p| + F_{p,q}~.
\end{equation}
Hence the function $F_\partial(\tau,\bar{\tau})$ contains information about the $S^3$ free energies of an infinite family of 3d Abelian gauge theories, namely all the theories obtained by applying $SL(2,\mathbb{Z})$ transformations to $T_{0,1}$.

We note in passing that the shift by $\frac 12 \log |q|$ in eq. \eqref{eq:transF} has a nice interpretation in terms of the $S^3$ free energy for a pure CS theory. Indeed, starting with a 4d gauge field with Neumann condition, applying the transformation $S T^{k}$, i.e. $\tau^\prime = -\frac{1}{\tau + k}$, and taking the decoupling limit $\tau^\prime \to \infty$, we are left with a pure CS theory at level $k$ on the boundary. Hence, the free energy $F_\partial$ in this limit should be the sum of the contribution of the decoupled 4d gauge field with Neumann boundary condition, and the contribution from the CS theory at level $k$, which is $\frac12 \log|k|$. This is precisely what eq. \eqref{eq:transF} gives. Similarly eq. \eqref{eq:transF2} can be interpreted by starting with a 4d gauge field with Dirichlet boundary condition, whose partition function is the left-hand side of  \eqref{eq:transF2}, applying $S T^{k} S$, i.e. $\tau^\prime = \frac{\tau}{ - k\tau +1}$, and going to the decoupling limit. Again, we find a decoupled 4d gauge field with Neumann boundary condition, and a CS theory at level $k$ on the boundary. The shift by $\frac 12 \log |p|$ in eq. \eqref{eq:transF2} precisely reproduces the $\frac12 \log|k|$ contribution of the CS theory.
 
\section{A Minimal Phase Transition}\label{sec:MinPhTr}
In this section we will study a non-trivial BCFT which conjecturally describes a second order (boundary) phase transition 
between two free boundary conditions $(p,q)$ and $(p',q')$ of the 4d gauge field, with $p q' - p' q=1$. In particular, the conjectural BCFT should have a single relevant boundary operator, which can be turned on to flow to either 
of these free boundary conditions in the IR, depending on the sign of the coupling. 
We will assume that this BCFT exists for all values of the gauge coupling $\tau$,
with no further phase transitions as a function of $\tau$. 

Without loss of generality, we can pick two canonical duality frames where the phase transition interpolates 
between Dirichlet and Neumann boundary conditions or viceversa. We can also pick two duality frames 
where the phase transition interpolates between Neumann and $(1,1)$ boundary conditions or viceversa.
\begin{itemize} 
\item If we go to weak coupling in the former duality frames, the boundary degrees of freedom should describe 
a phase transition between phases with spontaneously broken or unbroken $U(1)$ global symmetry. 
We expect that to be described by a critical $O(2)$ model.
\item If we go to weak coupling in the latter duality frames, the boundary degrees of freedom should describe 
a phase transition between two gapped phases with unbroken $U(1)$ global symmetry, but background Chern-Simons coupling 
which differs by one unit. We expect that to be described by a massless Dirac fermion.
\end{itemize}

Keeping track of the duality transformations between the different frames, we can assemble an overall picture.
\begin{itemize}
\item Denote as $\tau_{DN}$ the gauge coupling associated to the description as a phase transition between 
Dirichlet and Neumann boundary conditions, so that one ``$O(2)$ cusp'' is at $\tau_{DN} \to \infty$. 
\item Then $\tau_{ND}= -1/\tau_{DN}$ is the coupling which is weak at the other $O(2)$ cusp, at $\tau_{DN}\to0$. 
\item Shifting the $\theta$ angle by $2 \pi$ gives an alternative description 
as a transition between Dirichlet and $(1,-1)$ boundary conditions, with coupling $\tau_{DN''} = \tau_{DN}-1$.
Dually, we get a transition between Neumann and $(1,1)$ boundary conditions, with coupling 
$\tau_{N N'} = -\tau_{DN''}^{-1} = \frac{1}{1-\tau_{DN}}$ which is weak at the ``Dirac fermion'' cusp,
$\tau_{DN} \to 1$. 
\item If we had shifted the $\theta$ angle in the opposite way, 
we would arrive to a  transition between Neumann and $(1,-1)$ boundary conditions, with coupling 
$\tau_{N N''} = -\tau_{DN'}^{-1} = -\frac{1}{1+\tau_{DN}}$ which is weak at the second ``Dirac fermion'' cusp,
$\tau_{DN} \to -1$. 
\end{itemize}

\begin{figure}
\centering
\includegraphics[clip, height=7cm]{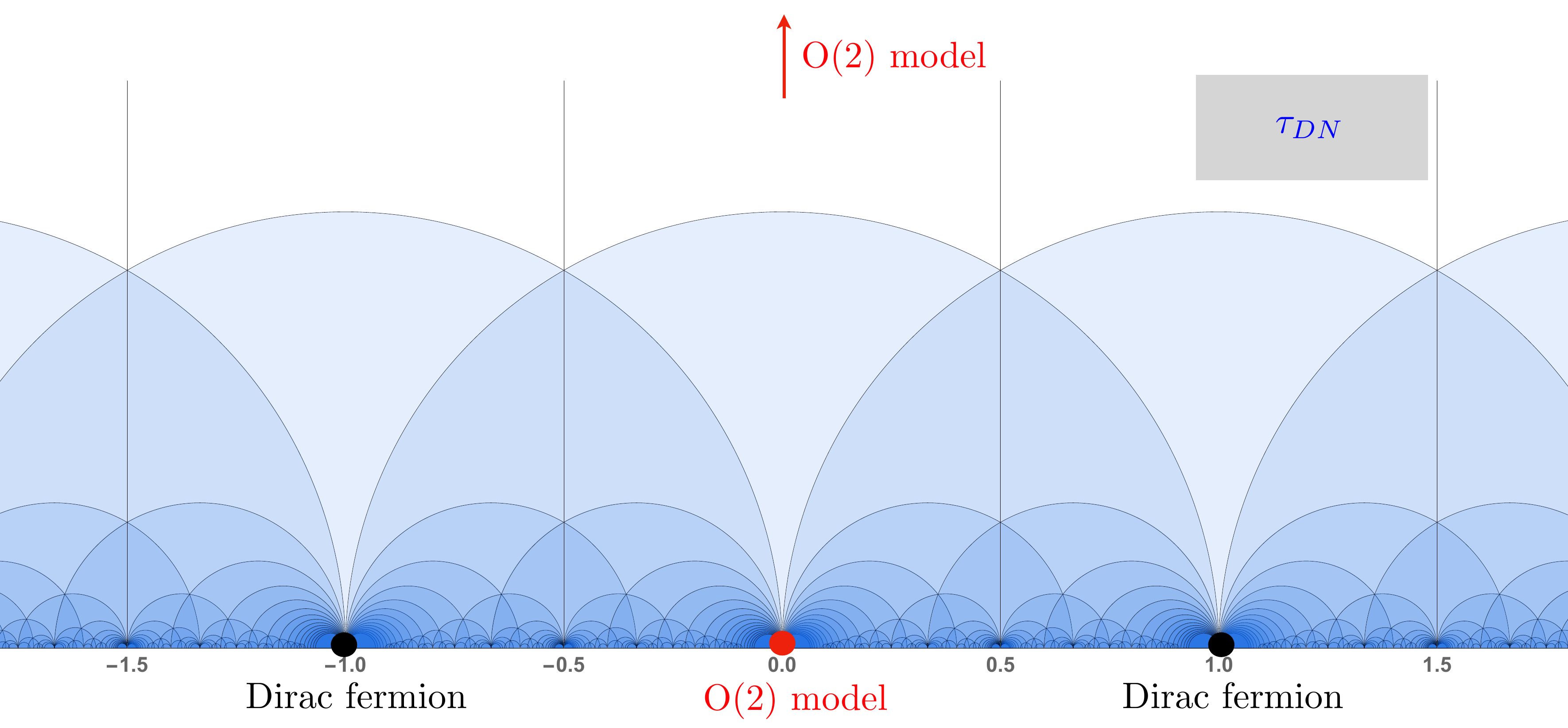}
\caption{The upper-half plane of the gauge coupling $\tau_{DN}$, i.e. in the duality frame in which at $\tau_{DN}\to \infty$ we find the $O(2)$ model on the boundary. Thanks to particle-vortex duality, the cusp in the origin $\tau_{DN} = 0$ also gives a decoupled $O(2)$ model on the boundary.  Thanks to the boson-fermion duality between $U(1)_1$ coupled to a critical scalar and a free Dirac fermion, the cusps at $\tau_{DN} = \pm 1$ give a free Dirac fermion. 
\label{fig:tauplaneO2}}
\end{figure}

\begin{figure}
\centering
\includegraphics[clip, height=7cm]{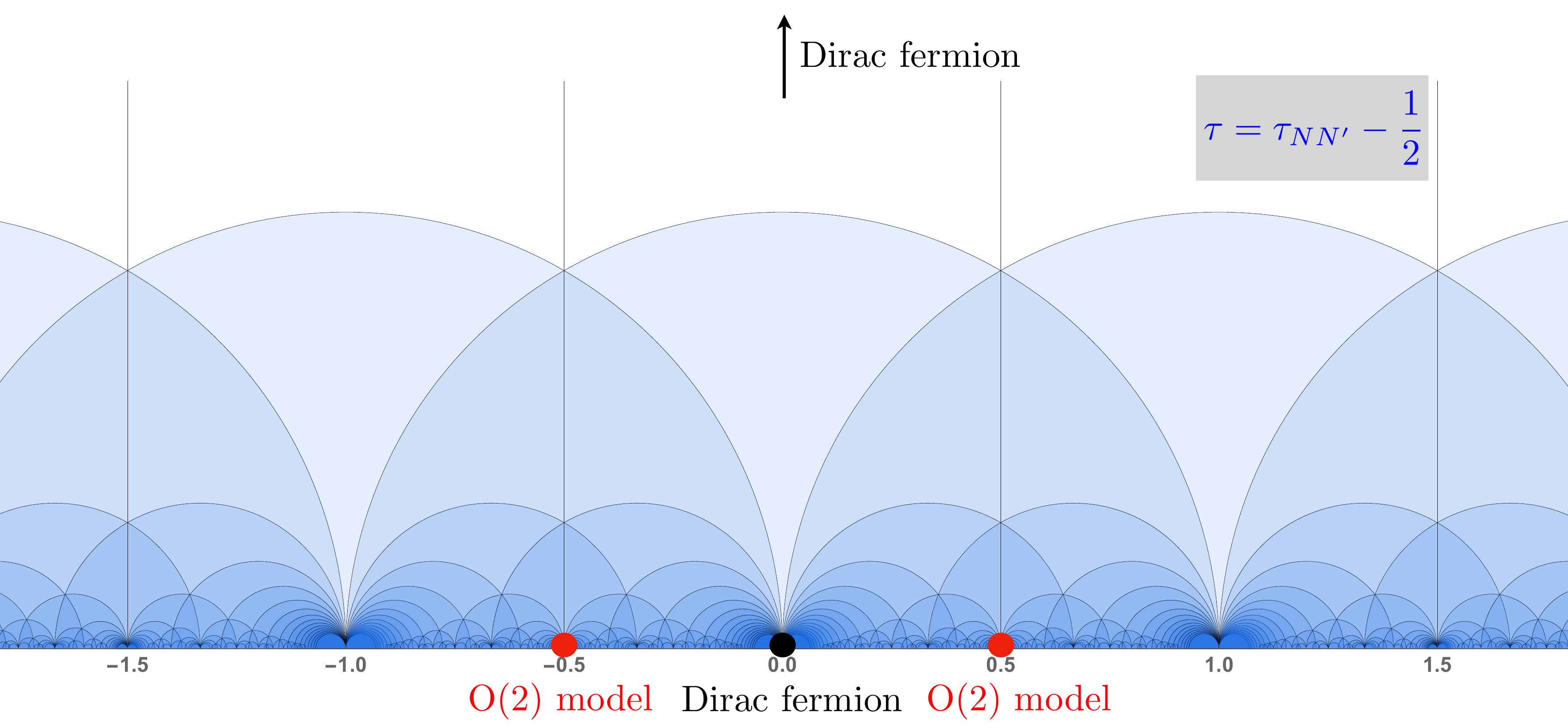}
\caption{The upper-half plane of the gauge coupling $\tau = \tau_{NN'} -\frac 12$, i.e. in the duality frame in which at $\tau\to \infty$ we find a free Dirac fermion on the boundary. Thanks to fermionic particle-vortex duality, the cusp in the origin $\tau = 0$ also gives a free Dirac fermion on the boundary.  Thanks to the boson-fermion duality between $U(1)_{\frac 12}$ coupled to a Dirac fermion and the $O(2)$ model, the cusps at $\tau= \pm \frac12$ give the $O(2)$ model.  \label{fig:tauplane}}
\end{figure}

In the following we will do most of our calculations in a perturbative expansion around a ``Dirac fermion'' cusp.
The correct boundary theory is actually a Dirac fermion dressed by half a unit of background Chern-Simons coupling \cite{AlvarezGaume:1984nf, Witten:2015aba}. It is convenient to absorb that background Chern-Simons coupling into an improperly-quantized shift of the $\theta$ angle, so that the gauge coupling is denoted as $\tau = \tau_{N N'} -\frac12 = \frac12 \frac{1+\tau_{DN}}{1-\tau_{DN}}$. Therefore, denoting with $\psi$ the Dirac fermion, the action that we consider is 
\begin{equation}
S[A, \tau +\tfrac 12] + \int_{y=0} d^3\vec{x}\, i\bar{\psi}\slashed{D}_A \psi~.
\label{eq:action1fermion}
\end{equation}
The second Dirac fermion cusp is at $\tau \to 0$ and the $O(2)$ cusps are at $\tau = \pm \frac12$. 
See fig.s \ref{fig:tauplaneO2}-\ref{fig:tauplane}.


Essentially by construction, the picture is compatible with a well-known duality web of particle-vortex, fermion-boson and fermion-fermion dualities \cite{Seiberg:2016gmd}, which inspired this investigation. In particular, thanks to the particle-vortex duality between the $O(2)$ model and the gauged $O(2)$ model \cite{Peskin:1977kp, Dasgupta:1981zz}, or equivalently thanks to its fermionic version \cite{Son:2015xqa}, in this case we have a $\mathbb{Z}_2$ subgroup of $SL(2,\mathbb{Z})$ that is a duality of $B(\tau,\bar{\tau})$, i.e. it leaves  invariant both the bulk and the boundary condition. This subgroup acts on $\tau = \tau_{NN'} -\frac 12$ as
\begin{equation}
\tau \to -\frac{1}{4\tau}~.
\end{equation}
It is interesting to note that the self-dual point $\tau = \frac{i}{2}$, i.e. $\tau_{DN} = i$, is an extreme of $F_{\partial}$. In our formalism, this is a straightforward consequence of the differential equations \eqref{eq:onepoint1}-\eqref{eq:onepoint2}, once we set $c_{11}=c_{22}=\frac{2}{\pi\,{\rm Im}\tau}$ and $c_{12}=0$  -- as dictated by self-duality and equation \eqref{eq:Idconstr}.\footnote{Alternatively, we can implement the reasoning of \cite{Baggio:2017mas} to show that this property follows from the emergent $\mathbb{Z}_2$ symmetry of the system at the self-dual point.}

Before proceeding, let us mention some of the previous literature on this theory, and comment on the relation to the results that we will present in the rest of this section. The interplay between the 3d dualities and the 4d electric-magnetic duality in the setup with a 3d Dirac fermion coupled to a bulk gauge field was investigated in \cite{Seiberg:2016gmd, Metlitski:2015eka, Wang:2015qmt, Hsiao:2017lch, Hsiao:2018fsc}. In particular \cite{Hsiao:2017lch, Hsiao:2018fsc} studied the transport properties of the boundary theory at the self-dual point. For the theory with an even number of Dirac fermions on the boundary, the two-loop two-point function of the boundary current $\hat{J}$ was obtained in \cite{Teber:2012de} (see also \cite{Kotikov:2013kcl, Teber:2014ita, Teber:2016unz, Kotikov:2016yrn}) while the Weyl anomalies (or equivalently the two- and three-point functions of the displacement operator) were computed to next-to-leading order in \cite{Herzog:2017xha, Herzog:2017kkj} (for the supersymmetric version of the theory see \cite{Herzog:2018lqz}). The point of view of boundary conformal field theory was first adopted in this theory in \cite{Herzog:2017xha, Herzog:2017kkj}, but these papers do not consider the action of electric-magnetic duality and the existence of multiple decoupling limits. Besides the transport coefficients and the Weyl anomalies, other boundary observables such as scaling dimensions of operators, or the hemisphere free-energy, were not studied before. Since the duality explained above only exists for the theory with one Dirac fermion, we will first concentrate on this case. Later we will also consider the theory with an even number $2N_f$ of fermions, both at large $N_f$ and in the special case $2 N_f = 2$.

\subsection{Perturbative Calculation of Scaling Dimensions}\label{eq:pertscaldim}

We will compute the anomalous dimensions of the first two fermion bilinear operators $O_s$ of spin $s$, namely
\begin{align}
O_0 & = \bar{\psi} \psi~, \\
(O_2)_{ab} & = i\big(\bar{\psi} \gamma_{(a}\overset\leftrightarrow{D}_{b)}\psi -\text{trace}\big)~,
\end{align}
up to two-loop level. Note that in the limit $\tau\to\infty$ of decoupling between bulk and boundary $(O_2)_{ab}$ becomes a conserved current, namely the stress-tensor of the 3d free-fermion CFT.

The anomalous dimension can be obtained from the renormalization of the 1PI correlator of the composite operator with two elementary fields 
\begin{equation}
\langle O_s(q=0) \psi(-p) \bar{\psi}(p)\rangle_{\rm 1PI}~.\label{eq:3point}
\end{equation}
We employ dimensional regularization and minimal subtraction, i.e. we set $d = 3-2\epsilon$ and keep the codimension fixed $=1$, expand the dimensionally-continued loop integrals around $\epsilon = 0$, and reabsorb the poles in the renormalization constants
\begin{align}
O_B & = Z_O O~, \\
\psi_B & = Z_{\psi} \psi~, 
\end{align}
where the subscript $B$ denotes the bare operators. 

Even though the correlator in \eqref{eq:3point} involves the operator $\psi$ that is not gauge-invariant, it is still sensible to renormalize it. The resulting renormalized correlator, as well as the renormalization constant $Z_{\psi}$, both depend on the choice of gauge-fixing, but the renormalization constant $Z_O$ of the gauge-invariant operator does not, hence we can extract physical information from it.

The renormalization constants admit the loopwise expansion (at small $g^2$ with fixed $\gamma$)
\begin{equation}
Z = 1+ \delta Z = 1+\sum_n \left(\frac{g^2}{1+\gamma^2}\right)^n\delta Z^{(n)}~, 
\end{equation}
where $\delta Z^{(n)}$ is a polynomial in $\gamma$ of degree $\leq n$, and furthermore by invariance under space reflections only even powers of $\gamma$ are present. The n-loop term $\delta Z^{(n)}$ contains divergences up to $\epsilon^{-n}$, but the familiar RG argument constrains all the coefficients in terms of the ones at lower loop order, except that of the $\epsilon^{-1}$ divergence. 

The anomalous dimension is then given by 
\begin{equation}
\gamma_O = \frac{d \log Z_O}{d\log \mu}~.
\end{equation}
The dependence on the renormalization scale $\mu$ is through the $d$-dimensional coupling 
\begin{equation}
g_B  = \mu^{\epsilon} g~.
\end{equation}
In the latter equation we do not need to include a renormalization of the coupling because, as we explained in section \ref{sec:Interaction}, $g$ does not run. Therefore we can rewrite
\begin{equation}
\gamma_O = -\epsilon\frac{\partial \log Z_O}{\partial \log g}~.
\end{equation}

\begin{figure}[htb]
 	\centering
 	\includegraphics[width=0.8\linewidth]{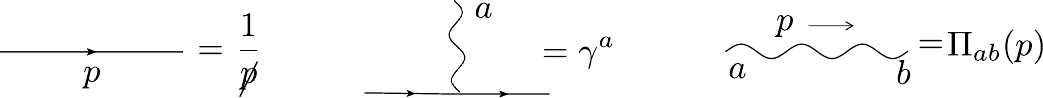}
 	\caption{Feynman rules. $\Pi_{ab}$ is defined in \eqref{eq:prop}.}
 	\label{fig:Fermionfeynrule1}
 \end{figure}

\begin{figure}[h]
	\centering
	\includegraphics[width=0.9\linewidth]{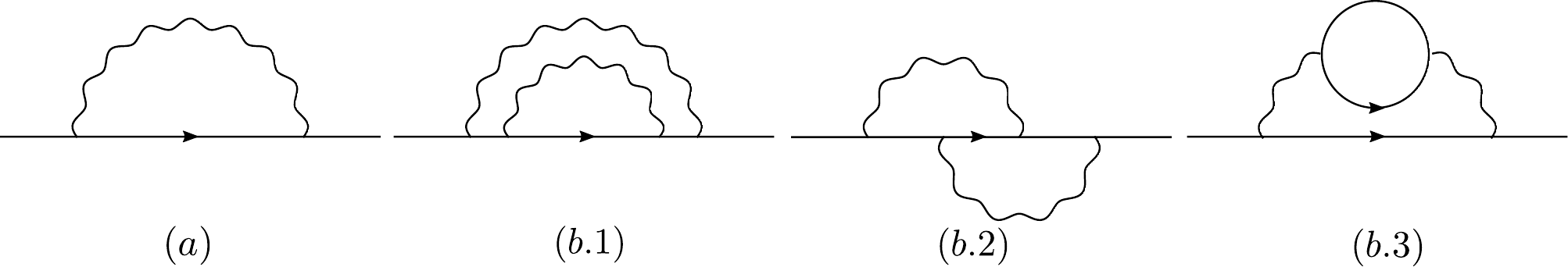}
	\caption{One loop and two loops diagrams. We sum over all possible insertions of the composite operators on the internal fermion lines, and also on vertices in the case of $O_2$.}
	\label{fig:loopdiagram}
\end{figure}

To compute \eqref{eq:3point} we use the Feynman diagrams in fig.\ref{fig:loopdiagram}, computed in momentum space, and for simplicity we take the composite operator to carry zero momentum. The Feynman rules given in fig.\ref{fig:Fermionfeynrule1}. 
 \begin{figure}[htb]
	\centering
	\includegraphics[width=0.8\linewidth]{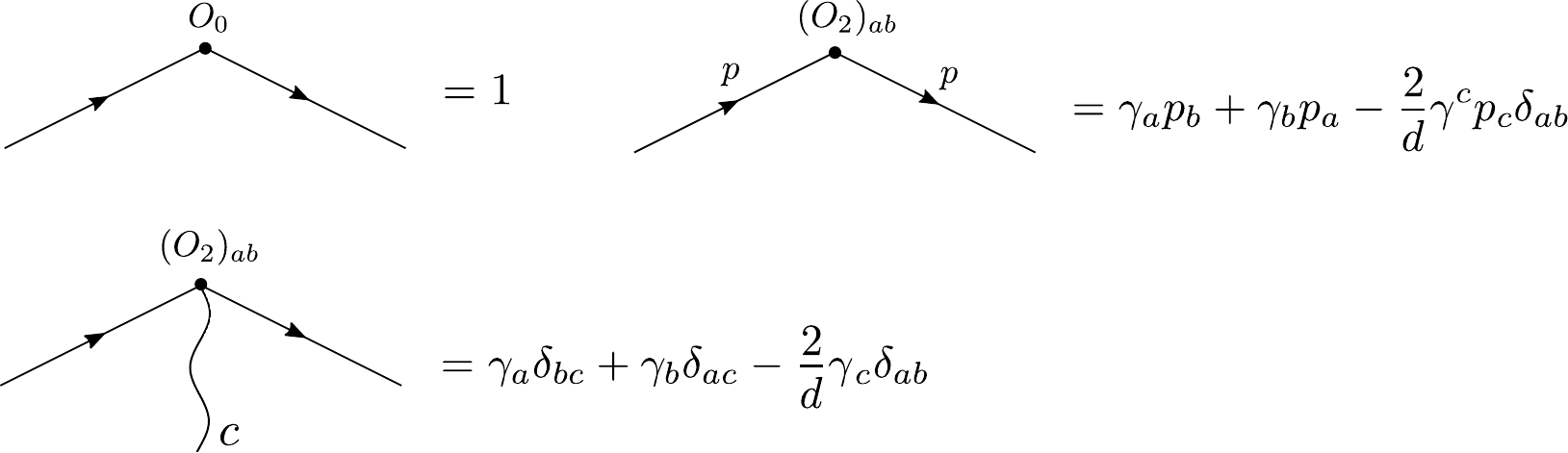}
	\caption{Feynman rules for the zero-momentum insertions of the composite operators. Note that there are two vertices associated to $O_2$.}
	\label{fig:Fermionfeynrule_ops}
\end{figure} 

We performed the calculation up to two loops. See appendix \ref{app:FeynInt} for more details about the computation of the two-loop Feynman diagrams. The resulting renormalization constants are 
\begingroup
\allowdisplaybreaks
\begin{align}
\delta Z_\psi& = \frac{g^2}{1+\gamma^2}\frac{2-3\xi}{24\pi^2\epsilon}  + \left(\frac{g^2}{1+\gamma^2}\right)^2 \left[\frac{(2-3\xi)^2}{1152 \pi^4 \epsilon^2} -\frac{9(1-2\gamma^2) \pi^2 + 16 }{3456\pi^4 \epsilon} \right] +\mathcal{O}(g^6)~.\label{eq:wf}\\
\delta Z_0& = -\frac{g^2}{1+\gamma^2} \frac{2}{3\pi^2 \epsilon}  + \left(\frac{g^2}{1+\gamma^2}\right)^2 \left[\frac{2}{9\pi^4\epsilon^2}+\frac{9\pi^2(1-2\gamma^2)-8}{108\pi^4 \epsilon}\right]+\mathcal{O}(g^6)~.\label{eq:Z0}\\
\delta Z_2& =\frac{g^2}{1+\gamma^2}  \frac{2}{5\pi^2 \epsilon} + \left(\frac{g^2}{1+\gamma^2}\right)^2 \left[\frac{2}{25 \pi^4 \epsilon^2} - \frac{75 \pi ^2 + 16}{3000 \pi ^4 \epsilon}\right]+\mathcal{O}(g^6)~,\label{eq:Z2}
\end{align}
\endgroup
where we denoted $\delta Z_s \equiv \delta Z_{O_s}$. Note that indeed $\delta Z_0$ and $\delta Z_2$ do not depend on the gauge-fixing parameter. As a check, we also verified that the operator $O_1 = \bar{\psi}\gamma^a \psi$ does not get renormalized, i.e. we explicitly computed the renormalization up to two-loop order and found $\delta Z_1=0$, as expected for a conserved current. On the other hand, note that $\delta Z_2\neq0$. This is a manifestation of the fact that the boundary degrees of freedom do not define a local 3d theory once we couple them to the bulk: the conservation of the boundary would-be stress-tensor is violated at $g\neq 0$, and the system only admits a stress-tensor in the bulk. This means that the short operator of spin 2 must recombine into a long conformal multiplet. In the appendix \ref{app:fakestress}, we show that this mechanism can be used to compute the one-loop anomalous dimension, and we use this to check the Feynman diagram calculation. 

The resulting anomalous dimensions, expressed as a function of $\tau$ are
\begin{align}
\gamma_0 & = -\frac{8 }{3\pi }\frac{{\rm Im} \tau}{|\tau|^2}+\frac{36\pi^2-32}{27\pi^2} \frac{({\rm Im} \tau)^2}{|\tau|^4}-\frac{8}{3}\frac{({\rm Re} \tau)^2}{|\tau|^4}+\mathcal{O}(|\tau|^{-3})~, \\
\gamma_2 & = \frac{8}{5\pi}\frac{{\rm Im} \tau}{|\tau|^2} -\frac{150 \pi^2+32}{375\pi^2}\frac{({\rm Im} \tau)^2}{|\tau|^4}  +\mathcal{O}(|\tau|^{-3})~\label{eq:anomalousO2}.
\end{align}
From these result we can immediately recover the anomalous dimensions for the 3d gauge theory $U(1)_k$ coupled to a Dirac fermion at large $k$ as explained in section \ref{sec:largeNlargek}. Since this is a local 3d theory, we expect $\gamma_2 = 0$ and indeed this is what we obtain from \eqref{eq:anomalousO2}. For the anomalous dimension of the scalar bilinear, that starts at two-loop order in this theory, we find
\begin{equation}
\gamma_0 = -\frac{8}{3 k^2} + \mathcal{O}(k^{-4})~,
\end{equation}
in agreement with \cite{Alves:1998mb}.\footnote{In \cite{Chen:1992ee} there appears to be a sign mistake in the two-loop diagram that we denoted with (b.2) in fig. \ref{fig:loopdiagram}. This mistake leads to the different result for this anomalous dimension given in \cite{Chen:1993cd}. Upon correcting that sign, we find perfect agreement with our result. We thank E. Stamou for helping us with this check.}

\subsection{Perturbative $F_\partial$}\label{sec:perturbativeFdelta}
\begin{figure}
	\centering
	\includegraphics[width=0.5\linewidth]{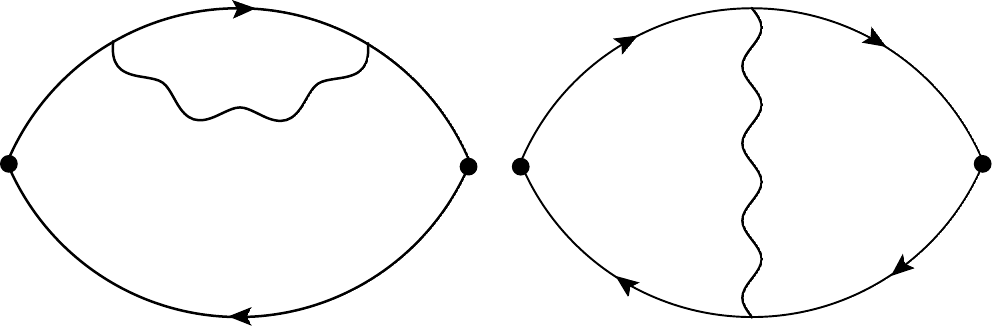}
	\caption{Leading corrections to the boundary current two-point function for the Dirac fermion.}
	\label{fig:currentcurrent}
\end{figure}

Thanks to the differential equation \eqref{eq:onepoint1}-\eqref{eq:onepoint2}, and to the relations derived in appendix \ref{app:current2ptfunctions}, the computation of the hemisphere free energy is reduced to the computation of the two-point functions of the boundary current $\hat{J}$. Up to next-to-leading order, we already wrote the universal formula \eqref{eq:Fcuspinf} for the hemisphere free energy in terms of the two-point function of the current $\hat{J}_{\text{CFT}}$ of the unperturbed CFT. In this particular setup where the boundary theory at $\tau \to \infty$ is a free Dirac fermion we can do better without much effort, because the correction to the current two-point function, given by the two diagrams in fig. \ref{fig:currentcurrent}, already exists in the literature. For the parity even part of the two-point function, we can either extract the value of these diagrams from the large-$N_f$ calculation of \cite{Giombi:2016fct}, using the similarities between the two perturbative expansions that we explained in \ref{sec:largeNlargek}, or alternatively use the computation performed directly in the mixed-dimensional setup in \cite{Teber:2012de, Teber:2016unz}.\footnote{In comparing with \cite{Teber:2012de, Teber:2016unz} one needs to take into account that they consider a 3d interface with the gauge field propagating on both sides, rather than a boundary. The propagator of the photon restricted to an interface has a factor of $\frac 12$ compared to the case of the boundary.} The parity-odd part can be obtained from the large-$k$ calculation in \cite{Spiridonov:1991ki}.
The sum of the diagrams in fig. \ref{fig:currentcurrent} is the next-to-leading order correction for the one-photon irreducible two-point function, which we denote by $\Sigma$, see appendix \ref{app:current2ptfunctions} for more details. Due to the shift in the real part of $\tau$, i.e. $\tau = \tau_{NN'} -\frac 12$, we have that $\kappa_\Sigma $ vanishes at leading order in perturbation theory, or equivalently $\kappa_J^{(0)} = 0$. The results mentioned above give 
\begin{align}
c_\Sigma & = \frac{1}{8\pi^2} + \frac{92-9\pi^2}{144\pi^3}\frac{{\rm Im}\tau}{|\tau|^2}   +\mathcal{O}(|\tau|^{-2})~,\\
\kappa_\Sigma & = \frac{4 + \pi^2}{16}\frac{{\rm Re}\tau}{|\tau|^2}   +\mathcal{O}(|\tau|^{-2})~.
\end{align}
Using \eqref{eq:cJfromS}-\eqref{eq:kJfromS} to obtain the total two-point function of $\hat{J}$, we find
\begin{align}
c_J & = \frac{1}{8\pi^2} +\frac{368-45\pi^2}{576 \pi^3}\frac{{\rm Im}\tau}{|\tau|^2}+\mathcal{O}(|\tau|^{-2})~,\\
\kappa_J & = \frac{16 + 5\pi^2}{64} \frac{{\rm Re}\tau}{|\tau|^2}+\mathcal{O}(|\tau|^{-2})~.
\end{align}
Plugging these values in the formulas \eqref{eq:V2V2}-\eqref{eq:V1V2} for $c_{22}$ and $c_{12}$, and solving the differential equations \eqref{eq:onepoint1}-\eqref{eq:onepoint2} we obtain
\begin{align}\label{FpartialWeak}
F_\partial & = -\frac{1}{4}\log\left[\frac{2\,{\rm Im}\tau}{|\tau|^2}\right]+ F_{\text{Dirac}}\nonumber \\& ~~~~~+\frac{\pi}{16}\frac{{\rm Im}\tau}{|\tau|^2}+ \frac{(368 -45 \pi ^2 )({\rm Im}\tau)^2 + (144 + 45 \pi^2 ) ({\rm Re}\tau)^2}{2304 |\tau|^4}+\mathcal{O}(|\tau|^{-3})~.
\end{align}
We fixed the integration constant by comparing with the decoupling limit, so that $F_{\text{Dirac}}$ is the $S^3$ free energy for a free Dirac fermion (two complex components) \cite{Klebanov:2011gs}
\begin{equation}
F_{\text{Dirac}} = \frac{\log 2}{4} + \frac{3 \zeta(3)}{8 \pi^2}~.
\end{equation}

\subsection{Extrapolations to the $O(2)$ Model}

We can now attempt to extrapolate the perturbative results obtained above around the Dirac fermion cusp to the $O(2)$ cusp (see fig. \ref{fig:tauplane}), to obtain the data of the $O(2)$ model. The $O(2)$ model, while being strongly coupled, is a well-studied theory via a variety of techniques, so that we can compare our extrapolations to the known data. Even though so far we only obtained the first two orders in perturbation theory, and one might be wary to already attempt an extrapolation, we will see that the results we obtain are compatible with the known data. We view this as an encouraging indication that the perturbative technique that we are presenting here can indeed be a useful tool to obtain data of 3d Abelian gauge theories, and as a motivation to try to obtain more precise predictions by going to higher orders.

In order to extrapolate, we need to apply a resummation technique. The nice property of our setup is the duality $\tau \leftrightarrow \tau'=-\frac{1}{4\tau}$, which means that the perturbative expansions obtained above also tell us about the behavior of the observables around $\tau' \to \infty$, i.e. the second Dirac fermion cusp. To leverage on this, the idea is to use a ``duality-improved'' Pad\'e approximant, i.e. a function with a number of free parameters that we can fix by matching to the perturbative result at $\tau\to\infty$, and that is invariant under a duality transformation.

Similar resummations were studied in the context of perturbative string theory \cite{Sen:2013oza} and $\mathcal{N}=4$ super Yang-Mills (SYM) in \cite{Beem:2013hha}. In particular \cite{Beem:2013hha} introduced Pad\'e-like approximants with the property of being invariant under a subgroup of $SL(2,\mathbb{Z})$, and we will borrow their method. Note that the perturbative results of the previous subsections, expressed in terms of $g_s = g^2$ and $\theta$, and expanded for small $g_s$ with $\theta$ fixed take the form
\begin{align}
\gamma_{0} & = -\frac{4 }{3 \pi ^2}g_s - \frac{8- 9 \pi ^2}{27 \pi ^4} g_s^2 +\mathcal{O}\left(g_s^3,g_s^3\theta^2\right)~,\\
\gamma_{2} & = \frac{4 }{5 \pi ^2}g_s -\frac{ 16 + 75 \pi ^2}{750 \pi ^4} g_s^2+\mathcal{O}\left(g_s^3,g_s^3\theta^2\right)~,\\
f_{\partial} &=\frac{1}{32}g_s +  \frac{368-45\pi^2}{9216\pi^2}g_s^2 +\mathcal{O}\left(g_s^3,g_s^3\theta^2\right)~,
\end{align}
and $f_\partial$ is the boundary free energy where the contributions from free gauge field as well as the constant term have been subtracted. The expressions above all have the pattern 
\begin{equation} 
a~g_s(1+ b~g_s+ \mathcal{O}(g_s^2,g_s^2\theta^2))~,
\end{equation} 
which can be approximated by the manifestly duality-invariant interpolation functions written in \cite{Beem:2013hha}. At this order, there are two of their functions that we can use, the integral-power Pad\'{e} $F_1(g_s,\theta)$ and half-integral-power Pad\'{e} $F_2(g_s,\theta)$, defined by
\begin{align}
F_1(g_s,\theta) &= \frac{h_1}{g_s^{-1} + (\mathrm{S}\cdot g_s)^{-1} - h_2}~,\label{eq:F1}\\
F_2(g_s,\theta) &= \frac{h_3 \left(g_s^{-1/2}+(\mathrm{S}\cdot g_s)^{-1/2}\right)}{g_s^{-3/2}+(\mathrm{S}\cdot g_s)^{-3/2}+h_4 \left(g_s^{-1/2}+(\mathrm{S}\cdot g_s)^{-1/2}\right)}~.\label{eq:F2}
\end{align}
where $\mathrm{S}\cdot g_s$ is the new gauge coupling under the transformation $\tau \rightarrow -\frac{1}{4 \tau}$, which reads explicitly
\begin{equation}
\mathrm{S}\cdot g_s = \frac{g_s^2 \theta ^2+16 \pi ^4}{\pi ^2 g_s}~.
\end{equation}
The unconventional negative power in the above two Pad\'{e} approximant was devised in \cite{Beem:2013hha} to remove the $\theta$ dependence in the Taylor expansion. This is appropriate to match our perturbative expansion up to the order we are considering, because the $\theta$-dependence starts at the subleading order $g_s^3$. On the other hand, while the perturbative expansion of $\mathcal{N}=4$ SYM is independent of $\theta$ to all orders in perturbation theory, and therefore in that context it is desirable to have an ansatz whose Taylor expansion does not contain $\theta$, in our setup observables do depend on $\theta$ even in perturbation theory. Indeed, by taking a different scaling such as $g_s$ small with $\gamma = \frac{\theta g_s}{4 \pi^2}$ fixed, rather than $\theta$ fixed, we would have a non-trivial dependence on $\gamma$ already at the order we are considering, and with this scaling we could not match the observables with the Taylor expansion of the approximants \eqref{eq:F1}-\eqref{eq:F2}. The upshot is that in order to use the duality-improved approximants from \cite{Beem:2013hha} we are forced to consider the expansion at small $g_s$ with $\theta$ fixed, and doing so we throw away some of the information contained in the perturbative calculation, namely the $\frac{ g_s^2\gamma^2}{(1+\gamma^2)^2} = (2\pi)^2\frac{({\rm Re }\tau)^2}{|\tau|^4}$ terms. It would be desirable to find an ansatz that is: (i) duality invariant; (ii) has a final limit to the real $\tau$ axis (or at least to the $O(2)$ cusp); and (iii) can be matched with the perturbative expansion at small $g_s$ and $\gamma$ fixed (at least up to the order $g_s^2$ at which the observables are currently known). 

By matching the coefficients in the expansion up to the order $g_s^2$, we find the unknown coefficients $h_i$ to be
\begin{align}
h_1 = a, \quad h_2 =  b, \quad h_3 = a, \quad h_4 = \frac{1}{4\pi} - b
\end{align}
In the table \ref{tab:values} we show the resulting values of the approximant extrapolated at the $O(2)$ point. 
\begin{table}
	\centering
	\begin{tabular}{|c|c|c|c|}
		\hline 
		& $2+ \gamma_1$ & $3 + \gamma_2$ & $f_\partial$ \\ 
		\hline 
		$\epsilon$ expansion & 1.494  & --- & 0.124  \\ 
		\hline 
		Bootstrap & $1.5117(25)$ & --- & --- \\
		\hline 
		$F_1(g_s = \infty,\theta = \pi)$ & 1.406 & 3.635 & 1.039  \\ 
		\hline 
		$F_2(g_s = \infty,\theta = \pi)$ & 1.560 & 3.391 &  0.166 \\ 
		\hline 
	\end{tabular}
	\caption{Comparison of the extrapolations with the known data: for the energy operator we are quoting the value from the conformal bootstrap \cite{Kos:2015mba}, and from the $\epsilon$-expansion \cite{Kleinert:1991rg}. For the sphere free energy we are comparing to the value from the $\epsilon$-expansion in \cite{Fei:2015oha}.}\label{tab:values}
\end{table}
The fermion-mass operator is mapped to the energy operator of the $O(2)$ model, whose dimension can be obtained for instance from the conformal bootstrap \cite{Kos:2015mba}, or from the $\epsilon$-expansion \cite{Kleinert:1991rg}. The spin 2 operator is expected to approach the conserved stress-energy tensor on the boundary in the decoupling limit, hence the dimension should approach the protected value $\Delta_2|_{\text{ cusp}} = 3$. As for the hemisphere free energy, one needs to subtract a finite contribution coming from the decoupled gauge field at the $O(2)$ cusp, and the remaining constant gives the sphere free energy of the $O(2)$ model. To our knowledge this has only been computed using $\epsilon$-expansion \cite{Fei:2015oha}. 

\begin{figure}[htb]
	\centering
	\includegraphics[width=0.46\linewidth]{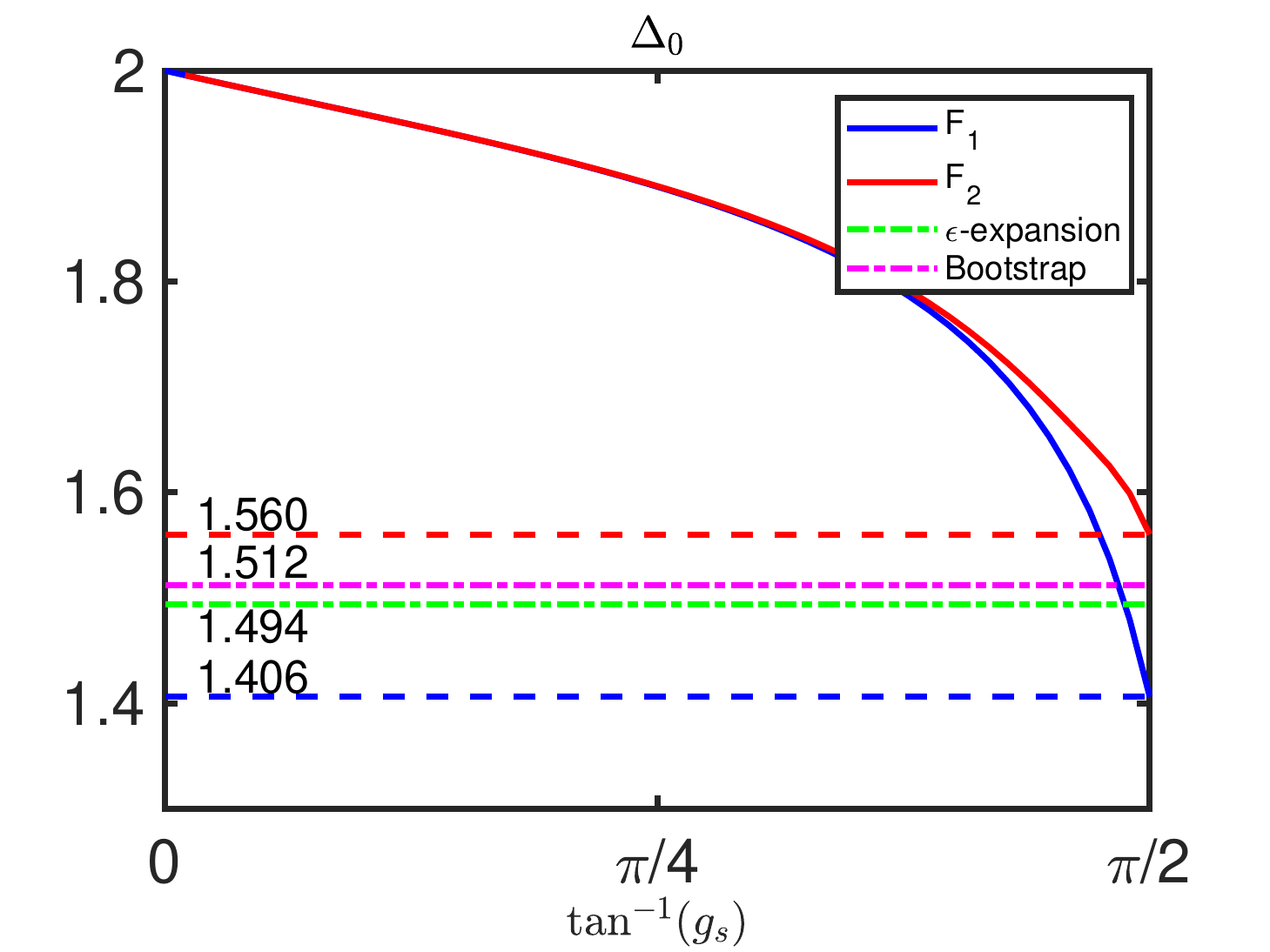}
	\includegraphics[width=0.46\linewidth]{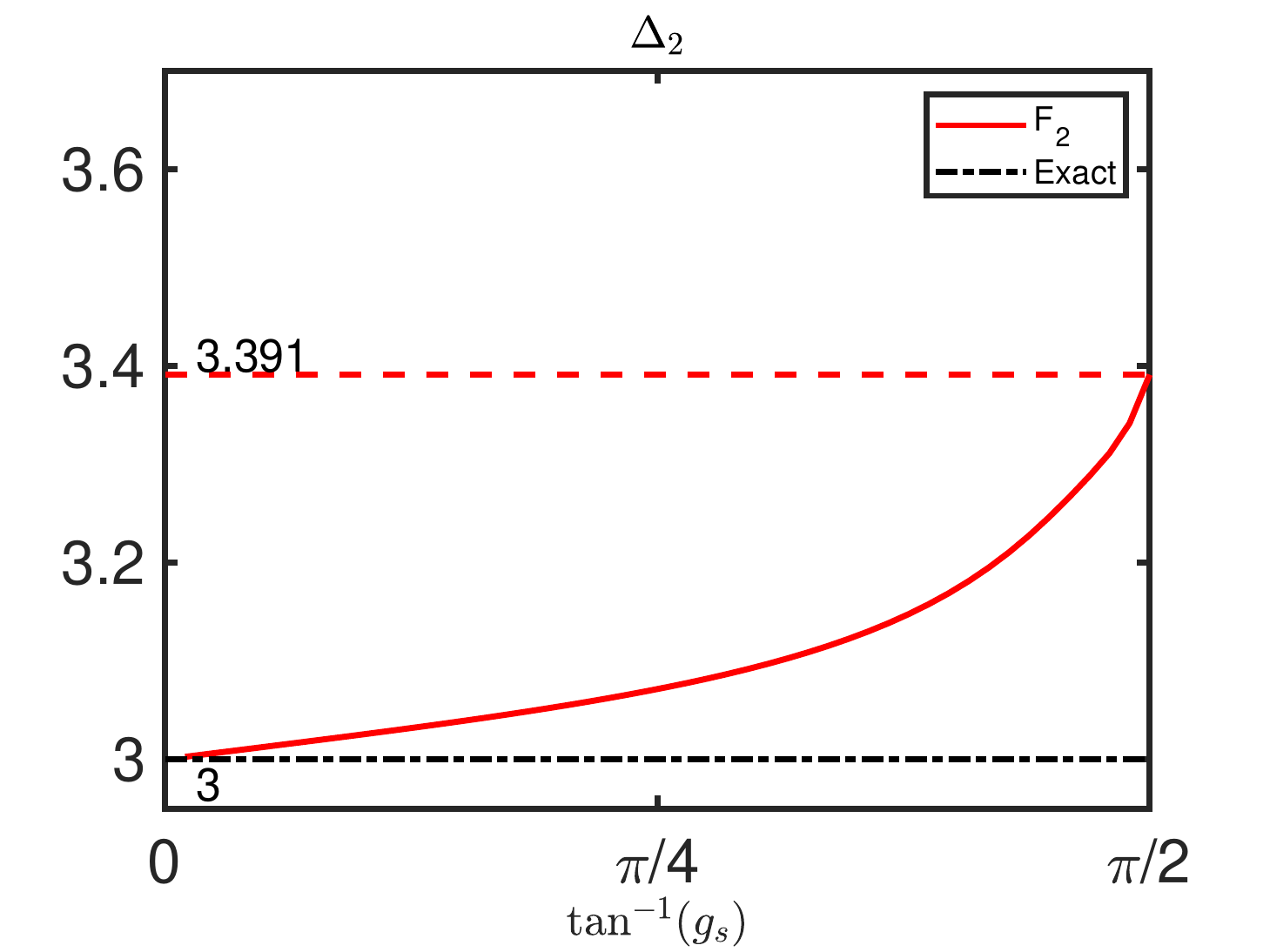}
	\caption{Extrapolations of the scaling dimensions from the Dirac fermion point ($\tan^{-1}(g_s) = 0$) to the $O(2)$ point ($\tan^{-1}(g_s) = \pi/2$).}
	\label{fig:Pade}
\end{figure}

We see that both ansatzes give good approximations for the dimension of the energy operator, and in particular $F_2$ is quite close to the values obtained with the other methods. For the other two observables, we see that the ansatz $F_2$ also gives compatible results, while $F_1$ is not as good. In fig.\ref{fig:Pade} we show the plots of the approximants at $\theta = \pi$, i.e. the value of the $O(2)$ cusp, as a function of $g_s$ from $0$ to $\infty$. 
\begin{figure}[htb]
	\centering
	\includegraphics[width=0.46\linewidth]{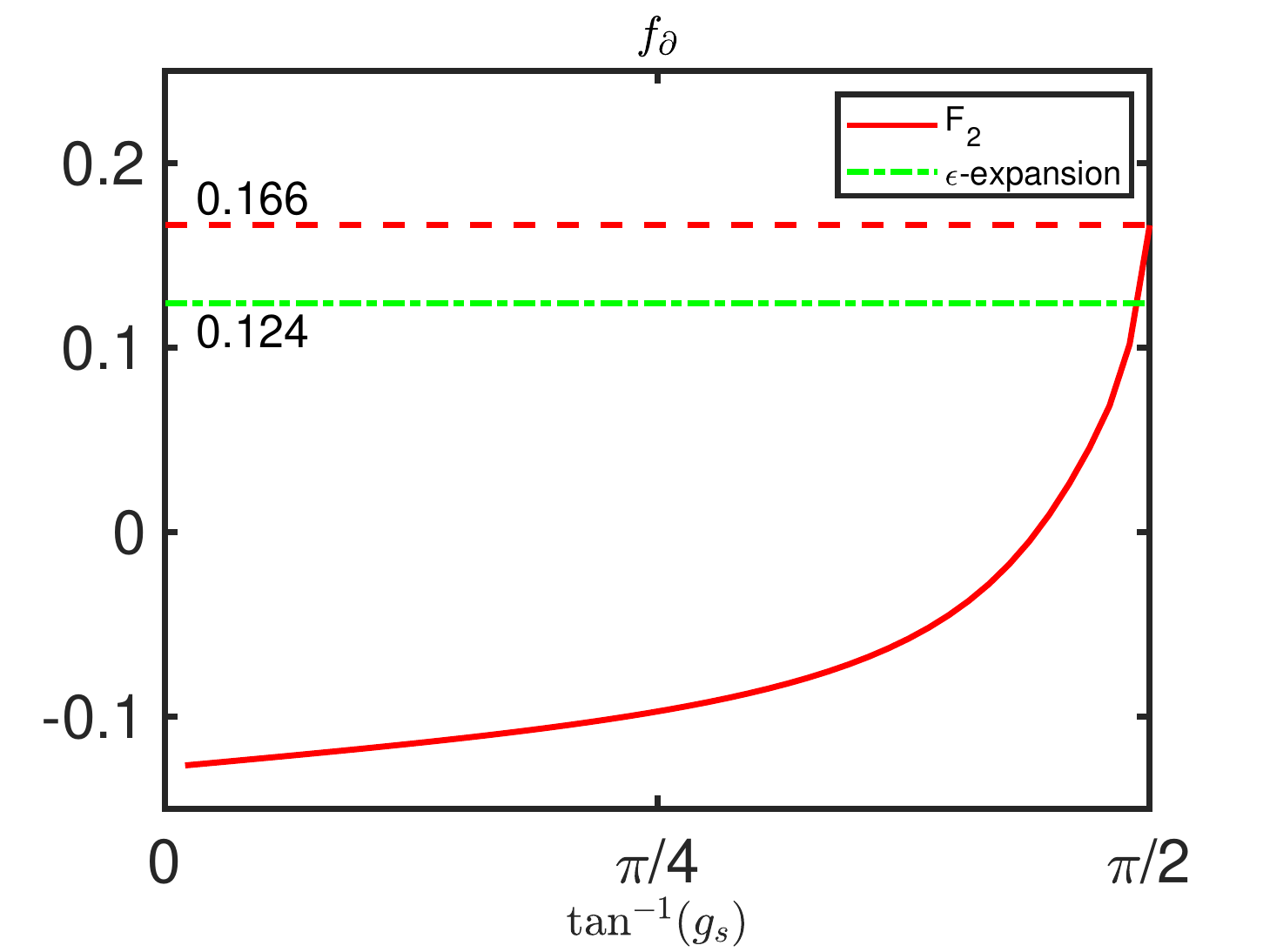}
	\caption{Extrapolations of the free energy from the Dirac fermion point ($\tan^{-1}(g_s) = 0$) to the $O(2)$ point ($\tan^{-1}(g_s) = \pi/2$).}
	\label{fig:Pade_freeenergy}
\end{figure}

\section{Other Examples}
\label{sec:exa}

\subsection{$2N_f$ Dirac fermions at large $N_f$}

In this section we consider the coupling of $2N_f$ Dirac fermions to the bulk gauge fields, all with the same charge $q=1$, and we take the limit of large $N_f$ with $\lambda = g^2 N_f$ fixed. For simplicity we take $\theta = 0$. We will see that computing observables in $1/N_f$ expansion, and later taking the limit $\lambda \to \infty$, one can recover the $1/N_f$ expansion in QED$_3$. This would be the expected result if we would take $g^2 \to \infty$ first, obtaining the decoupling limit in which on the boundary we have QED$_3$ with $2N_f$ flavors, and later take $N_f$ large. Hence, the observation here is that these two limits commute. This is interesting because order by order in $1/N_f$ we can explicitly follow observables as exact functions of $\lambda$, and see how they interpolate from the ``ungauged cusp'' at $\lambda = 0$ to the ``gauged cusp'' at $\lambda =\infty$.

\begin{figure}
\center
\includegraphics[scale=0.75]{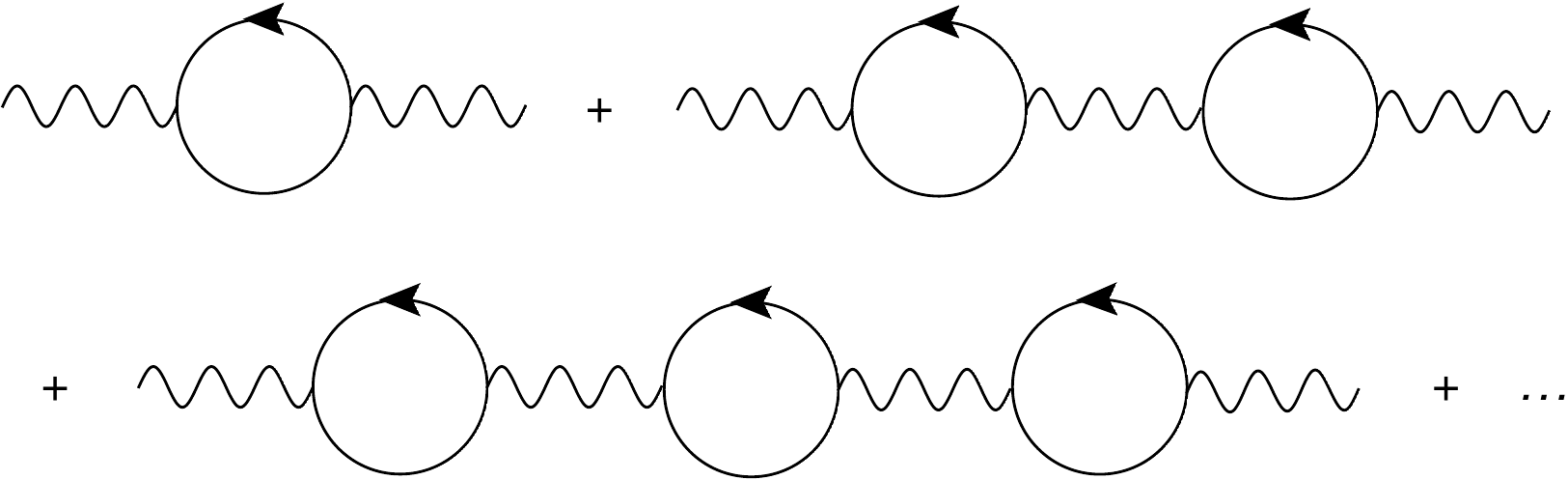}
\caption{The diagrams that contribute to the boundary propagator of the photon in the limit $N_f \to \infty$ with $\lambda = g^2 N_f$ fixed.}
\label{fig:LargeNf}
\end{figure}

To derive that the limits commute, it is sufficient to observe that in the limit of large $N_f$ with $\lambda = g^2 N_f$ fixed we can obtain an effective propagator for the photon by resumming the fermionic bubbles, see Fig. \ref{fig:LargeNf}, obtaining (up to gauge redundancy)
\begin{align}
\Pi^{(1/N_f)}_{a b}(\vec{p}) = & \frac{1}{N_f |\vec{p}\,|} \lambda \sum_{k=0}^{\infty}\left(- \frac{\lambda}{8}\right)^k\left(\delta_{ab} - \frac{p_a p_b}{\vec{p}^{~2}}\right)  \\
= & \frac{8 }{N_f |\vec{p}\,| }\frac{\lambda}{\lambda + 8} \left(\delta_{ab} - \frac{p_a p_b}{\vec{p}^{~2}}\right) ~.\label{eq:effphpr}
\end{align}
In the limit $\lambda \to \infty$ the propagator becomes
\begin{equation}
\Pi^{(1/N_f)}_{a b}(\vec{p}) \underset{\lambda\to\infty}{\longrightarrow} \frac{8}{N_f |\vec{p}\,| } \left(\delta_{ab} - \frac{p_a p_b}{\vec{p}^{~2}}\right) ~,
\end{equation}
which coincides with the effective propagator in QED$_3$ at large $N_f$. It follows that compared to the large-$N_f$ expansion of QED$_3$, in this setup the diagrams that compute $1/N_f$ corrections are simply dressed by a factor $\lambda/(\lambda + 8)$ for each photon propagator. In particular the $1/N_f$-expansion of observables, e.g. boundary scaling dimensions, will approach the corresponding value in large-$N_f$ QED$_3$ upon taking the limit $\lambda \to\infty$. However, recall that in the $1/N_f$-expansion diagrams that contribute at the same order might have different number of internal photon lines, so we cannot just replace $1/N_f$ with $1/N_f\times\lambda/(\lambda + 8)$ everywhere to obtain the exact dependence on $\lambda$ of a certain observable.

Let us now consider the two-point function of the boundary current $\hat{J}$, and obtain from it the hemisphere free energy at large $N_f$. We can obtain the $1/N_f$ correction to the one-photon irreducible two-point function of $\hat{J}$ ---computed by the diagrams in Fig. \ref{fig:currentcurrent} with the effective photon propagator \eqref{eq:effphpr}--- by taking the result of the large-$N_f$ calculation in \cite{Giombi:2016fct} and dressing it by the factor due to the single photon propagator, with the result
\begin{align}
c_\Sigma=\frac{N_f}{4 \pi ^2} \left(1+\frac{1}{N_f}\frac{\lambda}{\lambda+8}\frac{184 - 18 \pi^2}{9 \pi ^2}+{\mathcal{O}}\left({N_f^{-2}}\right)\right)~.
\end{align}
Correspondingly, from equation \eqref{eq:V2V2} and \eqref{eq:V1V1} we have $c_{12}=0$ and
\begin{align}
c_{22} = \frac{16 }{\pi ^2 N_f}\frac{\lambda}{\lambda + 8}-\frac{32 \left(92 - 9 \pi ^2\right)}{9 \pi ^4 N_f^2}\frac{\lambda^3}{(\lambda + 8)^3}+\mathcal{O}\left(N_f^{-3}\right)~.
\end{align}
We can now plug $c_{22}$ in the differential equation \eqref{eq:onepoint1}, appropriately rewritten in terms of the variable $\lambda$. Solving for $F_\partial(\lambda)$ up to the order $1/N_f$ we find
\begin{align}\label{eq:Flambda}
F_\partial(\lambda )= \frac{1}{4}\log \left[\frac{\pi N_f (\lambda + 8)^2}{64 \lambda}\right] + 2N_f
F_{\text{Dirac}}+\frac{\left(92-9 \pi ^2\right) }{18 \pi ^2 N_f }\frac{\lambda^2}{(\lambda + 8)^2} +{\mathcal{O}}\left({N_f^{-2}}\right)~.
\end{align}
Recall that the arbitrary integration constant is fixed by matching with the decoupling limit. In the decoupling limit $F_\partial$ is the sum of a contribution from the free fermions on the boundary, namely $2N_f F_{\text{Dirac}}$, and a contribution from the boundary value of the gauge field with Neumann condition, that we discussed in section \ref{sec:FreeEnergy}. The latter contribution is only a function of $g^2$, and when rewritten in terms of $\lambda$ it gives a $\log(N_f)$ constant term. Hence we need to include such a dependence on $N_f$ in the integration constant, and this is how we obtain the $\log(N_f)$ term in \eqref{eq:Flambda}. Similarly, we find that a $\lambda$-independent term of order $1/N_f$ needs to be included in the integration constant, to ensure that the $1/N_f$ correction vanishes when $\lambda = 0$. The general lesson here is that when we integrate the equation in the $\lambda$ variable, the integration constant required to reproduce the decoupling limit will be a non-trivial function of the parameter $N_f$.

From the $\lambda \to \infty$ limit of \eqref{eq:Flambda} we can extract the sphere free-energy QED$_3$ at large $N_f$. More specifically, the latter is obtained by subtracting to the $\lambda\rightarrow \infty$ limit of the free energy the contribution of the Neumann boundary condition of the bulk gauge field computed at $(g')^2=\frac{4\pi^2}{g^2}$, namely
\begin{align}
F_{\text{QED}_3}& = \lim_{\lambda \to \infty}\left(F_\partial(\lambda)  + \left. \frac{1}{4}\log \left[\frac{(g')^2}{\pi}\right]\right\vert_{(g')^2 = \frac{4\pi^2 N_f}{\lambda}}\right) \\
& =2N_f
F_{\text{Dirac}}+\frac{1}{2} \log \left(\frac{\pi  N_f}{4}\right)+\frac{92 -9\pi^2}{18 \pi ^2}\frac{1}{N_f}+{\mathcal{O}}\left({N_f^{-2}}\right)~.\label{eq:largeNfF}
\end{align}
Both the logarithmic and the constant terms reproduce perfectly the result of \cite{Klebanov:2011td}. To our knowledge, the ${\mathcal{O}}\left({N_f^{-1}}\right)$ correction was not computed before.

As we will now briefly review, the free-energy as a function of $N_f$ can be used to diagnose the IR fate of QED$_3$. For $N_f$ smaller than a critical value $N_f^c$ the theory is conjectured to flow to a flavor-symmetry breaking phase rather than to the conformal phase that exists at large $N_f$. A possible scenario for the transition is that the IR scaling dimension of singlet four-fermion operators would cross marginality \cite{Braun:2014wja, DiPietro:2015taa, DiPietro:2017kcd}, implying that the IR fixed point that exists at large $N_f$ merges at $N_f = N_f^c$ with a second fixed point in which the quartic operators are turned on, and they both disappear \cite{Giombi:2015haa, Gukov:2016tnp}. After the transition they can still be interpreted as complex fixed points \cite{Kaplan:2009kr, Gorbenko:2018ncu}. This scenario was recently investigated in \cite{Benvenuti:2018cwd,Benvenuti:2019ujm} using large $N_f$ techniques and in \cite{Li:2018lyb} using the conformal bootstrap. This merger/annihilation scenario, together with the monotonicity of the sphere free-energy along RG, was used in \cite{Giombi:2015haa} to constrain $N_f^c$: assuming that $F_{\text{QED}_3}$ can still be interpreted as the free-energy of the nearby complex fixed point when $N_f < N_f^c$, the existence of the RG flow from the vicinity of the complex fixed point towards the symmetry breaking phase requires that $F_{\text{QED}_3} > F_{\text{G.B.}}$ for $N_f < N_f^c$. Here $F_{\text{G.B.}} = (2 N_f^2 +1) F_{\text{scalar}} $ is the free energy of the Goldstone bosons in the symmetry breaking phase. As an application of the calculation above, we can now run this argument using the large-$N_f$ approximation for $F_{\text{QED}_3}$ in eq. \eqref{eq:largeNfF}. It turns out that the coefficient of the $1/N_f$ term is numerically very small, i.e. $\sim 0.02$, so for the interesting values of $N_f$ of order 1 it does not affect significantly this test, and the resulting estimate is $N_f^c \sim 4.4$. For this value of $N_f$, the $1/N_f^2$ corrections that we are neglecting in \eqref{eq:largeNfF} are quite small, and assuming that the smallness of the coefficients persists at higher orders this suggests that the estimate might be reliable. 

\subsection{Complex Scalar}\label{sec:Scalar}

In section \ref{sec:MinPhTr} we studied the case a free fermion on the boundary, and we saw that one of the gauged cusps correspond to the $O(2)$ Wilson-Fisher model. This is a consequence of the boson/fermion dualities that relate a gauged fermion to a critical scalar, or a gauged critical scalar to a free fermion \cite{Seiberg:2016gmd}. These dualities can be seen as the low-rank analogue of the large-$N$ regular fermion/critical scalar dualities in CS-matter theories \cite{Giombi:2011kc, Aharony:2012nh, GurAri:2012is, Aharony:2015mjs}. Besides the Wilson-Fisher fixed point, the scalars also admit the Gaussian fixed point consisting of $N$ free complex scalars. Likewise the theory of $N$ Dirac fermions is conjectured to have a second fixed point with quartic interactions turned on, i.e. the UV fixed point of the Gross-Neveu model. The corresponding CS-matter theories are also conjectured to be dual in a level-rank duality fashion, giving the so-called regular boson/critical fermion duality. There is a large amount of evidence for this duality at large $N$, and its extension to finite $N$ was recently studied in \cite{Aharony:2018pjn, Dey:2018ykx}. It is not clear whether the duality still holds when $N=1$. Assuming it does, it would have a nice manifestation in our setup: by starting with a free complex scalar on the boundary, one would find that the cusp at $\tau =1$ corresponds to the Gross-Neveu CFT with 1 Dirac fermion.\footnote{The Gross-Neveu CFT is expected to exist also for a small number $N$ of Dirac fermion, the UV completion being provided by a Yukawa theory. See \cite{Fei:2016sgs} for a recent study in $\epsilon$-expansion.} One crucial new ingredient of the regular boson/critical fermion dualities is the existence of a additional sextic couplings that are classically marginal and potentially lead to multiple fixed points that can be mapped across the duality. 

With this motivation in mind, we will now consider the setup with a free complex scalar on the boundary, coupled to the bulk gauge field. The action is
\begin{equation}
	S[A,\tau] + \int_{y = 0} d^3\vec{x} \ \big(|D_A\phi|^2 + \rho(|\phi|^2)^3\big)~. 
\end{equation}
The couplings $|\phi|^2$ and $|\phi|^4$ are fine-tuned to zero. This fine-tuning might need to be adjusted as a function of the bulk gauge coupling. At least for $\tau$ large enough, these operators are relevant and correspondingly the beta function is linear in the couplings, so this adjustment is possible. On the other hand, the beta function for the classically marginal operator $|\phi|^6$ will start quadratically in $\rho$ and we need to check the existence of (real) fixed points. 

\begin{figure}[h]
	\centering
	\includegraphics[width=0.8\linewidth]{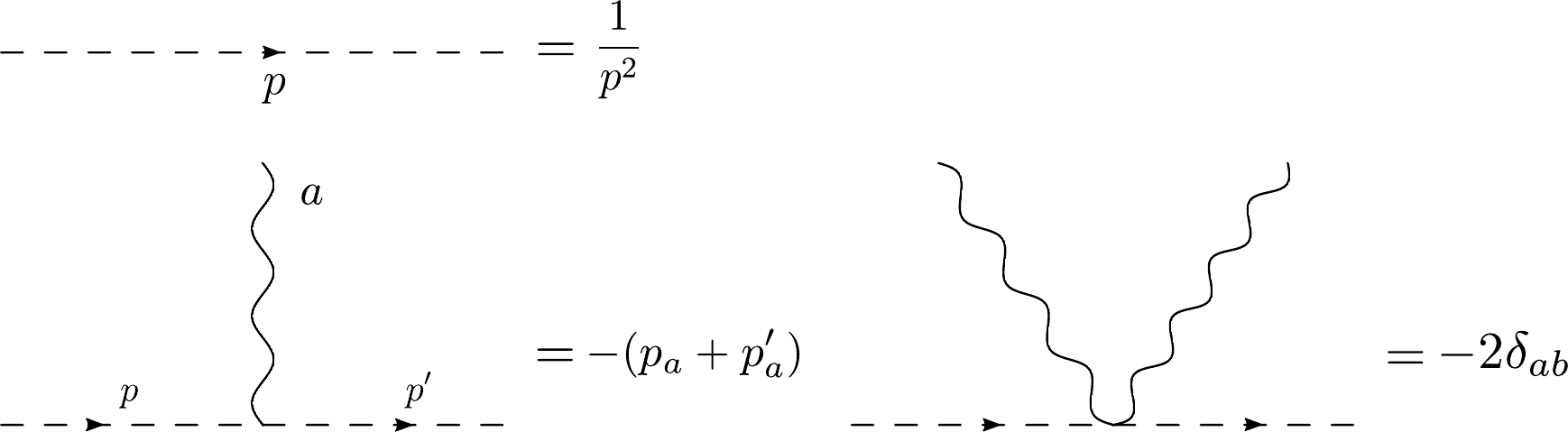}
	\caption{Feynman rules with the complex scalar on the boundary}
	\label{fig:scalarfeynrule}
\end{figure}

We list the Feynman rules in the Fig.~\ref{fig:scalarfeynrule}.

\begin{figure}[h]
	\centering
	\includegraphics[width=0.8\linewidth]{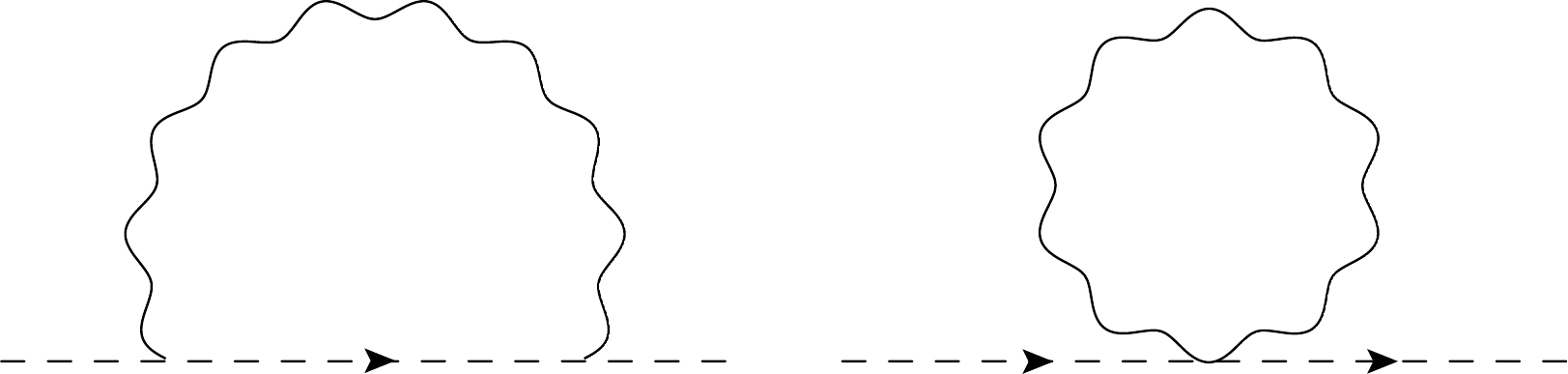}
	\caption{One loop diagrams that contribute to the wave-function renormalization. }
	\label{fig:scalarZphioneloop}
\end{figure}

To compute the $\beta$ function of $\rho$ we need to renormalize the six-point vertex. We use the same approach as in the fermion case, i.e. we dimensionally regularize by continuing the dimension of the boundary to $d = 3-2\epsilon$, keeping the codimension fixed $=1$. The boundary action in renormalized variables is
\begin{equation}
\int_{y=0} d^d \vec{x}\ |D\phi_B|^2 + \rho_B |\phi_B|^6 = \int_{y=0} d^d \vec{x} \ Z_\phi^2|D\phi|^2 + Z_\rho \rho \mu^{4\epsilon}|\phi|^6 ~,\end{equation}
where the subscript $B$ denotes the bare variables. 
Fig. \ref{fig:scalarZphioneloop} shows the diagrams that contribute to the wavefunction renormalization of the field $\phi$, from which we obtain 
\begin{equation}
\delta Z_\phi = -\frac{ (3 \xi -8)}{24 \pi ^2 \epsilon }\frac{g^2}{1+\gamma^2} + \mathcal{O}(g^4)~.\label{eq:anodimphi}
\end{equation}

\begin{figure}[htb]
	\centering
\includegraphics[width = \linewidth]{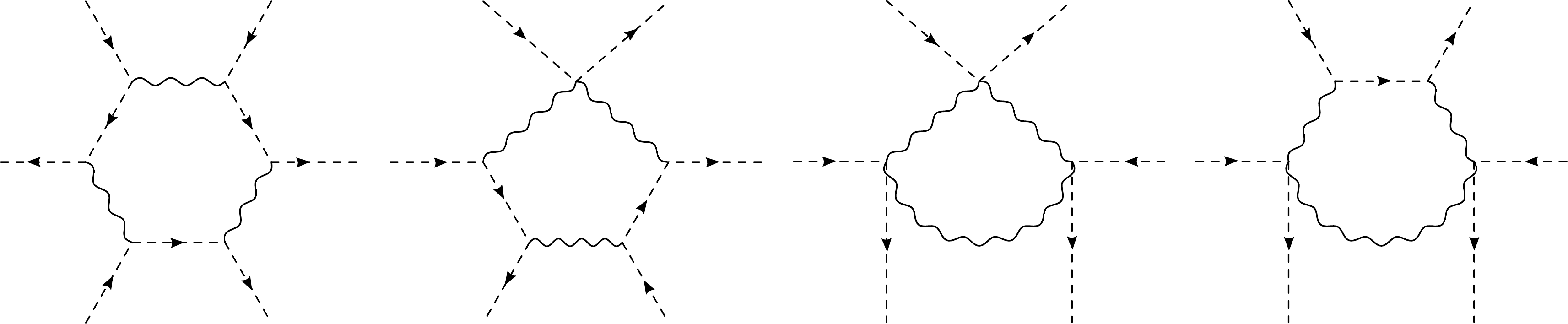}
	\caption{Diagrams contributing to $\mathcal{O}(g^6)$ in $\beta_\rho$.}
	\label{fig:scalarZrhoa06}
\end{figure}

\begin{figure}[htb]
	\centering
	\includegraphics[width=0.75\linewidth]{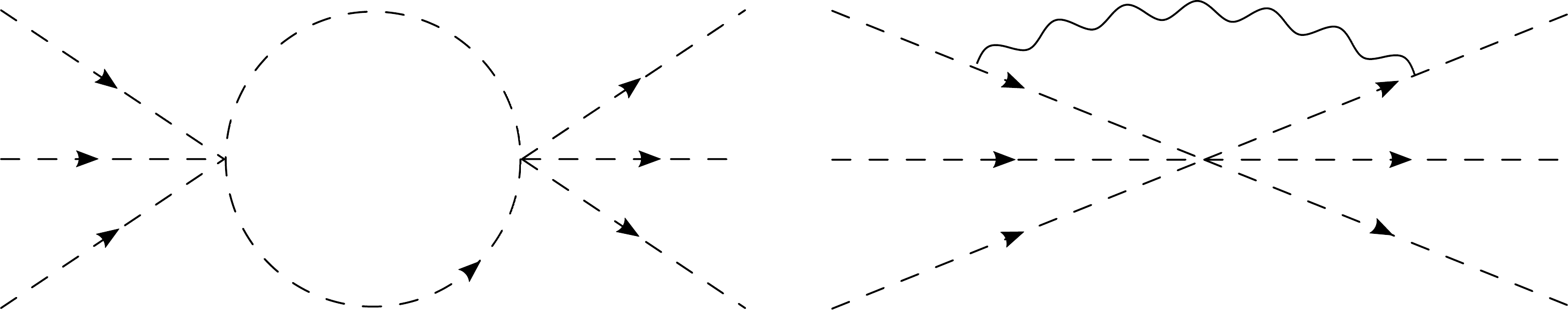}
	\caption{Diagrams contributing to $\mathcal{O}(\rho^2)$ and $\mathcal{O}(\rho g^2)$ in $\beta_\rho$.}
	\label{fig:scalarZrhoa12}
\end{figure}

There are three types of diagram contributing to the six-point vertex counterterm, showed in Fig. \ref{fig:scalarZrhoa06} and \ref{fig:scalarZrhoa12}, from which we can compute
\begin{equation}
\rho\delta Z_\rho = \frac{15}{8 \pi ^2 \epsilon } \rho^2 - \frac{3}{4\pi^2\epsilon}\frac{g^2}{1+\gamma ^2}\xi\,\rho  -\frac{24(1-3\gamma^2)}{\pi ^2 \epsilon }\left(\frac{g^2}{1+\gamma ^2}\right)^3 + \mathcal{O}(\rho^3, \rho^2 g^2, \rho g^4, g^8)~.
\end{equation}

The $\beta$ function is 
\begin{align}
\beta_\rho(\rho,g)  &=\left.\left(- 4\epsilon\rho - \rho \frac{\partial \log Z_\rho/Z_{\phi}^6}{\partial \log\mu} \right)\right\vert_{\epsilon=0}\\
& = \frac{15}{2\pi^2} \rho^2 -\frac{4}{\pi^2}\rho\frac{g^2}{1+\gamma^2}  -\frac{48(1-3\gamma^2)}{\pi^2}\left(\frac{g^2}{1+\gamma^2}\right)^3 + \mathcal{O}(\rho^3, \rho^2 g^2, \rho g^4, g^8)~.
\end{align}
Up to this order we find: a zero at $\rho=\rho_*^{+} > 0$ from the first two terms, and since $\rho_*^{+} = \mathcal{O}(g^2)$ the third term is negligible; and a zero at $\rho=\rho_*^{-}$ from the second and the third therm, and since $\rho_*^{-} = \mathcal{O}(g^4)$ the first term is negligible. The positions of the zeroes are
\begin{equation}
\rho_*^{+} = \frac{8}{15}\frac{g^2}{1+\gamma^2} + \mathcal{O}(g^4)~, \qquad \rho_*^{-} = -12(1-3\gamma^2)\left(\frac{g^2}{1+\gamma^2}\right)^2+ \mathcal{O}(g^6)~. 
\end{equation}
The derivative of $\beta_\rho$ is positive at $\rho_*^+$ and negative at $\rho_*^-$. Hence we have found that perturbatively around large $\tau$ there exists a fixed point $\rho=\rho_*^+$ which is IR stable, and gives a scalar potential bounded from below. The fixed point $\rho_*^-$ on the other hand is only physical for $1-3\gamma^2<0$, because otherwise it gives the wrong sign of the scalar potential, and it is unstable in the RG sense.

Having checked the existence of the fixed point in perturbation theory, we proceed to consider the anomalous dimension of boundary operators in this theory, similarly to what we did in section \eqref{eq:pertscaldim} for the fermion case. We consider the mass-squared operator $O = |\phi|^2$. Its anomalous dimension can be obtained from the renormalization of the 1PI correlator of the composite operator with two elementary fields 
\begin{equation}
\langle O(q=0) \phi(-p) \bar{\phi}(p)\rangle_{\rm 1PI}~. \label{eq:scalar3pointfn}
\end{equation}

\begin{figure}[h]
	\centering
	\includegraphics[width=\linewidth]{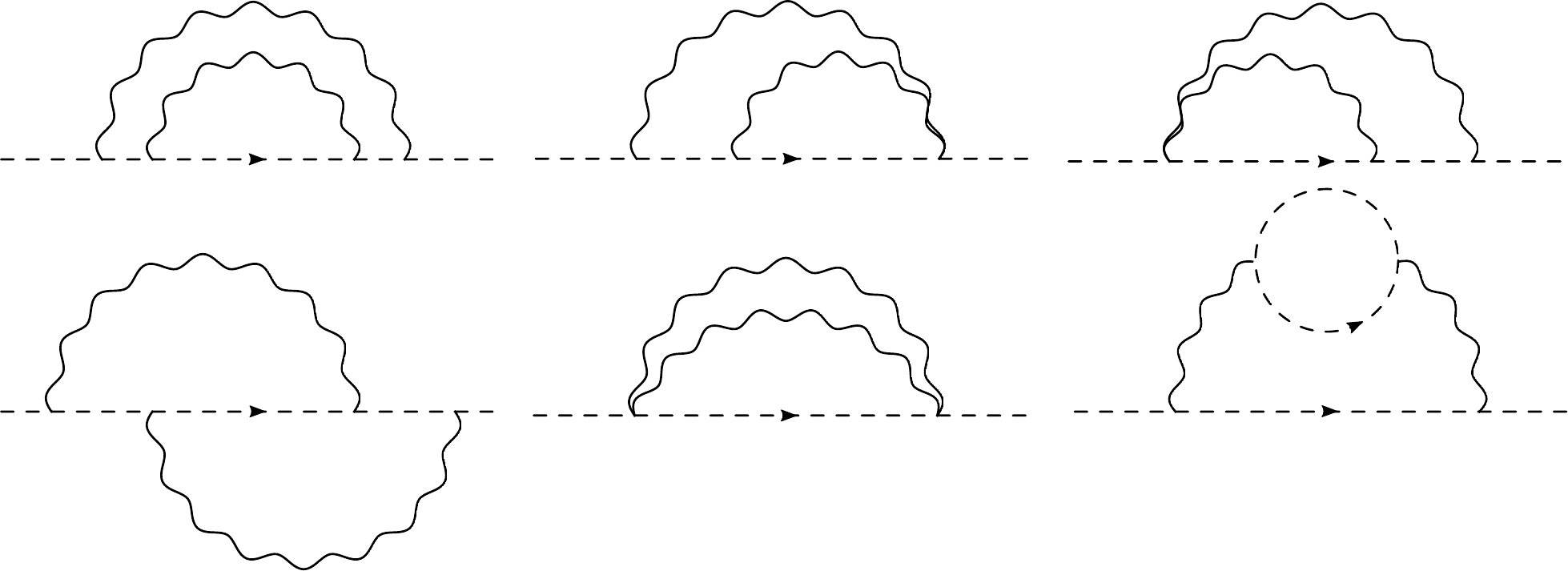}
	\caption{One loop and two loops diagrams}
	\label{fig:scalarZmTwoloop}
\end{figure}

The one-loop (two-loop) diagrams contributing to the three-point function \eqref{eq:scalar3pointfn} are showed in Fig.\ref{fig:scalarZphioneloop} (Fig.\ref{fig:scalarZmTwoloop}, respectively).

At one loop, using \eqref{eq:anodimphi}, the renormalization constant of the operator is found to be
\begin{equation}
\delta Z_{O} = -\frac{2}{3\pi^2\epsilon} \frac{g^2}{1+\gamma^2} + \mathcal{O}(g^4)~,
\end{equation}
and correspondingly the anomalous dimension is
\begin{equation}
\gamma_O = -\frac{4}{3\pi^2}  \frac{g^2}{1+\gamma^2}+ \mathcal{O}(g^4)~.
\end{equation}

Differently from the fermion case, we were not able to evaluate all of the dimensionally-regularized integrals coming from the two-loop diagrams of Fig. \ref{fig:scalarZmTwoloop}. See the appendix \ref{app:FeynInt} for the details. Knowing the two-loop anomalous dimension would enable an extrapolation to $\tau=1$ that could be compared with the known estimates of the mass operator in the Gross-Neveu CFT. This is therefore an interesting direction left for the future.

\subsection{QED$_3$ with Two Flavors}\label{sec:QEDtwoFlavors}

In this section we will discuss a realization in our setup of QED$_3$ coupled to two Dirac (complex two-component) fermions of charge $1$. There are several reasons why this is an interesting theory: it is conjectured to describe the easy-plane version of the ``deconfined'' N\'eel-VBS quantum phase transition in antiferromagnets \cite{PhysRevB.70.144407}, and enjoy an emergent $O(4)$ symmetry \cite{Hsin:2016blu, Cordova:2017kue}; while initially believed to be a second-order transition, recent evidences from simulations of the spin system on the lattice \cite{Qin:2017cqw} and from the conformal bootstrap \cite{IliesiuTALK} suggest that this is actually a weakly first-order transition, which can still be compatible with the QED description if the latter has a complex fixed point with $O(4)$ symmetry (see section 5 of \cite{Gorbenko:2018ncu} and \cite{Benvenuti:2018cwd}); it is conjectured to enjoy a self-duality \cite{Hsin:2016blu, Cordova:2017kue, Xu:2015lxa, Karch:2016sxi} and a fermion-boson duality \cite{Wang:2017txt}. 

A simple way to realize QED with two flavors in our setup would be to put the CFT of two Dirac fermions on the boundary, and couple a bulk gauge field to the $U(1)$ symmetry that gives charge 1 to both of them. However in this case we only expect a weakly coupled cusp at $\tau \to \infty$. For the purpose of attempting an extrapolation from weak coupling, it would be desirable to have additional weakly coupled cusps, as in the example of section \ref{sec:MinPhTr}. With this idea in mind, a more promising approach is to consider a generalization of the former set-up in which we have two Maxwell gauge fields in the bulk and two Dirac fermions on the boundary, namely two decoupled copies of the theory of section \ref{sec:MinPhTr}. By performing an $S$-duality for either of the two gauge fields separately we find again two free Dirac fermions on the boundary. On the other hand using the larger electric-magnetic duality group that exists for a theory of two gauge fields, we can also go to a duality frame where in the decoupling limit we have precisely QED with two flavors on the boundary. 

In the rest of this section we will first review electric-magnetic duality for multiple Maxwell fields, and then show how to get QED with two flavors starting with two copies of a bulk gauge field coupled to a boundary Dirac fermion. The task of performing perturbative calculations of observables in this theory is left for the future.

\subsubsection{Multiple Maxwell Fields}

The action of free bulk $U(1)^n$ gauge theory is determined in terms of $n$ Abelian gauge fields $A^{I}$, such that $F^I=d A^I$ and an $n \times n$ symmetric matrix of complexified gauge couplings $\tau_{IJ}$
\begin{align}
S[A^I,\tau_{IJ}] &= \int_{y\geq 0} d^4x \left(\frac{1}{4g^2_{IJ}} F^I_{\mu\nu} F^{J,\mu\nu} +\frac{i\theta_{IJ}}{32\pi^2} \epsilon_{\mu\nu\rho\sigma}F^{I,\mu\nu} F^{J,\rho\sigma}\right)\\
&= -\frac{i}{8\pi} \int_{y\geq 0} d^4x (\tau_{IJ} F^{-I}_{\mu\nu}F^{-J,\mu\nu}-\bar{\tau}_{IJ}F^+_{I\mu\nu}F^{+J\mu\nu}),
\end{align}
where $\tau_{IJ}=\frac{\theta{IJ}}{2\pi}+\frac{2\pi i}{g_{IJ}^2}$ and we introduced $F^{\pm,I}_{\mu\nu} = \frac12 (F^I_{\mu\nu} \pm \frac12 \epsilon_{\mu\nu\rho\sigma} F^{I,\rho\sigma})$. This theory enjoys an $Sp(2n,\bZ)$ duality group
\begin{equation}
\tau'_{IJ}= 
({A_I^K\tau_{LM} + B_{IM}})({C^{JN}\tau_{NM} + D^J_M})^{-1},
\end{equation}
where 
\begin{equation}
M =  \left(\begin{array}{cc}
A & B \\ 
C & D
\end{array} \right) \in Sp(2n,\bZ)~.
\end{equation}
This duality group is generated by the three types of transformations obtained in \cite{hua_generators_1949,Dimofte:2011ju}, which we reproduce here \footnote{More precisely, these elements generate $Sp(2n,\bZ)/\sim$, where we identify $S \sim -S$.} 

\begin{align}
\text{T-type:}\, & \left(\begin{array}{cc}
I & B \\ 
0 & I
\end{array} \right),\,
&
\begin{array}{ll}
\text{where $I$ the $n\times n$ identity and $B$ is a symmetric}\\
\text{matrix that generates $\tau' = \tau+B$,}
\end{array}
\end{align}

\begin{align}
\text{S-type:}\, & \left(\begin{array}{cc}
I-J & -J \\ 
J & I-J
\end{array} \right),\,
&
\begin{array}{ll}
\text{where $J= \text{diag}(j_1,j_2,\dots,j_n)$ and $j_i \in \{0,1\}$.}\\
\text{This gauges those $A_i$'s that have $j_i =1$.}
\end{array}
\end{align}

\begin{align}
\text{GL-type:}\, & \left(\begin{array}{cc}
U & 0 \\ 
0 & U^{-1T}
\end{array} \right),\,
&
\begin{array}{ll}
\text{where $U\in SL(n,\bZ)$ generate the rotations $A' = U^{-1T} A$.}
\end{array}
\end{align}

In the rest of this section we will be focusing on the case of $n=2$. Following \cite{hua_generators_1949} we define the generators of $Sp(4,\bZ)$ as
\begin{equation}
T = 
\begin{pmatrix}
\begin{matrix}
1 &  \\
& 1
\end{matrix}
& \rvline & \begin{matrix}
1 &  \\
& 0
\end{matrix} \\
\hline
\bigzero & \rvline &
\begin{matrix}
1 &  \\
& 1
\end{matrix}
\end{pmatrix}~,
\qquad  S = 
\begin{pmatrix}
\begin{matrix}
0&  \\
&1 
\end{matrix}
& \rvline &  \begin{matrix}
-1&  \\
&0 
\end{matrix} \\
\hline
\begin{matrix}
1&  \\
&0 
\end{matrix} & \rvline &
\begin{matrix}
0&  \\
&1 
\end{matrix}
\end{pmatrix}~,
\end{equation} 
\begin{equation}
R_1 = 
\begin{pmatrix}
\begin{matrix}
& 1 \\
1 & 
\end{matrix}
& \rvline & \bigzero \\
\hline
\bigzero & \rvline &
\begin{matrix}
&1  \\
1 & 
\end{matrix}
\end{pmatrix}~, \qquad  R_2 = 
\begin{pmatrix}
\begin{matrix}
1 & 1 \\
0 & 1 
\end{matrix}
& \rvline & \bigzero \\
\hline
\bigzero & \rvline &
\begin{matrix}
1&0  \\
-1 &1 
\end{matrix}
\end{pmatrix}~.
\end{equation}
Furthermore we use the succinct notation $S[1,0]$, $S[0,1]$ to denote the gauging of $A^1$, $A^2$ (respectively) and $T[m,n]$ for the introduction of the Chern-Simons terms $m A^1dA^1 +n A^2dA^2$.

\subsubsection{Targeting two-flavor QED}
We now have all the tools to obtain two-flavour QED$_3$ via an $Sp(4,\bZ)$ action from a theory of two free fermions. The action of two-flavour QED$_3$ is \cite{Cordova:2017kue}\footnote{
	Here we are using a different charge normalization compared to \cite{Cordova:2017kue}. For example, the lowest charged gauge invariant operator is the meson $\bar{\psi}_i\psi_j$, which has charge $1$ under gauge field $A'_1$ in our case but charge $2$ under the gauge field $X$ in \cite{Cordova:2017kue}. Our choice is necessary if we want to start from \eqref{eq:free_two_flavors}, because $Sp(4,\bZ)$ respects the charge normalization. The difference between the charge-two theory and charge-one theory is that the former has fewer monopole operators. Starting with the charge-one theory, we can gauge $\bZ_2 \subset U(1)_J$, where $U(1)_J$ is the magnetic $U(1)$ global symmetry. This has the effect of changing the gauge group $G = U(1)$ to $\tilde{G}$ such that $\tilde{G}/\bZ_2 = G$. For example, in this case $G = U(1)$, and we gauge $\bZ_2\subset U(1)_J$, then the new gauge group is $\tilde{G} = U(1)$ but with the replacement of the gauge field $A_\mu \rightarrow 2A_\mu$, namely all the particle charges are multiplied by 2 \cite{Komargodski:2017keh,Gaiotto:2014kfa}. In this way we obtain the charge-two theory.}

\begin{equation}
S[A'^I,\tau'_{IJ}] + \int_{y=0} \left(i\bar{\psi}_1\cancel{D}_{a} \psi_1 +i\bar{\psi}_2\cancel{D}_{a+A'^1} \psi_2 +\frac{1}{4\pi} ada  +\frac{1}{2\pi} adA'^2 -\frac{1}{4\pi} A'^2dA'^2 \right) +2{\rm CS}_g~,
\label{eq:two_flavors}
\end{equation}
where $A'^{I=1,2}$ are bulk U(1) gauge fields while $a$ is a 3d spin$_c$ connection. The gravitational term ${\rm CS}_g$ is needed because 
\begin{equation}
\int_{\partial M} \frac{1}{4\pi} ada + 2{\rm CS}_g=2\pi \int_M \left(-\frac{1}{48} \frac{\Tr R\wedge R}{(2\pi)^2} + \frac{1}{8\pi^2}f\wedge f\right)~,
\end{equation} 
which is well-defined for a spin$_c$ connection $a$.\footnote{In the sense that this combination of boundary CS term is independent of the choices of different extensions of the boundary into bulk mod $2\pi \bZ$}

We want to target this action via an $Sp(4,\bZ)$ transformation from
\begin{equation}
S[A^I,\tau_{IJ}] +\int_{\partial M} (i\bar{\psi}_1\cancel{D}_{A^1} \psi_1 +i\bar{\psi}_2\cancel{D}_{A^2} \psi_2)~,
\label{eq:free_two_flavors}
\end{equation}
where $A^{I=1,2}$ are spin$_c$ connections. To this end, we can start from a rotation of the gauge fields by performing a GL-type transformation with 
\begin{equation}
U= \left(\begin{array}{cc}
1 & -1 \\ 
0 & 1
\end{array} \right)~,
\end{equation}
and then act with $T[1,0](-1)S[1,0]T[-1,0]$.\footnote{We follow the notation in \cite{Seiberg:2016gmd} that the minus sign in $S^2 = -1$ denotes charge conjugation.} The resulting relation between $\tau$ and $\tau'$ is
\begin{equation}\label{eq:twoflavortau}
\left(\begin{array}{cc}
\tau_{11} & \tau_{12} \\ 
\tau_{21} & \tau_{22}
\end{array}\right)=\left(
\begin{array}{cc}
-\tau'_{12}+\tau'_{22}-\frac{(\tau'_{12}+1) (-\tau'_{11}+\tau'_{21}+2)}{\tau'_{11}-1} & \frac{-\tau'_{12} (\tau'_{21}+1)+(\tau'_{11}-1) \tau'_{22}}{\tau'_{11}-1} \\
\frac{-(\tau'_{12}+1) \tau'_{21}+(\tau'_{11}-1) \tau'_{22}}{\tau'_{11}-1} & \frac{(\tau'_{11}-1) \tau'_{22}-\tau'_{12} \tau'_{21}}{\tau'_{11}-1} \\
\end{array}
\right)~.
\end{equation}
The decoupling limit of \eqref{eq:two_flavors} is
\begin{eqnarray}
\left(\begin{array}{cc}
\tau_{11}' & \tau_{12}' \\ 
\tau_{21}' & \tau_{22}'
\end{array}\right)=  \left(\begin{array}{cc}
\infty &0\\
0&\infty
\end{array}\right)~,
\end{eqnarray}
which according to \eqref{eq:twoflavortau} corresponds to 
\begin{equation}
\left(\begin{array}{cc}
\tau_{11} & \tau_{12} \\ 
\tau_{21} & \tau_{22}
\end{array}\right) = \left(\begin{array}{cc}
1+\infty & \infty \\ 
\infty & \infty
\end{array}  \right)~,\label{eq:QEDtwoflav}
\end{equation}
by which we mean $\tau_{12} - \tau_{22} = \tau_{21} - \tau_{22}= \tau_{11} -1-\tau_{22} = 0$ is satisfied while taking the limit $\tau_{22} \rightarrow \infty$.

Let us also write down explicitly the self-dualities of the theory \eqref{eq:free_two_flavors}.\footnote{Note that here we are not shifting the definition of the bulk coupling $\tau$ by $1/2$ as we did in \eqref{eq:action1fermion}. So the transformation is the same as the one presented in \cite{Seiberg:2016gmd} instead of the transformation $\tau' = -1/4\tau$ that we had in the previous section.} 
Recall from section \ref{sec:MinPhTr} that 
\begin{equation}
 S[A,\tau]+ \int_{y=0} i \bar{\psi}\cancel{D}_A\psi~,
\end{equation}
and
\begin{equation}
S[A',\tau'] + \int_{y=0} i \bar{\chi} \cancel{D}_{A'} \chi~,
\end{equation}
are equivalent when $\tau' = ST^{-2}ST^{-1}\circ\tau = (\tau-1)/(2\tau-1)$. Applying this to either $A^1$ or $A^2$ in \eqref{eq:free_two_flavors}, we obtain that the decoupling limits in the two following duality frames also correspond to two free Dirac fermions 
\begin{align}
\tau''_{IJ} &= S[1,0]T[-2,0]S[1,0]T[-1,0]\circ \tau_{IJ}~, \\
\tau'''_{IJ} &= S[0,1]T[0,-2]S[0,1]T[0,-1]\circ \tau_{IJ}~. 
\end{align}
Hence, in the variable $\tau_{IJ}$ the theory \eqref{eq:free_two_flavors} has weakly coupled cusps at 
\begin{equation}
\left(\begin{array}{cc}
\tau_{11} & \tau_{12} \\ 
\tau_{21} & \tau_{22}
\end{array}\right) =\left(\begin{array}{cc}
\infty & 0 \\ 
0 & \infty
\end{array} \right)~,\quad\left(\begin{array}{cc}
\pm\frac{1}{2} & 0 \\ 
0 & \infty
\end{array} \right)~,\quad 
\left(\begin{array}{cc}
\infty & 0 \\ 
0 & \pm\frac{1}{2}
\end{array} \right)~.\label{eq:weaklycoup}
\end{equation}

To summarize, we showed that the theory \eqref{eq:free_two_flavors} of two bulk gauge fields coupled to two Dirac fermions has two additional duality frames \eqref{eq:weaklycoup} in which the boundary theory is still the free theory of two Dirac fermions, and a duality frame \eqref{eq:QEDtwoflav} in which the boundary theory is QED$_3$ with two flavors. Clearly, additional duality frames corresponding to QED$_3$ with two flavors can be obtained by applying the transformation \eqref{eq:twoflavortau} to either of the additional free-fermions points. This is a promising setup to study QED$_3$ with two flavors via an extrapolation from the weakly-coupled points.

\section{Future Directions}\label{sec:conc}
We conclude by discussing some directions for future investigation. 
\begin{itemize}
\item{A universal feature of the setup considered in this paper is the existence of bulk line operators, whose endpoints may be attached to boundary charged operators. It is possible to assign conformal dimensions to the local operators at the location where the line defect ends on the boundary, and these dimensions can be computed perturbatively. Similarly to cusp anomalous dimensions, they are functions of the angle between the defect and the boundary. Starting with the dimensions of the endpoints of 't Hooft lines (and 't Hooft-Wilson lines) around $\tau\to\infty$ with a certain CFT on the boundary, it would be interesting to attempt an extrapolation to the cusps on the real axis, where they approach the dimensions of local monopole operators in the gauged version of the initial CFT. Concretely, in the example of section 4, from the dimension of the endpoint of a 't Hooft line around the Dirac fermion point one can attempt to recover the scaling dimension of the spin operator of the $O(2)$ model.}
\item{It would be interesting to perform perturbative calculations of anomalous dimensions and of the free energy in the theory with two bulk gauge fields presented in section \ref{sec:QEDtwoFlavors}, and attempt an extrapolation to QED$_3$ with two flavors. In particular, it is possible to use our setup to test whether this theory exists as a real CFT, by studying the dimension of four-fermion operators and checking whether they cross marginality before we reach the QED cusp, leading to the ``phase-transition'' described in section \ref{sec:Strong}.}
\item{In the model considered in section \ref{sec:MinPhTr} we have only used the two-sided extrapolations to give estimates for the $O(2)$ model. However there are infinitely many other cusps on the real axis where strongly-coupled  CFTs live, and they are of course amenable to the same extrapolation technique. These theories typically take the form of QED-CS theories, and they also describe interesting phase transitions \cite{Lee:2018udi}. A direction for the future would be to use our method to give estimates for the observables of these theories.}
\item{Finally, dualities analogous to the one considered in this paper exist for $\mathcal{N}=2$ gauge theory. One of the simplest examples is the so-called triality \cite{Intriligator:1996ex,deBoer:1996ck,deBoer:1997kr,Aharony:1997bx} generated by $ST$ transformation \cite{Dimofte:2011ju,Dimofte:2011py}, with $(ST)^3 = 1$. It would be interesting to see how the triality can improve the extrapolation. Thanks to supersymmetric localization the boundary free energy and dimensions of chiral endpoints of line operators are exactly computable \cite{Gaiotto:2014gha}. For many other interesting observables, such as the conformal dimensions of operators analogous to $O_0$, which are non-protected, one has to resort to Feynman diagrams.}
\end{itemize}

\bigskip

\bigskip

\section*{Acknowledgements }

We thank the Simons Collaboration on the Non-perturbative Bootstrap for organizing many stimulating conferences and workshops where part of this work was carried out. We are grateful to M. Baggio, N. Bobev, S. Chester, S. Cremonesi, M. Del Zotto, M. Meineri, G. Tartaglino-Mazzucchelli, E. Stamou, E. Trevisani and B. van Rees for stimulating discussions. EL thanks the Perimeter Institute for Theoretical Physics for hospitality. Research at Perimeter Institute is supported by the Government of Canada through Industry Canada and by the Province of Ontario through the Ministry of Research \& Innovation.
EL is supported by the Simons Foundation grant $\#$488659 (Simons Collaboration on the non-perturbative bootstrap).

\medskip

\noindent 


\begin{appendices}

\appendix

\section{Method of Images}\label{app:Imag}

In this appendix we show how to compute the two-point function of $F_{\mu\nu}$ in the free theory using the method of images. 

Reflections about the boundary are implemented by the matrix 
\begin{equation}
R_\mu^{~\nu} = \delta_\mu^{~\nu} - 2n_\mu n^\nu~,
\end{equation}
where $n^\mu$ is the inward pointing vector normal to the boundary. Note that the reflection of the field strength  
\begin{equation}
F_{\mu\nu}^R(x) \equiv R_\mu^{~\mu'} R_\nu^{~\nu'}F_{\mu'\nu'}(R\,x)
\end{equation} 
has components $(F^R_{ya}(x),\,\widetilde{F^R_{ya}}(x)) = (-F_{ya}(R\,x),\,\tilde{F}_{ya}(R\,x)) $. Hence, the combination
\begin{equation}
\langle F_{\mu\nu}(x_1) F_{\rho\sigma}(x_2) \rangle_{\mathbb{R}^3 \times \mathbb{R}_+}  \equiv \langle F_{\mu\nu}(x_1) F_{\rho\sigma}(x_2) \rangle_{\mathbb{R}^4} - s \langle F_{\mu\nu}(x_1) F^R_{\rho\sigma}(x_2) \rangle_{\mathbb{R}^4}~,\label{eq:image}
\end{equation}
satisfies the equation of motion and Bianchi identity for $y \geq 0$, and  also satisfies the Dirichlet (Neumann with $\gamma = 0$) boundary condition upon choosing the sign $s=1$ ($s=-1$, respectively). Even though Bose symmetry is not manifest in \eqref{eq:image}, it is satisfied because $\langle F_{\mu\nu}(x_1) F^R_{\rho\sigma}(x_2) \rangle_{\mathbb{R}^4} = \langle F^R_{\mu\nu}(x_1) F_{\rho\sigma}(x_2) \rangle_{\mathbb{R}^4}$. We can then rewrite the image term using the cross-ratio $\xi$ and the vectors $X_{i\,\mu}$ by means of the following identity
\begin{equation}
R_\rho^{~\rho'} I_{\mu\rho'}(x_1 - R x_2) = I_{\mu\rho}(x_{12}) - 2 X_{1\,\mu} X_{2\,\rho} ~.
\end{equation}
In this way we find \eqref{eq:NDIm}.

In the more general case of Neumann boundary condition with $\gamma \neq 0$, consider the combination
\begin{align}
F'_{\mu\nu} & = F_{\mu\nu} + i \gamma \tilde{F}_{\mu\nu} = \mathcal{M}_{\mu\nu}^{~~~\mu'\nu'}F_{\mu'\nu'}~ \\
\mathcal{M}_{\mu\nu}^{~~~\mu'\nu'} & =\delta_{[\mu}^{\mu'}\delta_{\nu]}^{\nu'} + i \frac{\gamma}{2}\epsilon_{\mu\nu}^{~~~\mu'\nu'}~.
\end{align}
For $F'_{\mu\nu}$ the problem is reduced to the Neumann boundary condition with $\gamma =0$, so we have
\begin{equation}
\langle F'_{\mu\nu}(x_1) F'_{\rho\sigma}(x_2) \rangle_{\mathbb{R}^3 \times \mathbb{R}_+}  \equiv \langle F'_{\mu\nu}(x_1) F'_{\rho\sigma}(x_2) \rangle_{\mathbb{R}^4} + \langle F'_{\mu\nu}(x_1) (F')^R_{\rho\sigma}(x_2) \rangle_{\mathbb{R}^4}~.\label{eq:fprime}
\end{equation}
Note that 
\begin{align}
(F')^R_{\rho\sigma}(x) & = \overline{\mathcal{M}}_{\mu\nu}^{~~~\mu'\nu'}F^R_{\mu'\nu'}~,\\
\overline{\mathcal{M}}_{\mu\nu}^{~~~\mu'\nu'}& =\delta_{[\mu}^{\mu'}\delta_{\nu]}^{\nu'} - i \frac{\gamma}{2}\epsilon_{\mu\nu}^{~~~\mu'\nu'}~.
\end{align}
Multiplying both sides of \eqref{eq:fprime} by $\mathcal{M}^{-1}\otimes\mathcal{M}^{-1}$ we obtain
\begin{equation}
\langle F_{\mu\nu}(x_1) F_{\rho\sigma}(x_2) \rangle_{\mathbb{R}^3 \times \mathbb{R}_+}  =  \langle F_{\mu\nu}(x_1) F_{\rho\sigma}(x_2) \rangle_{\mathbb{R}^4} + (\mathcal{M}^{-1}\overline{\mathcal{M}})_{\rho\sigma}^{~~~\rho'\sigma'}\langle F_{\mu\nu}(x_1) F^R_{\rho'\sigma'}(x_2) \rangle_{\mathbb{R}^4}~.
\end{equation}
Finally we use that
\begin{equation}(\mathcal{M}^{-1}\overline{\mathcal{M}})_{\rho\sigma}^{~~~\rho'\sigma'} = \frac{1-\gamma^2}{1+\gamma^2} \delta_{[\rho}^{\rho'}\delta_{\sigma]}^{\sigma'} - i \frac{\gamma}{1+\gamma^2} \epsilon_{\rho\sigma}^{~~~\rho'\sigma'}~,
\end{equation}
to write the final result for the two-point function in terms of the parameter $\gamma$ and the covariant structures $G$ and $H$, thus obtaining \eqref{eq:imagesFF}.

\section{Defect OPE of $F_{\mu\nu}$}
\label{app:detailsBulkDefOPE}
Let us consider what can appear as a primary inside the bulk-to-boundary OPE of the field strength $F_{\mu\nu}$. By spin selection rules only vectors are admitted, with two possible structures, namely
\begin{align}\label{dOPEAgen}
F_{\mu\nu}(\vec{x}, y) \underset{y\to 0}{\sim}\frac{1}{y^{2-\widehat{\Delta}_1}}\hat{V}_1^a(\vec{x})2\delta_{a[\mu}\delta_{\nu]y} - \frac{1}{y^{2-\widehat{\Delta}_2}} i \epsilon^{abc}\hat{V}_{2\,c}(\vec{x})\delta_{a [\mu }\delta_{\nu] b}+\dots
\end{align}
and the ellipsis denotes contributions from descendants. Using the bulk eom and Bianchi identity, we have that
\begin{align}
\partial_y F_{y a}\sim \frac{ (\widehat{\Delta}_1-2)}{y^{3-\widehat{\Delta}_1}}\hat{V}_{1\,a}(\vec{x})+\dots ~,\nonumber\\
\partial_y \tilde{F}_{y a}\sim -i \frac{ (\widehat{\Delta}_2-2)}{y^{3-\widehat{\Delta}_1}}\hat{V}_{2\,a}(\vec{x})+\dots ~,
\end{align}
must be boundary descendants. This requires $\widehat{\Delta}_1=\widehat{\Delta}_2=2$. We conclude that the only allowed boundary primaries are conserved currents.

To obtain the complete form of the bulk-to-boundary OPE of $F$ (including all the descendants) we first need the exact $\langle F\hat{V} \rangle $ correlator. This can be easily computed using the techniques of \cite{Lauria:2018klo} to find
\begin{align}\label{FtoV}
 \langle F_{ya}(x) \hat{V}_{i\,c}(0) \rangle & =\frac{1}{x^{4}}\,\left[\left(\frac{2 y^2 \delta_{ac}}{x^2}-I_{ac}(x)\right)c_{1i}(\tau)-  2i\, c_{2i}(\tau)\frac{y}{x^2} \epsilon_{acd}x^d\right]~,\nonumber\\
 \langle F_{ab}(x)\hat{V}_{i\,c}(0) \rangle&=\frac{1}{x^{4}}\,\left[i\left(\frac{2 y^2 \epsilon_{abc}}{x^2}-\epsilon_{abd}I^d_c(x)\right)c_{2i}(\tau)-2c_{1i}(\tau)\frac{y}{x^2} (\delta_{ac}x_b-\delta_{bc}x_a)\right]~,
  \end{align}
where $c_{ij}(\tau)$ are defined in eq. \eqref{eq:VVco}. The bulk-to-boundary OPE of $F$ can now be obtained by expanding both sides of \eqref{FtoV} to find
 \begin{align}\label{defectOPE}
 F_{ab}(\vec{x},y)=&\sum_{n=0}^\infty(-1)^{n}\left[(\delta_{ac} \delta_{bd}-\delta_{ad} \delta_{bc})y\partial^d\frac{(y^2 \vec{\partial}^{\,2})^n}{(2 n+1)!}\hat{V}_1^c(\vec{x})-i  \epsilon_{abc}\frac{(y^2 \vec{\partial}^{\,2})^n}{2n!}\hat{V}_2^c(\vec{x})\right]~,\nonumber\\
 F_{ya}(\vec{x},y)=&\sum_{n=0}^\infty(-1)^{n}\left[- \frac{(y^2 \vec{\partial}^{\,2})^n}{2n!}\hat{V}_{1\,a}(\vec{x})+i \epsilon_{abd} \, y\partial^d \frac{(y^2 \vec{\partial}^{\,2})^n}{(2n +1)!}\hat{V}_2^b(\vec{x})\right]~.
 \end{align}
 With the bulk-to-boundary OPE above, it is straightforward to obtain the $\langle F F \rangle$ 2-point function in terms of the defect CFT data as in \eqref{eq:FFfull}.

\section{Bulk OPE Limit of $\langle F_{\mu\nu} F_{\rho\sigma}\rangle$}
\label{bootapp}
Here we present some details of the bootstrap analysis presented in Section \ref{sec:resultsBoot}. To simplify computations, it is convenient to start from a configuration where the two bulk operators lie at the same parallel distance from the defect, i.e. $\vec{x}_{12}=0$. In this case some expressions in \eqref{eq:FFfull} simplify considerably, e.g.
\begin{align}
& G_{ay,by}(\vec{x}_{12}=0,y_1-y_2)  = -\frac{\delta_{ab}}{(y_1-y_2)^4}~,\\ & H_{ay,by}(\vec{x}_{12}=0,y_1,y_2)  = \frac{2X_{1\,y}X_{2\,y}\delta_{ab}}{(y_1-y_2)^4}\underset{y_1\to y_2}{\sim} -\frac{2 \delta_{ab}}{(y_1-y_2)^4}~,\\
& G_{ab,cy}(\vec{x}_{12}=0,y_1-y_2) =0 = H_{ab,c y}(\vec{x}_{12}=0,y_1,y_2)~, \\
& v^4\vert_{\vec{x}_{12}=0}  = \frac{(y_1-y_2)^4}{(y_1 + y_2)^4} \underset{y_1\to y_2}{\sim} \frac{(y_1-y_2)^4}{16 y_2^4}~.
\end{align}
It is now a simple exercise to derive the bulk OPE limit of \eqref{eq:FFfull}
\begin{align}\label{FFfull1}
\langle F_{ab}(\vec{x},y_1) F_{cy}(\vec{x},y_2) \rangle  &\underset{y_1\to y_2}{\sim} -\frac{i \alpha_3}{16 y_2^4} \epsilon_{abc} +\dots, \nonumber\\
\langle F_{ay}(\vec{x},y_1) F_{by}(\vec{x},y_2) \rangle &\underset{y_1\to y_2}{\sim} -\left(\frac{\alpha_1 }{(y_1-y_2)^4}+\frac{\alpha_2}{16 y_2^4}\right)\delta_{ab}+\dots
\end{align}
where the ellipsis denote contributions from descendants. On the other hand from \eqref{FFbulk} one finds
\begin{align}\label{2ptFromBulkOPE}
\langle F_{ab}(\vec{x},y_1) F_{cy}(\vec{x},y_2)\rangle&\underset{y_1\to y_2}{\sim}\frac{1}{12}\frac{a_{F\tilde{F}}(\tau,\bar{\tau})}{y_2^4}\epsilon_{abc}+\dots~, \nonumber\\
\langle F_{ay}(\vec{x},y_1) F_{by}(\vec{x},y_2)\rangle&\underset{y_1\to y_2}{\sim}-\left(\frac{g^2}{\pi^2}\frac{1}{(y_1-y_2)^4}-\frac{1}{12}\frac{a_F^2(\tau,\bar{\tau})}{y_2^4}\right)\delta_{ab}+\dots~.
\end{align}
Crossing symmetry now implies that \eqref{2ptFromBulkOPE} and \eqref{FFfull1} must match, therefore
\begin{align}
\alpha_1=\frac{g^2}{\pi^2},\quad a_{F^2}(\tau,\bar{\tau})=-\frac{3}{4}\alpha_2,\quad  a_{F\tilde{F}}(\tau,\bar{\tau})=-i\frac{3}{4} \alpha_3.  
\end{align}
From the solution above, upon using \eqref{eq:alphadef} one obtains
\begin{align}\label{FperpFperpComp}
c_{11}(\tau,\bar{\tau})+c_{22}(\tau,\bar{\tau})=\frac{2g^2}{\pi^2},\quad a_{F^2}(\tau,\bar{\tau})=\frac{3}{8}(c_{22}(\tau,\bar{\tau})-c_{11}(\tau,\bar{\tau})),\quad  a_{F\tilde{F}}(\tau,\bar{\tau})=i\frac{3}{4} c_{12}(\tau,\bar{\tau}).  
\end{align}

\section{Current Two-Point Functions}\label{app:current2ptfunctions}

In this appendix derive some useful relations between the two-point functions of the conserved boundary currents. 
The two-point functions of the currents $\hat{V}_i^a$ -- see \eqref{dOPEF} -- in momentum space are
\begin{equation}
\langle \hat{V}_i^a(p) \hat{V}_j^b(-p)\rangle= - \frac{\pi ^2}{2} c_{ij} p \left(\delta^{ab}-\frac{p^a p^b}{p^2}\right)+\frac{\kappa_{ij}}{2\pi} \epsilon^{abc}p_c~.
\end{equation}
The main goal is to express the coefficients $c_{ij}$ --that enter directly in the expression of the bulk two-point and one-point functions-- in terms of the two-point correlator of the current $\hat{J}^a$, which is more natural to compute in perturbation theory at large $\tau$.

In perturbation theory it is convenient to define a two-point function of $\hat{J}^a$ that cannot be disconnected by cutting a photon line, which we will call one-photon irreducible and denote with the symbol $\Sigma$
\begin{equation}
\langle \hat{J}^a(p) \hat{J}^b(-p)\rangle\vert_{\text{one-photon irr.}}\equiv \Sigma^{ab}(p)= - \frac{\pi^2}{2}c_\Sigma(\tau,\bar{\tau}) p \left(\delta^{ab}-\frac{p^a p^b}{p^2}\right)+\frac{\kappa_\Sigma(\tau,\bar{\tau})}{2\pi} \epsilon^{abc}p_c~.\label{eq:Sigma}
\end{equation}
Clearly this two-point function reduces to the two-point function of the current of the 3d CFT as $\tau\to \infty$. 
\begin{figure}
\centering
\includegraphics[clip, height=2cm]{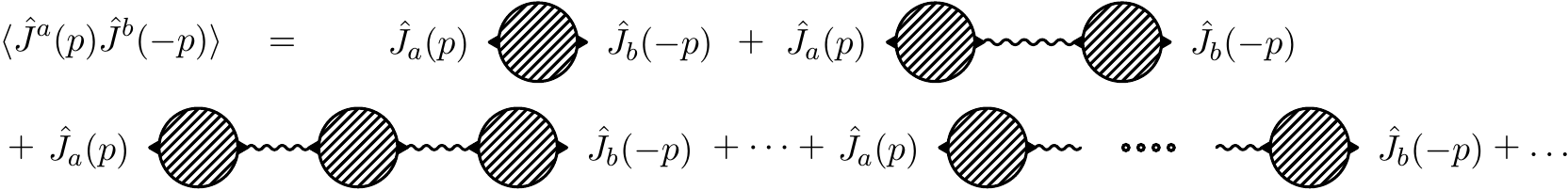}
\caption{The two-point function of the boundary current $\hat{J}$. The shaded blob represents the one-photon irreducible two-point function $\Sigma(p)$, by which we mean the sum of all the diagrams that cannot be disconnected by cutting a photon line. The full two-point function can be obtained in terms of $\Sigma$, via the geometric sum shown in the figure. \label{fig:geometric}}
\end{figure}
By resumming the diagrams in fig. \ref{fig:geometric} we obtain 
\begin{align}
\langle \hat{J}^a(p) \hat{J}^b(-p)\rangle & = \left(\Sigma(p)\cdot (\mathds{1} - \Pi(p) \cdot \Sigma(p))^{-1}\right)^{ab} \\
& =- \frac{\pi^2}{2}c_J(\tau,\bar{\tau}) p \left(\delta^{ab}-\frac{p^a p^b}{p^2}\right)+\frac{\kappa_J(\tau,\bar{\tau})}{2\pi} \epsilon^{abc}p_c~, \label{eq:JJ}
\end{align}
where $\Pi$ is the boundary propagator of the photon (see eq. \eqref{eq:prop}) and
\begin{align}
\frac{\pi^2}{2}c_J & =\frac{\frac{\pi ^2}{2} c_\Sigma \left( \frac{\pi ^2}{2} c_\Sigma  g^2+ \gamma ^2+1\right)+ \frac{g^2 \kappa_\Sigma^2}{4\pi^2}}{\left(\frac{\pi^2}{2} c_\Sigma g^2 + 1\right)^2+\left(\gamma +\frac{g^2 \kappa_\Sigma}{2\pi} \right)^2}~, \label{eq:cJfromS}\\
\frac{\kappa_J}{2\pi} & = \frac{\frac{ \gamma}{g^2}   \left(\frac{\pi ^2}{2} c_\Sigma g^2\right)^2+ \frac{\kappa_\Sigma}{2\pi} \left(\gamma ^2+\gamma  \frac{g^2 \kappa_\Sigma}{2\pi} +1\right)}{\left(\frac{\pi^2}{2} c_\Sigma g^2 + 1\right)^2+\left(\gamma +\frac{g^2 \kappa_\Sigma}{2\pi} \right)^2}~.\label{eq:kJfromS}
\end{align}

\begin{figure}
\centering
\includegraphics[clip, height=1.5cm]{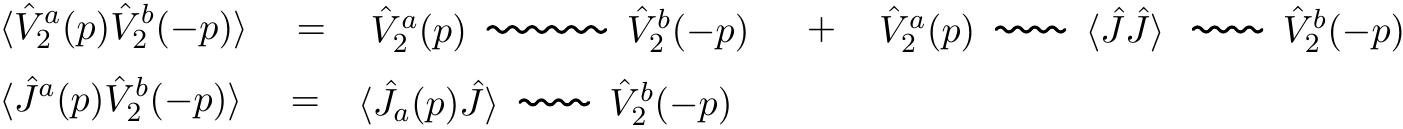}
\caption{Relations between the two-point functions involving the current $V_2$ and the two-point function $\langle J J\rangle$. The relation in the second line is only true up to a contact term. \label{fig:V2V2}}
\end{figure}

We will also need the mixed two-point function $\langle \hat{J} \hat{V}_2\rangle$ which similarly can be  parametrized as
\begin{equation}
\langle \hat{J}^a(p) \hat{V}_2^b(-p)\rangle = - \frac{\pi ^2}{2} c_{J2} p \left(\delta^{ab}-\frac{p^a p^b}{p^2}\right)+\frac{\kappa_{J2}}{2\pi} \epsilon^{abc}p_c~.
\end{equation}
Since $\hat{V}_2^a=\frac{i}{2}\, \epsilon^{abc}F_{bc}\vert_{y=0}$, we can readily express the two-point function of $\hat{V}_2$ and the mixed two-point function of $\hat{V}_2$ and $\hat{J}$ in terms of the two-point function of $\hat{J}$ and the boundary propagator of the photon, using the relations depicted in fig. \ref{fig:V2V2}. 
We obtain
\begin{align}\label{eq:V2V2}
\frac{\pi^2}{2}c_{22} & = \frac{g^2}{1+\gamma ^2}+\left(\frac{g^2}{1+\gamma ^2}\right)^2 \left(\left(\gamma ^2-1\right)\frac{\pi^2}{2} c_J-2 \gamma  \frac{\kappa _J}{2\pi}\right)~,\\
\frac{\kappa_{22}}{2\pi} & = -\frac{g^2}{1+\gamma ^2}\gamma+\left(\frac{g^2}{1+\gamma ^2}\right)^2 \left(\gamma \pi^2 c_J + \left(\gamma ^2-1\right)  \frac{\kappa _J}{2\pi}\right)~,\\
\frac{\pi^2}{2}c_{J2} & = \frac{g^2}{1+\gamma^2}\left(-\gamma \frac{\pi ^2}{2} c_J+\frac{\kappa_J}{2\pi}\right)~,\\
\frac{\kappa_{J2}}{2\pi} & =1- \frac{g^2}{1+\gamma^2}\left(\frac{\pi ^2}{2} c_J +\gamma \frac{\kappa_J}{2\pi}\right)~.
\end{align}
Finally, using that $\hat{V}_1 = -g^2 \hat{J} - \gamma \hat{V}_2$, we obtain that 
\begin{align}\label{eq:V1V1}
\frac{\pi^2}{2}c_{11} & =\frac{\pi^2}{2}\left( g^4 c_J + 2 g^2 \gamma c_{J2} +\gamma^2 c_{22}\right) \nonumber\\&=\frac{g^2}{1+\gamma ^2} \gamma ^2-\left(\frac{g^2}{1+\gamma ^2}\right)^2 \left( \left(\gamma ^2-1\right) \frac{\pi^2}{2}c_J-2 \gamma  \frac{\kappa_J}{2\pi} \right)~,\\
\frac{\kappa_{11}}{2\pi} & =  g^4 \frac{\kappa_J}{2\pi} + 2 g^2 \gamma \frac{\kappa_{J2}}{2\pi} +\gamma^2 \frac{\kappa_{22}}{2\pi} \nonumber\\& = \frac{g^2}{1+\gamma ^2} \gamma  \left(\gamma ^2+2\right)-\left(\frac{g^2}{1+\gamma ^2}\right)^2 \left(\gamma \pi^2 c_J + \left(\gamma ^2-1\right)  \frac{\kappa _J}{2\pi}\right)~,\\
\frac{\pi^2}{2}c_{12} & =- \frac{\pi^2}{2}\left( g^2 c_{J2} + \gamma c_{22} \right)\nonumber\\& =- \frac{g^2}{1+\gamma ^2} \gamma+\left(\frac{g^2}{1+\gamma ^2}\right)^2 \left(\gamma \pi^2 c_J + \left(\gamma ^2-1\right)  \frac{\kappa _J}{2\pi}\right)~,\label{eq:V1V2}\\
\frac{\kappa_{12}}{2\pi} & = - g^2 \frac{\kappa_{J2}}{2\pi} - \gamma \frac{\kappa_{22}}{2\pi}\nonumber\\& =-\frac{g^2}{1+\gamma ^2} - \left(\frac{g^2}{1+\gamma ^2}\right)^2  \left(\left(\gamma ^2-1\right)\frac{\pi^2}{2} c_J-2 \gamma  \frac{\kappa _J}{2\pi}\right)~.
\end{align}
We see that all the coefficients $c_{ij}$ can be expressed in terms of the functions of the coupling $c_J$ and $\kappa_J$ (or equivalently $c_\Sigma$ and $\kappa_\Sigma$). As a check, note that the first identity in \eqref{eq:Bootsres}, that was derived from the contribution of the identity in the bulk OPE and relates $c_{11}$ and $c_{22}$, is identically satisfied.

\section{Calculation of $\langle\hat{V}_i \hat{V}_j \hat{D}\rangle$}\label{app:VVD}

We start by computing the three-point function 
\begin{align}
\langle F_{\mu\nu}(x_1)F_{\rho\sigma}(x_2)\hat{D}(\vec{x}_3)\rangle~.
\end{align}
using the boundary channel. At leading order in the boundary OPE limit the three-point function becomes
\begin{align}
\langle\hat{V}_i^a(\vec{x}_1)\hat{V}_j^b(\vec{x}_2)\hat{D}(\vec{x}_3)\rangle~,
\end{align}
which upon placing the displacement operator at infinity simplifies to \cite{Costa:2011mg, Dymarsky:2017xzb}
\begin{align}\label{eq:VVD3pt}
\langle \hat{V}_i^a(\vec{x}_1)\hat{V}_j^b(\vec{x}_2)\hat{D}(\infty)\rangle\equiv\lim_{\vec{x}_3\rightarrow \infty}|\vec{x}_3|^8\langle \hat{V}_i^a(\vec{x}_1)\hat{V}_j^b(\vec{x}_2)\hat{D}(\vec{x}_3)\rangle= \, \lambda_{ij\hat{D}+}^{(1)}\,\, \delta^{ab}+\lambda_{ij\hat{D}-}^{(1)}\,\, {\hat{x}_{12}^c}{}\epsilon^{abc}~.
\end{align}
From the boundary OPE-channel we find 
\begin{align}\label{eq:FFDeq}
\langle F_{ay}(x_1)F_{by}(x_2)\hat{D}(\infty)\rangle& = \, \lambda_{11\hat{D}+}^{(1)}\,\, \delta_{ab}+\lambda_{11\hat{D}-}^{(1)}\,\,( {\hat{x}_{12}^f}{}\epsilon_{abf}+\dots)~,\\
\langle F_{ay}(x_1)F_{bc}(x_2)\hat{D}(\infty)\rangle& = -i\,\epsilon_{bc}{}^{e} (\lambda_{12\hat{D}+}^{(1)}\,\, \delta_{ae}+\lambda_{12\hat{D}-}^{(1)}\,\, ({\hat{x}_{12}^f}{}\epsilon_{aef}+\dots))~,\\
\langle F_{ab}(x_1)F_{cd}(x_2)\hat{D}(\infty)\rangle& = -\, \epsilon_{ab}{}^{e}\epsilon_{cd}{}^{g}(\lambda_{22D+}^{(1)}\delta_{eg}+\lambda_{22D-}^{(1)}\,\, (\hat{x}_{12}^f\epsilon_{egf}+\dots))~,
\end{align}
where the ellipses denote the descendant contributions from the second term of \eqref{eq:VVD3pt}, which are proportional to $\lambda_{ij\hat{D}-}^{(1)}$ and will not play any role in the following.  

Next, we compute the three-point function using the bulk OPE channel. The Lorentz spin and scaling dimensions of the full set of operators appearing in the OPE of two $F$'s can be found in \cite{Beccaria:2014zma} -- see eq. (2.12) therein -- where they are discussed in the context of the so-called minimal type-C higher spin theory on AdS$_5$, the bulk dual to the free Maxwell CFT$_4$. All the operators with scaling dimension $\Delta >4$ in this OPE are higher-spin conserved currents (there is both a family of symmetric traceless tensors and a family of mixed-symmetry ones), and in addition there is the identity operator and a few operators of scaling dimension $\Delta = 4$: the scalar operators $F^2$ and $F\tilde{F}$, the stress tensor $T_{\mu\nu} = (\frac{1}{g^2}F_{\mu \rho}F^{~\rho}_\nu - {\rm trace})$, and a non-conserved operator in the representation $(2,0)\oplus(0,2)$ of rotations, i.e. a tensor with four indices and the same symmetry and trace properties of a Weyl tensor, for this reason we will denote it as $W_{\mu\nu\rho\sigma}$. The three-point function in the bulk OPE channel is written as a sum of the bulk-boundary two-point functions between these operators and the displacement operator. Let us analyze which of these two-point functions can contribute. First of all, it is easy to see that two-point function between the conserved higher-spin currents and the displacement operator must vanish. This is an instance of the more general statement that in boundary CFTs bulk conserved currents $J$ can only have non-zero two-point functions with a scalar boundary operator $\hat{O}$ that has the same scaling dimension. The latter statement can be easily proved by placing the boundary operator at infinity, because in this case invariance under scaling and parallel translations force the two-point function to take the schematic form
\begin{equation}
\langle J(y, \vec{x}) \hat{O}(\infty) \rangle = b_{J\hat{O}}\frac{1}{y^{\Delta_J -\Delta_{\hat{O}}}} ~(\text{structure})~,
\end{equation}
where ``structure'' denotes an appropriate tensor built out of the $\delta^{\mu\nu}$, the unit normal vector $n^\mu$ and possibly epsilon tensors. Clearly when $\Delta_J \neq \Delta_{\hat{O}}$ this two-point function cannot be compatible with current conservation unless the coefficient $b_{J\hat{O}}$ vanishes. Moreover, rotational invariance \eqref{eq:DWI} implies that also the operator $W_{\mu\nu\rho\sigma}$ has vanishing two-point function with the displacement.\footnote{To see this, consider the projector on the $(2,0)$ representation
\begin{align}
(P^{(2,0)})_{\mu\nu\rho\sigma}^{~~\mu'\nu'\rho'\sigma'} \equiv \frac{1}{2}P_{\mu\nu}^{+~\mu'\nu'}P_{\rho\sigma}^{+~\rho'\sigma'} + \frac{1}{2} P_{\rho\sigma}^{+~\mu'\nu'}P_{\mu\nu}^{+~\rho'\sigma'} - \frac{1}{3} P_{\mu\nu,\rho\sigma}^+ P^{+\,\mu'\nu',\rho'\sigma'}~.
\end{align}
Since the two-point function between $W_{\mu\nu\rho\sigma}(x)$ and $\hat{D}(\infty)$ is a constant, the allowed structures are obtained by acting with this projector on constant four-tensors built out of $\delta$ and $\epsilon$, such as: $\delta_{\mu'\rho'}\delta_{\nu'\sigma'}$, $\delta_{\mu'\rho'}\delta_{\nu' y}\delta_{\sigma' y}$, $\epsilon_{\mu'\nu'\rho'\sigma'}$, $\epsilon_{\mu'\nu'\rho' y}\delta_{\sigma' y}$. Applying the projector to any of these structures we find 0.
 }
Therefore, the only bulk operators that can contribute to the three-point function are the scalar operators and the stress-tensor. When the displacement is placed at infinity, the corresponding two-point functions are
\begin{align}
\langle F^2(x)\hat{D}(\infty)\rangle & =b_{F^2, \hat{D}}~,\label{eq:twoptD1}\\
\langle F\tilde{F}(x)\hat{D}(\infty)\rangle & =b_{F\tilde{F},\hat{D}}~,\label{eq:twoptD2}\\
\langle T_{\mu\nu}(x)\hat{D}(\infty)\rangle & =b_{T, \hat{D}}\,\left( \delta_{\mu y}\delta_{\nu y}-\frac14\delta_{\mu\nu}\right)~.\label{eq:twoptD3}
\end{align}
Using the OPE \eqref{eq:TOPED} and the Ward identity \eqref{eq:DWI} we can express the above two-point function coefficients in terms of the one-point function of the scalar operators, and of the coefficient $C_{\hat{D}}$ in the two-point function of the displacement, namely \cite{McAvity:1995zd,Gliozzi:2015qsa,Billo:2016cpy} 
\begin{align} 
b_{F^2, \hat{D}} & = -\frac{32a_{F^2}}{\pi^2}~,\label{eq:twoCoeff1}\\
b_{F\tilde{F}, \hat{D}} & = -\frac{32a_{F\tilde{F}}}{\pi^2}~,\label{eq:twoCoeff2}\\
b_{T, \hat{D}} & = \frac{4\,C_{\hat{D}}}{3}~.\label{eq:twoCoeff3}
\end{align}
Since the two-point functions are constant, we can simply plug in the three-point function the leading bulk OPE, ignoring the descendants (and also ignoring the singular contribution from the identity that drops from the three-point function)
\begin{align}
F_{\mu\nu}(x) F^{\rho\sigma}(0)\underset{x\to 0}{\sim} \frac{1}{12}(\delta_\mu^\rho\delta_\nu^\sigma-\delta_\nu^\rho\delta_\mu^\sigma)F^2(0)+\frac{1}{12}\epsilon_{\mu\nu}^{\rho\sigma}F\tilde{F}\,(0) + 2g^2 \delta_{[\mu}^{[\rho}T_{\nu]}^{\sigma]}(0)~.
\end{align}
Using eq.s \eqref{eq:twoCoeff1}-\eqref{eq:twoCoeff2} in the two-point functions, we find
\begin{align}\label{eq:FFDbulk}
\langle F_{ay}(x_1)F_{by}(x_2)\hat{D}(\infty)\rangle =&-\left(\frac{8}{3\pi^2}a_{F^2}-\frac{g^2}{3}C_{\hat{D}}\right)\delta_{ab}~,\\
\langle F_{ab}(x_1)F_{cd}(x_2)\hat{D}(\infty)\rangle =& -\left(\frac{8}{3\pi^2}a_{F^2}+\frac{g^2}{3}C_{\hat{D}}\right)\epsilon_{abe}\epsilon_{cde}~,\\
\langle F_{ay}(x_1)F_{bc}(x_2)\hat{D}(\infty)\rangle =&-\frac{8}{3\pi^2}a_{F\tilde{F}}\,\,\epsilon_{abc}~.
\end{align}
Finally, by comparing \eqref{eq:FFDbulk} with \eqref{eq:FFDeq} we find \eqref{eq:3ptfinal}. 

\section{Dimension of the Boundary Pseudo Stress Tensor}\label{app:fakestress}
In section \ref{sec:MinPhTr} we mentioned that the conservation of the stress tensor of the 3d CFT is violated at $g\neq 0$ due to multiplet recombination. At $g\neq 0 $ we will call this operator boundary pseudo stress tensor. This is expected from the Ward identities derived in \cite{Billo:2016cpy}. In this Appendix we exploit this idea, to reproduce the one loop result of \eqref{eq:anomalousO2}. We start from the boundary Lagrangian of a 3d Dirac fermion $\psi$ 
\begin{align}
\mathcal{L}= i \,\bar{\psi}\cancel{D}_A \psi,
\end{align}
where $D_a \psi= (\partial_a- i A_a) \psi$ and $D_a \bar\psi= (\partial_a+ i A_a) \bar\psi$. The algebra of gamma matrices is $\{\gamma_a,\gamma_b\}=2\delta_{ab}$. The pseudo boundary stress tensor is
\begin{equation}
(O_2)_{ab}=\frac{i}{2}[\bar{\psi}\gamma_{(a}D_{b)}\psi-D_{(a}\bar{\psi}\gamma_{b)}\psi],
\end{equation}
where the symmetrization includes a factor of $1/2$. Note that the above operator is traceless as a consequence of the equations of motion:
\begin{equation}
\gamma^a D_a \psi=0\, \quad  D_a\bar{\psi} \gamma^a=0.
\end{equation}
Using $[D_a,D_b]\psi=-i F_{ab}$ we obtain
\begin{equation}\label{eq:Odef}
\partial_a O_2^{ab}= F^{ab}\bar{\psi}\gamma_a \psi,
\end{equation}
In the decoupling limit $g\rightarrow 0$ the two-point function of $F_{ab}$ vanishes, hence effectively the right-hand side of \eqref{eq:Odef} is 0 and the operator $O_2^{ab}$ becomes a proper stress tensor for the boundary theory, with conformal dimension $\Delta_{2} = 3$. Upon turning on $g$, this dimension must be lifted from the unitarity bound, i.e. $\Delta_{2} (g)=3+g^2 \Delta^{(2)}_{2}+O(g^4)$. The two-point function of ${O_2}$ is fixed by 3d conformal invariance to be
\begin{align}\label{eq:tautau}
\langle {O_2}^{ab}(\vec{x}) {O_2}^{cd}(0)\rangle & =\frac{C_{2}(g)}{|\vec{x}|^{2\Delta_{2} (g)}}I^{ab,cd}(\vec{x})~,\nonumber\\
I^{ab,cd}(\vec{x})= \frac{1}{2}[I^{\rm 3d\,ac}(\vec{x})I^{\rm 3d\,bd}(\vec{x}) & +I^{\rm 3d\,ad}(\vec{x})I^{\rm 3d\,bc}(\vec{x})]-\frac{1}{3}\delta_{ab}\delta_{cd}~,
\end{align}
with $I^{\rm 3d\,ac}(\vec{x})$ defined in \eqref{eq:I3d} and $C_{2}(g)=c_{2}^{(0)}+g^2 c_{2}^{(2)}+O(g^4)$, being $c_{2}^{(0)}=\frac{3}{16\pi^2}$ the central charge for a single free 3d Dirac fermion \cite{Osborn:1993cr}. Furthermore the recombination rule \eqref{eq:Odef} tells us
\begin{equation}\label{Anselmi}
  \langle \partial_a {O_2}^{ab}(\vec{x})\, \partial_c O_{2}^{cd}(0)\rangle = \langle (F^{ab}\bar{\psi}\gamma_a \psi) (\vec{x})  (F^{cd}\bar{\psi}\gamma_{c} \psi) (0) \rangle.
\end{equation}
On one hand, the r.h.s. of \eqref{Anselmi} can be computed at three level using \eqref{eq:twopbpos} with the result
\begin{align}\label{rhs}
\langle (F^{c a}\bar{\psi}\gamma_c \psi) (\vec{x})  (F^{d b}\bar{\psi}\gamma_{d} \psi) (0) \rangle=\frac{{\color{black} 4} g^2 c_J^{(0)}}{\pi^2}\frac{I^{\rm 3d\,ab}(\vec{x})}{|\vec{x}|^8}+O(g^4),
\end{align}
where $c_J^{(0)}= \frac{1}{8\pi^2}$ is the central charge for the $U(1)$ conserved current $\hat{J}_a=\bar{\psi}\gamma_a \psi$ of a free 3d Dirac fermion \cite{Osborn:1993cr}. 

On the other hand, taking two derivatives of \eqref{eq:tautau} and expanding to the lowest non trivial order in $g$ gives
\begin{align}\label{lhs}
\langle \partial_c {O_2}^{ca}(\vec{x})\, \partial_d {O_2}^{db}(0)\rangle=\frac{10}{3}g^2c_{2}^{(0)} \Delta^{(2)}_{2} \frac{I^{\rm 3d\,ab}(\vec{x})}{|\vec{x}|^8}+O(g^4).
\end{align}
 Hence the above result, together with \eqref{Anselmi} and \eqref{lhs} fixes the anomalous dimension of $O_2$ up to $O(g^4)$ terms to be
\begin{equation}
\Delta_{2}(g)=3+\frac{\color{black} 6}{5 \pi^2}\frac{c_J^{(0)}}{c_{2}^{(0)}} g^2+O(g^4)=3+\frac{\color{black} 4}{5 \pi^2} g^2+O(g^4),
\end{equation}
in agreement with \eqref{eq:anomalousO2}.

\section{Two-loop Integrals} \label{app:FeynInt}

In the perturbative calculations of anomalous dimensions we encountered two-loop diagrams with operator insertions at zero-momentum and two external legs. After performing tensor reduction to get rid of the numerators, the resulting integrals always take the form of a two-loop massless two-point integral, namely
\begin{align}
&G(n_1,n_2,n_3,n_4,n_5) \equiv (4\pi)^d(k^2)^{n_1+n_2+n_3+n_4+n_5 -d} \nonumber \\ 
&~~~~~~~~~~~~~~~\times\int \frac{d^d p}{(2\pi)^d} \frac{d^d q}{(2\pi)^d}\frac{1}{(p^2)^{n_1}(q^2)^{n_2}((k+p)^2)^{n_3}((k+q)^2)^{n_4}((p-q)^2)^{n_5}}~.
\end{align}
$k$ here is the external momentum associated to the two external legs, and $p$ and $q$ are the loop momenta. The powers $n_i$ depend on the diagram we are considering (and in fact each diagrams will give rise to a linear combination of $G$'s with several different sets of $n_i$'s after reducing the numerators). In order to extract the two-loop renormalization constants we need to find the $1/\epsilon^2$ and $1/\epsilon$ poles in the $\epsilon\to 0$ expansion of the constants $G(n_1,n_2,n_3,n_4,n_5)$, evaluated at $d=3-2\epsilon$. (The coefficient of $1/\epsilon^2$ are fixed by one-loop data, so they do not contain new information.) 

The function $G(n_1,n_2,n_3,n_4,n_5)$ enjoys a large group of symmetries \cite{Barfoot:1987kg} that allows to relate its values at different sets of quintuples of powers. Some of the symmetries are manifest from the definition, e.g. $G(n_1,n_2,n_3,n_4,n_5)=G(n_2,n_1,n_4,n_3,n_5)=G(n_3,n_4,n_1,n_2,n_5)=G(n_4,n_3,n_2,n_1,n_5)$. When one or more of the $n_i$'s vanish, there is a closed expression for $G(n_1,n_2,n_3,n_4,n_5)$ in terms of gamma functions. When all of the $n_i$'s are integer, the strategy to compute $G(n_1,n_2,n_3,n_4,n_5)$ is to use integration-by-parts identities \cite{Chetyrkin:1981qh,Tkachov:1981wb} to lower the positive $n_i$'s, until the result is reduced to a linear combination of $G$'s with at least one vanishing entry. However, due to the $1/|p|$ ``non-local'' propagator of the photon restricted to the boundary, in our setup we encounter diagrams in which two of the $n_i$'s are half-integer, and the remaining three are integer.\footnote{Specifically, this happens for the diagrams that compute the coefficient of $\frac{({\rm Im}\tau)^2}{|\tau|^2}$ in the two-loop anomalous dimensions. The diagrams that compute the coefficient of $\frac{({\rm Re}\tau)^2}{|\tau|^2}$ do have only integer powers, and in fact they are the same as the diagrams in large-$k$ perturbation theory of CS-matter theories that compute the leading corrections to parity-even observables.} In this case it might be impossible to reduce to the case of a vanishing power using integration-by-parts, and a further input is needed. The paper \cite{Broadhurst:1996ur} derived a closed formula for $G(n_1,n_2,n_3,1,1)$ (and symmetry-related cases), with generic real $n_1,n_2,n_3$, in terms of the generalized hypergeometric function ${}_3F_2$. To recover the $1/\epsilon^2$ and $1/\epsilon$ poles from the result of \cite{Broadhurst:1996ur}, one needs to perform a Taylor expansion of the ${}_3F_2$ in its parameters. This is typically hard to do analytically, but the algorithm of \cite{Huang:2012qz} can be used to expand numerically to very high precision. 

The strategy that we used is then to reduce all of the integrals that we encountered to a small number of ``master integrals'' using integration-by-parts identities. These master integrals have the property that they can be evaluated with the formula in \cite{Broadhurst:1996ur}, and that using the numerical expansion we can easily recognize the values of the coefficients. To compute anomalous dimensions in the fermion theory of section \ref{sec:MinPhTr} we used the following two master integrals
\begin{align}
G(1,\tfrac12,\tfrac12,1,1) & \underset{\epsilon\to0}{\sim} \frac{0}{\epsilon^2} +  \frac{0}{\epsilon} + \mathcal{O}(1)~,\\
G(1,\tfrac32,\tfrac12,1,1) & \underset{\epsilon\to0}{\sim} \frac{0}{\epsilon^2} +  \frac{4}{\pi \epsilon} + \mathcal{O}(1)~.
\end{align}
We never needed the $1/\epsilon$ coefficient of the master integral in the second line, and the only case in which we needed its $1/\epsilon^2$ coefficient is in the check that the gauged current does not get any anomalous dimension. So all of our non-trivial results only depend on the master integral in the first line. In the scalar theory of section \ref{sec:Scalar} we also encountered the integral $G(1,\tfrac12,\tfrac12,1,2)$, which we were not able to compute with this strategy.

We will now give the result that we found for the contribution of each diagram to the renormalization constants. We make reference to the labeling of the diagrams in figure \ref{fig:loopdiagram}. In the two-loop calculation we also need to consider the one-loop diagram with the insertions of one-loop counterterms for the vertex or for the internal fermion lines, and we refer to this contribution as ``$\text{c.t.}$''. We denote $L \equiv  \log (\pi  \mu^2) -\gamma_E  $ where $\gamma_E$ is the Euler constant and $\mu$ is the scale introduced by dimensional regularization. Locality of counterterms requires that the $L$-dependence must cancel from the coefficient of the $1/\epsilon$ pole when all the diagrams are summed up, but generically it will be present in single diagrams. The cancelation of the $L$-dependence (and also the cancelation of $\xi$ in the gauge-invariant quantities) in the sum of all the diagrams is a check of the calculation.
\begin{itemize}
	\item{Wavefunction renormalization of the fermion: denoting the external momentum running on the fermion line with $k$, all the diagrams are proportional to $\slashed{k}$, with coefficients 
		\begin{align}
		(a) & =\frac{g^2}{1+\gamma^2}  \frac{(2-3 \xi )}{12 \pi ^2 \epsilon }~,\\
		(b.1) & =\frac{g^4}{(1+\gamma^2)^2}\left(\frac{(2-3 \xi)^2}{288 \pi ^4 \epsilon ^2}(1+2 \epsilon L)
		+\frac{63 \xi ^2-90 \xi +32 }{432 \pi ^4 \epsilon } + \frac{\gamma^2}{96 \pi^2 \epsilon}\right)~,\\
		(b.2) & = \frac{g^4}{(1+\gamma^2)^2}\left(-\frac{(2-3 \xi)^2}{144 \pi ^4 \epsilon ^2}(1+2\epsilon L)
		-\frac{117 \xi ^2-168 \xi +64}{432 \pi ^4 \epsilon } - \frac{\gamma^2}{192 \pi^2 \epsilon} \right)~,\\
		(b.3) & =-\frac{g^4}{(1+\gamma^2)^2} \frac{1-\gamma^2}{192 \pi ^2 \epsilon } ~,\\
		\text{c.t.} & = \frac{g^4}{(1+\gamma^2)^2}\left(\frac{ (2-3 \xi )^2}{144 \pi ^4 \epsilon ^2}(1+\epsilon L)+\frac{  54 \xi ^2-78 \xi +28}{432 \pi ^4 \epsilon }\right)~.
		\end{align}
		Requiring the divergence to cancel with $-\delta((Z_\psi)^2) \slashed{k}$, we obtain eq. \eqref{eq:wf}.}
	\item{Anomalous dimension of $O_0$: summing over all possible insertions in the given topology, the diagrams give
		\begin{align}
		(a) & =\frac{g^2}{1+\gamma^2}\frac{2+ \xi}{4 \pi ^2 \epsilon }~,\\
		(b.1) & =\frac{g^4}{\left(\gamma ^2+1\right)^2} \left(\frac{(2+ \xi ) (10-3 \xi)}{96 \pi ^4 \epsilon ^2}(1+2\epsilon L)
		-\!\frac{27 \xi ^2-86 \xi -232}{144 \pi ^4\epsilon }+\! \frac{\gamma^2}{32 \pi^2 \epsilon}\right)~,\\
		(b.2) & = \frac{g^4}{\left(\gamma ^2+1\right)^2} \left(\!\!-\frac{(2+\xi) (2-3\xi)}{48 \pi ^4 \epsilon ^2}(1+2\epsilon L)
		+\!\frac{63 \xi ^2+40 \xi -112 }{144 \pi ^4\epsilon }+\!\frac{3\gamma^2}{64 \pi^2 \epsilon}\right)~,\\
		(b.3) & =-\frac{g^4}{\left(\gamma ^2+1\right)^2}\frac{5-5\gamma^2}{64 \pi ^2 \epsilon }~,\\
		\text{c.t.} & = \frac{g^4}{(1+\gamma^2)^2}\left(-\frac{(2+\xi)^2}{16 \pi ^4 \epsilon ^2}(1+\epsilon L)-\frac{2\xi^2 + 7\xi +6}{8 \pi ^4\epsilon}\right)~.
		\end{align}
		Requiring the divergence to cancel with $\delta((Z_\psi)^2 Z_0)$, we obtain eq. \eqref{eq:Z0}.}
	\item{Anomalous dimension of $O_2$: we sum over all possible insertions in the given topology. The diagrams are proportional to the tree-level insertion of $O_2$ (see fig. \ref{fig:Fermionfeynrule_ops}) with the following coefficients
		\begin{align}
		(a) & =-\frac{g^2}{1+\gamma^2}  \frac{34-15\xi}{60 \pi ^2 \epsilon }~,\\
		(b.1) & =\frac{g^4}{(1+\gamma^2)^2} \left(-\frac{225 \xi ^2-300 \xi +4}{7200 \pi ^4 \epsilon ^2}(1+2\epsilon L)
		\right. \nonumber \\ &~~~~~~~~~~~~~~~~~~~~~~~~~~~~~~~~~~~~~~~~~~\left.-\frac{5175 \xi ^2-12690 \xi +4096 }{54000 \pi ^4\epsilon }- \frac{\gamma^2}{240\pi^2 \epsilon} \right)~,\\
		(b.2) & = \frac{g^4}{(1+\gamma^2)^2}\left(\frac{45 \xi ^2-132 \xi +116}{720 \pi ^4 \epsilon ^2}(1+2\epsilon L)
		\right. \nonumber \\ &~~~~~~~~~~~~~~~~~~~~~~~~~~~~~~~~~~~~~~~~~~\left.+\frac{1305 \xi ^2-6432 \xi +8416}{10800 \pi ^4\epsilon }- \frac{\gamma^2}{960\pi^2 \epsilon}\right)~,\\
		(b.3) & =\frac{g^4}{(1+\gamma^2)^2} \frac{29-5\gamma^2}{960 \pi ^2 \epsilon } ~,\\
		\text{c.t.} & = \frac{g^4}{(1+\gamma^2)^2} \left(-\frac{(15 \xi -34)^2}{3600 \pi ^4 \epsilon ^2}(1+\epsilon L)-\frac{675 \xi ^2-9735 \xi +18598}{27000 \pi ^4\epsilon }\right)~.
		\end{align}
		Requiring the divergence to cancel with $\delta((Z_\psi)^2 Z_2)$, we obtain eq. \eqref{eq:Z2}.}
\end{itemize}

\end{appendices}

\cleardoublepage

\bibliography{references}
\bibliographystyle{JHEP}

\end{document}